\documentclass[journal]{IEEEtran}
\usepackage{latexsym}
\usepackage{cite}
\usepackage{amssymb}
\usepackage{bbm}
\usepackage{bm}
\usepackage{comment}
\usepackage[usenames,dvipsnames]{color}
\usepackage{multicol}
\usepackage{lipsum}
\usepackage{dsfont}
\usepackage[dvips]{graphics}
\usepackage{graphicx}
\usepackage{amsmath}
\usepackage{psfrag}
\usepackage{lipsum}
\usepackage{subfig}
\usepackage{setspace}
\usepackage{algorithm}
\usepackage{algpseudocode}
\usepackage{float}
\usepackage{mathtools}
 \allowdisplaybreaks

\newtheorem{theorem}{Theorem}

\newtheorem{definition}{Definition}

\newcommand{\supp}{\mbox{supp}}
\newtheorem{lemma}{Lemma}
\newtheorem{remark}{Remark}

\newtheorem{corollary}[theorem]{Corollary}

\newcommand\numberthis{\addtocounter{equation}{1}\tag{\theequation}}
\catcode`~=11 
\newcommand{\urltilde}{\kern -.15em\lower .7ex\hbox{~}\kern .04em}
\catcode`~=13 

\begin{document}

\makeatletter
\newcommand{\vasti}{\bBigg@{3}}
\newcommand{\vast}{\bBigg@{4}}
\newcommand{\Vast}{\bBigg@{5}}
\makeatother
\newcommand{\be}{\begin{equation}}
\newcommand{\ee}{\end{equation}}
\newcommand{\ba}{\begin{align}}
\newcommand{\ea}{\end{align}}
\newcommand{\baa}{\begin{align*}}
\newcommand{\eaa}{\end{align*}}
\newcommand{\bea}{\begin{eqnarray}}
\newcommand{\eea}{\end{eqnarray}}
\newcommand{\beaa}{\begin{eqnarray*}}
\newcommand{\eeaa}{\end{eqnarray*}}
\newcommand{\p}[1]{\left(#1\right)}
\newcommand{\pp}[1]{\left[#1\right]}
\newcommand{\ppp}[1]{\left\{#1\right\}}
\newcommand{\ber}{$\ \mbox{Ber}$}
\newcommand\aatop[2]{\genfrac{}{}{0pt}{}{#1}{#2}}

\title{Broadcast Channels with Privacy Leakage Constraints}

\author{Ziv Goldfeld, \emph{Student Member, IEEE}, Gerhard Kramer, \emph{Fellow, IEEE}, and Haim H. Permuter, \emph{Senior Member, IEEE}
\thanks{Z. Goldfeld and H. H. Permuter were supported in part by European Research Council under the European Union's Seventh Framework Programme (FP7/2007-2013)/ERC grant agreement n$^\circ$337752, in part by the Israel Science Foundation and in part by the Cyber Security Research Center within the Ben-Gurion University of the Negev. G. Kramer was supported by an Alexander von Humboldt Professorship endowed by the German Federal Ministry of Education and Research.
\newline Z. Goldfeld and H. H. Permuter are with the Department of Electrical and Computer Engineering, Ben-Gurion University of the Negev, Beer-Sheva, Israel (gziv@post.bgu.ac.il, haimp@bgu.ac.il). G. Kramer is with the Institute for Communications Engineering, Technical University of Munich, Munich D-80333, Germany (gerhard.kramer@tum.de).}}
\maketitle


\begin{abstract}
The broadcast channel (BC) with one common and two private messages with leakage constraints is studied, where leakage rate refers to the normalized mutual information between a message and a channel symbol string. Each private message is destined for a different user and the leakage rate to the other receiver must satisfy a constraint. This model captures several scenarios concerning secrecy, i.e., when both, either or neither of the private messages are secret. Inner and outer bounds on the leakage-capacity region are derived when the eavesdropper knows the codebook.  The inner bound relies on a Marton-like code construction and the likelihood encoder. A Uniform Approximation Lemma is established that states that the marginal distribution induced by the encoder on each of the bins in the Marton codebook is approximately uniform. Without leakage constraints the inner bound recovers Marton's region and the outer bound reduces to the UVW-outer bound. The bounds match for semi-deterministic (SD) and physically degraded (PD) BCs, as well as for BCs with a degraded message set. The leakage-capacity regions of the SD-BC and the BC with a degraded message set recover past results for different secrecy scenarios. A Blackwell BC example illustrates the results and shows how its leakage-capacity region changes from the capacity region without secrecy to the secrecy-capacity regions for different secrecy scenarios.
\end{abstract}


\begin{IEEEkeywords}
Broadcast channel, Marton's inner bound, Privacy Leakage, Secrecy, Physical-layer Security.
\end{IEEEkeywords}

%
\section{Introduction}\label{SEC:introduction}

\begin{figure}[t!]
    \begin{center}
        \begin{psfrags}
            \psfragscanon
            \psfrag{I}[][][0.83]{$\mspace{-40mu}(\mspace{-1.5mu}M_0\mspace{-1.5mu},\mspace{-2mu}M_1\mspace{-1.5mu},\mspace{-2mu}M_2\mspace{-1.5mu})$}
            \psfrag{J}[][][0.9]{\ \ \ \ \ \ \ Enc $f^{(n)}$}
            \psfrag{K}[][][1]{\ \ \ $\mathbf{X}$}
            \psfrag{T}[][][0.9]{\ \ \ \ \ \ \ \ Channel}
            \psfrag{M}[][][1]{\ \ \ $\mathbf{Y}_1$}
            \psfrag{N}[][][1]{\ \ \ $\mathbf{Y}_2$}
            \psfrag{O}[][][0.9]{\ \ \ \ \ \ \ Dec $\phi^{(n)}_1$}
            \psfrag{P}[][][0.9]{\ \ \ \ \ \ \ Dec $\phi^{(n)}_2$}
            \psfrag{Q}[][][0.85]{\ \ \ \ \ \ \ \ \ \ \ $(\hat{M}_0^{(1)},\hat{M}_1)$}
            \psfrag{R}[][][0.85]{\ \ \ \ \ \ \ \ \ \ \ $(\hat{M}_0^{(2)},\hat{M}_2)$}
            \psfrag{X}[][][0.9]{\ \ \ \ \ \ \ \ \ $W_{Y_1,Y_2|X}$}
            \psfrag{V}[][][0.85]{\ \ \ \ \ \ \ \ \ \ \ \ \ $\ell_2(c_n)\mspace{-1.5mu}\leq \mspace{-1.5mu} nL_2$}
            \psfrag{U}[][][0.85]{\ \ \ \ \ \ \ \ \ \ \ \ \ $\ell_1(c_n)\mspace{-1.5mu}\leq \mspace{-1.5mu} nL_1$}
            \includegraphics[scale = .37]{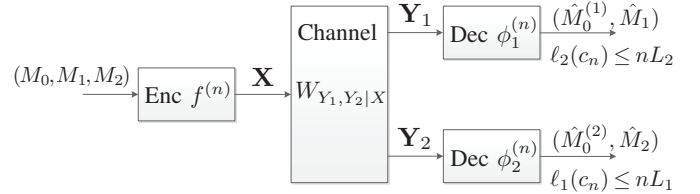}
            \caption{A BC with a common message and privacy leakage constraints $\ell_1(c_n)\triangleq I_{c_n}(M_1;\mathbf{Y}_2)\leq nL_1$ and $\ell_2(c_n)\triangleq I_{c_n}(M_2;\mathbf{Y}_1)\leq nL_2$, where $I_{c_n}$ denotes that the mutual information term are taken with respect to the distribution induced by the code $c_n=\left(f^{(n)},\phi_1^{(n)},\phi_2^{(n)}\right)$.} \label{FIG:general_BC_leakage}
            \psfragscanoff
        \end{psfrags}
     \end{center}
 \end{figure}

Public and confidential messages are often transmitted over the same channel. However, the underlying principles for constructing codes without and with secrecy are different. Without secrecy constraints, codes should use all available channel resources to reliably convey information to the destinations. Confidential messages, on the other hand, require that some channel resources are allocated to preserve security. We study relationships between the coding strategies and the fundamental limits of communication with and without secrecy. To this end we simultaneously account for secret and non-secret transmissions over a two-user broadcast channel (BC) by means of privacy leakage constraints (Fig.~\ref{FIG:general_BC_leakage}).



\subsection{Past Work}

Information theoretic secrecy was introduced by Shannon \cite{Shannon_Secrecy1949} who studied communication between a source and a receiver in the presence of an eavesdropper. Wyner modeled secret communication over noisy channels (also known as physical layer security) when he introduced the degraded wiretap channel (WTC) and derived its secrecy capacity \cite{Wyner_Wiretap1975}. Csisz{\'a}r and K{\"o}rner \cite{Csiszar_Korner_BCconfidential1978} extended Wyner's result to a general BC where the source also transmits a common message to both users. The development of wireless communication, whose inherent open nature makes it vulnerable to security attacks, has inspired a growing interest in the fundamental limits of secure communication.

Multiuser settings with secrecy were extensively treated in the literature. Broadcast and interference channels with two confidential messages were studied in \cite{BC_Confidential_Yates2008}, where inner and outer bounds on the secrecy-capacity region of both problems were derived. The secrecy-capacity region for the semi-deterministic (SD) BC was established in \cite{Semi-det_BC_secrect_two2009}. The capacity region of a SD-BC where only the message of the stochastic user is kept secret from the deterministic user was derived in \cite{Semi-det_BC_secrect_one2009}. The opposite case, i.e., when the message of the deterministic user is confidential, was solved in \cite{Goldfeld_strong_secrecy_cooperation2016}. Secret cooperative communication was considered in \cite{Ulukus_Cooperative_RBC2011}, where the authors derive inner and outer bounds on the rate-equivocation region of the relay-BC (RBC) with one or two confidential messages. Gaussian multiple-input multiple-output (MIMO) BCs and WTCs were studied in \cite{Poor_Gaussian_MIMO_BC_Secrecy2009,Liu_Shamai_MIMOWTC2009,Poor_Shamai_Gaussian_MIMO_BC_Secrecy2010,Khitsi_MIMOWTC2010,Ulukus_Gaussian_Wiretap2011,Hassibi_MINOWTC2011}, while \cite{Ulukus_External_Eve2009,Bagherikaram_Gaussin_External_Eve2009,Piantanida_External_Eve2015} focused on BCs with an eavesdropper as an external entity from which all messages are kept secret.

Many of the aforementioned achievability results were derived by combining Marton's coding for BCs \cite{Marton_BC1979,ElGamal_Martonbound_1981} and Wyner's wiretap coding \cite{Wyner_Wiretap1975,Csiszar_Korner_BCconfidential1978}. Marton coding usually uses a joint typicality encoder (JTE) whose success is guaranteed by invoking the Mutual Covering Lemma (MCL) \cite[Lemma 8.1]{ElGamal2011}. However, the JTE and the MCL have a cumbersome security analysis. Several past works avoid the complications by performing the security analysis without conditioning on the random codebook. This significantly simplifies the derivations, but one would like to have security even if the codebooks are known by the eavesdropper. 

\subsection{Model}

We study a two-user BC over which a common message for both users and a pair of private messages, each destined for a different user, are transmitted. A limited amount of rate of each private message may be leaked to the opposite receiver. The leaked rate is quantified as the normalized mutual information between the message of interest and the channel output sequence at the opposite user. Setting either leakage to zero or infinity reduces the problem to the case where the associated message is confidential or non-confidential, respectively. Thus, our problem setting specializes to all four scenarios concerning secrecy: when both, either or neither of the private messages are secret. We derive inner and outer bounds on the leakage-capacity region of the BC. The inner bound relies on a leakage-adaptive coding scheme that accounts for the codebook being known to the eavesdropper.


The derived bounds are tight for SD-BCs, physically degraded (PD) BCs, and BCs with a degraded message set, thus characterizing their leakage-capacity regions. Furthermore, we derive a condition for identifying the privacy leakage threshold above which the inner bound saturates.  Various past results are captured as special cases. By taking the leakage thresholds to infinity, our inner bound recovers Marton's inner bound with a common message \cite{GP_SemideterministicBC1980}, which is tight for every BC with a known capacity region. Making the leakage constraint inactive in our outer bound recovers the UVW-outer bound \cite{UVW_Outer2010} or the New-Jersey outer bound \cite{NJ_Outer2008}. These bounds are at least as good as previously known bounds (see \cite{Liang_PHD2005,Nair_ElGamal_Outer_Bound2007,Nair_outer2008}). The leakage-capacity region of the SD-BC reduces to each of the regions in \cite{Semi-det_BC_secrect_two2009,Semi-det_BC_secrect_one2009,GP_SemideterministicBC1980} and \cite{Goldfeld_Weak_Secrecy_ISIT2015} by discarding the common message and choosing the leakage constraints appropriately. The capacity result also recovers the optimal regions for the BC with confidential messages \cite{Csiszar_Korner_BCconfidential1978} and the BC with a degraded message set (without secrecy) \cite{Korner_BC_DegradedMessageSet1977}. Finally, a Blackwell BC (BW-BC) \cite{vanderMeulen_Blackwell1975,Gelfand_Blackwell1977} illustrates the results and visualizes the transition of the leakage-capacity region from the capacity region without secrecy to the secrecy-capacity regions for different secrecy scenarios.

\subsection{Organization}

This paper is organized as follows. Section \ref{SEC:preliminaries} establishes notation and preliminary definitions. In Section \ref{SEC:uniformity_lemma} we discuss the need for replacing the JTE with the likelihood encoder and state a Uniform Approximation Lemma. Section \ref{SEC:BC_leakage} describes the BC with privacy leakage constraints, states inner and outer bounds on the leakage-capacity region and characterize the optimal regions of several special cases. In Section \ref{SEC:special_cases} we discuss past results that are captured within our framework and Section \ref{SEC:example} visualizes the results by means of a BW-BC example. Finally, Section \ref{SEC:summary} summarizes the main achievements and insights of this work.


\section{Notations and Preliminary Definitions}\label{SEC:preliminaries}


\subsection{Notations}

 We use the following notations. As customary $\mathbb{N}$ is the set of natural numbers (which does not include 0), while $\mathbb{R}$ denotes the reals. We further define $\mathbb{R}_+=\{x\in\mathbb{R}|x\geq 0\}$ and $\mathbb{R}_{++}=\mathbb{R}\setminus\{0\}$. Given two real numbers $a,b$, we denote by $[a\mspace{-3mu}:\mspace{-3mu}b]$ the set of integers $\big\{n\in\mathbb{N}\big| \lceil a\rceil\leq n \leq\lfloor b \rfloor\big\}$. Calligraphic letters such are $\mathcal{X}$ denote sets, the complement of $\mathcal{X}$ is denoted by $\mathcal{X}^c$, while $|\mathcal{X}|$ stands for its cardinality. $\mathcal{X}^n$ denoted the $n$-fold Cartesian product of $\mathcal{X}$. An element of $\mathcal{X}^n$ is denoted by $x^n=(x_1,x_2,\ldots,x_n)$; whenever the dimension $n$ is clear from the context, vectors (or sequences) are denoted by boldface letters, e.g., $\mathbf{x}$. A substring of $\mathbf{x}\in\mathcal{X}^n$ is denoted by $x_i^j=(x_i,x_{i+1},\ldots,x_j)$, for $1\leq i\leq j \leq n$; when $i=1$, the subscript is omitted. We also define $x^{n\backslash i}=(x_1,\ldots,x_{i-1},x_{i+1},\ldots,x_n)$.


Let $\big(\mathcal{X},\mathcal{F},\mathbb{P}\big)$ be a probability space, where $\mathcal{X}$ is the sample space, $\mathcal{F}$ is the $\sigma$-algebra and $\mathbb{P}$ is the probability measure. Random variables over $\big(\mathcal{X},\mathcal{F},\mathbb{P}\big)$ are denoted by uppercase letters, e.g., $X$, with conventions for random vectors similar to those for deterministic vectors. The probability of an event $\mathcal{A}\in\mathcal{F}$ is denoted by $\mathbb{P}(\mathcal{A})$, while $\mathbb{P}(\mathcal{A}\big|\mathcal{B}\mspace{2mu})$ denotes the conditional probability of $\mathcal{A}$ given $\mathcal{B}_n$. We use $\mathds{1}_\mathcal{A}$ to denote the indicator function of $\mathcal{A}$. The set of all probability mass functions (PMFs) on a finite set $\mathcal{X}$ is denoted by $\mathcal{P}(\mathcal{X})$, i.e.,
\begin{equation}
    \mathcal{P}(\mathcal{X})=\left\{P:\mathcal{X}\to[0,1]\Bigg| \sum_{x\in\mathcal{X}}P(x)=1]\right\}.
\end{equation}
PMFs are denoted by the uppercase letters such as $P$ or $Q$, with a subscript that identifies the random variable and its possible conditioning. For example, for a discrete probability space $\big(\mathcal{X},\mathcal{F},\mathbb{P}\big)$ and two correlated random variables $X$ and $Y$ over that space, we use $P_X$, $P_{X,Y}$ and $P_{X|Y}$ to denote, respectively, the marginal PMF of $X$, the joint PMF of $(X,Y)$ and the conditional PMF of $X$ given $Y$. In particular, $P_{X|Y}$ represents the stochastic matrix whose elements are given by $P_{X|Y}(x|y)=\mathbb{P}\big(X=x|Y=y\big)$. Expressions such as $P_{X,Y}=P_XP_{Y|X}$ are to be understood as $P_{X,Y}(x,y)=P_X(x)P_{Y|X}(y|x)$, for all $(x,y)\in\mathcal{X}\times\mathcal{Y}$. Accordingly, when three random variables $X$, $Y$ and $Z$ satisfy $P_{X|Y,Z}=P_{X|Y}$, they form a Markov chain, which we denote by $X-Y-Z$. We omit subscripts if the arguments of a PMF are lowercase versions of the random variables. The support of a PMF $P$ and the expectation of a random variable $X\sim P$ are denoted by $\supp(P)$ and $\mathbb{E}_P\big[X\big]$, respectively; when the distribution of $X$ is clear from the context we write its expectation simply as $\mathbb{E}\big[X\big]$. Similarly, $H_P$ and $I_P$ denote entropy and mutual information that are calculated with respect to an underlying PMF $P$.

For a discrete measurable space $(\mathcal{X},\mathcal{F})$, a PMF $Q\in\mathcal{P}(\mathcal{X})$ gives rise to a probability measure on $(\mathcal{X},\mathcal{F})$, which we denote by $\mathbb{P}_Q$; accordingly, $\mathbb{P}_Q\big(\mathcal{A})=\sum_{x\in\mathcal{A}}Q(x)$, for every $\mathcal{A}\in\mathcal{F}$. For a random vector $X^n$, if the entries of $X^n$ are drawn in an independent and identically distributed (i.i.d.) manner according to $P_X$, then for every $\mathbf{x}\in\mathcal{X}^n$ we have $P_{X^n}(\mathbf{x})=\prod_{i=1}^nP_X(x_i)$ and we write $P_{X^n}(\mathbf{x})=P_X^n(\mathbf{x})$. Similarly, if for every $(\mathbf{x},\mathbf{y})\in\mathcal{X}^n\times\mathcal{Y}^n$ we have $P_{Y^n|X^n}(\mathbf{y}|\mathbf{x})=\prod_{i=1}^nP_{Y|X}(y_i|x_i)$, then we write $P_{Y^n|X^n}(\mathbf{y}|\mathbf{x})=P_{Y|X}^n(\mathbf{y}|\mathbf{x})$. The conditional product PMF $P_{Y|X}^n$ given a specific sequence $\mathbf{x}\in\mathcal{X}^n$ is denoted by $P_{Y|X=\mathbf{x}}^n$.

Let $\mathcal{X}$ and $\mathcal{Y}$ be finite sets. The empirical PMF $\nu_{\mathbf{x}}$ of a sequence $\mathbf{x}\in\mathcal{X}^n$ is
\begin{equation}
	\nu_{\mathbf{x}}(x)\triangleq\frac{N(x|\mathbf{x})}{n}
\end{equation}
where $N(x|\mathbf{x})=\sum_{i=1}^n\mathds{1}_{\{x_i=x\}}$. We use $\mathcal{T}_\delta^{n}(P_X)$ to denote the set of letter-typical sequences of length $n$ with respect to the PMF $P_X\in\mathcal{P}(\mathcal{X})$ and the positive number $\delta$ \cite[Chapter 3]{Massey_Applied}, i.e., we have
\begin{equation}
	\mathcal{T}_\delta^{n}(P_X)\mspace{-1mu}=\mspace{-1mu}\Big\{\mathbf{x}\in\mathcal{X}^n\Big|\mspace{5mu}\big|\nu_{\mathbf{x}}(x)-P_X(x)\big|\leq\delta P_X(x),\ \forall x\mspace{-1mu}\in\mspace{-1mu}\mathcal{X}\Big\}.\label{EQ:typical_set_def}
\end{equation}
Furthermore, for a joint PMF $P_{X,Y}\in\mathcal{P}(\mathcal{X}\times\mathcal{Y})$ a $\delta>0$ and a fixed sequence $\mathbf{y}\in\mathcal{Y}^n$, we define
\begin{equation}
\mathcal{T}_\delta^{n}(P_{X,Y}|\mathbf{y})=\Big\{\mathbf{x}\in\mathcal{X}^n\Big|(\mathbf{x},\mathbf{y})\in\mathcal{T}_\delta^{n}(P_{X,Y})\Big\}.
\end{equation}

Another notion used throughout this work is information density. Let $(\mathcal{X}\times\mathcal{Y},\mathcal{F},P_{X,Y})$ be a probability space, where $\mathcal{X}$ and $\mathcal{Y}$ are arbitrary sets. The \emph{information density} $i_P:\mathcal{X}\times\mathcal{Y}\to\mathbb{R}_{++}$ of $P_{X,Y}$ is given by
\begin{subequations}
\begin{equation}
i_P(x;y)=\log\frac{\mathsf{d}P_{X,Y}}{\mathsf{d}P_{X}P_{Y}}(x,y)\label{EQ:information_density_gen}
\end{equation}
where $\frac{\mathsf{d}P}{\mathsf{d}Q}$ is the Radon-Nikodym derivative of $P$ with respect to $Q$ and $P_X$ and $P_Y$ are the marginal probability measures induced by $P_{X,Y}$ on $\mathcal{X}$ and $\mathcal{Y}$, respectively. If $\mathcal{X}$ and $\mathcal{Y}$ are discrete and $P_{X,Y}\in\mathcal{P}(\mathcal{X}\times\mathcal{Y})$, then \eqref{EQ:information_density_gen} simplifies as
\begin{equation}
i_P(x;y)=\log\frac{P_{X,Y}(x,y)}{P_{X}(x)P_{Y}(y)}.\label{EQ:information_density}
\end{equation}
\end{subequations}
Whenever the underlying distribution is clear from the context, we drop the subscript $P$ from $i_P$.

\subsection{Measures of Distribution Proximity}


We measure the proximity between two distributions by using total variation (TV). 
\begin{subequations}
\begin{definition}[Total Variation]
	Let $(\mathcal{X},\mathcal{F})$ be a measurable space and $P$ and $Q$ be two probability measures on $\mathcal{F}$. The total variation between $P$ and $Q$ is
	\begin{equation}
		||P-Q||_{\mathsf{TV}}=\sup_{\mathcal{A}\in\mathcal{F}}\big|P(\mathcal{A})-Q(\mathcal{A})\big|.\label{EQ:total_variation_def}
	\end{equation}
	If the sample space $\mathcal{X}$ is countable and $P,Q\in\mathcal{P}(\mathcal{X})$, then \eqref{EQ:total_variation_def} reduces to
	\begin{equation}
		||P-Q||_{\mathsf{TV}}=\frac{1}{2}\sum_{x\in\mathcal{X}}\big|P(x)-Q(x)\big|.\label{EQ:total_variation_def_discrete}
	\end{equation}
\end{definition}
\end{subequations}

We also consider the fidelity between two distributions. 
\begin{subequations}
\begin{definition}[Fidelity]
	Let $(\mathcal{X},\mathcal{F})$ be a measurable space and $P$ and $Q$ be two probability measures on $\mathcal{F}$, such that $P\ll Q$, i.e., $P$ is absolutely continuous with respect to $Q$. The fidelity between $P$ and $Q$ is
	\begin{equation}
		\mathsf{F}(P,Q)=\mathbb{E}_Q\sqrt{\frac{\mathsf{d}P}{\mathsf{d}Q}}.\label{EQ:fidelity_gen_def}
	\end{equation}
	If the sample space $\mathcal{X}$ is countable and $P,Q\in\mathcal{P}(\mathcal{X})$, then \eqref{EQ:fidelity_gen_def} reduces to
	\begin{equation}
		\mathsf{F}(P,Q)=\sum_{x\in\mathcal{X}}\sqrt{P(x)Q(x)}.\label{EQ:fidelity_def}
	\end{equation}
\end{definition}
\end{subequations}
Fidelity satisfies $\mathsf{F}(P,Q)\in[0,1]$ and is related to the TV as follows \cite[Lemma 1]{Yassaee_fidelity_ISIT2015}.
\begin{lemma}[Fidelity and Total Variation]\label{LEMMA:fidelity_vs_TV}
For any two probability measures $P$ and $Q$ over the same measurable space $(\mathcal{X},\mathcal{F})$, we have
\begin{equation}
1-\mathsf{F}(P,Q)\leq ||P-Q||_{\mathsf{TV}} \leq \sqrt{1-F^2(P,Q)}.\label{EQ:fidelity_vs_TV}
\end{equation}
\end{lemma}

Via Jensen's inequality, the right-most inequality in \eqref{EQ:fidelity_vs_TV} extends to the expected values of the fidelity and the TV between two conditional distributions as follows \cite[Lemma 2]{Yassaee_fidelity_ISIT2015}, \cite[Lemma 7]{Yassaee_random_binnign2014}.
\begin{lemma}[Extension to Expected Values]\label{LEMMA:fidelity_vs_TV_expected}
Let $(\Omega,\mathcal{G},\mu)$ be a probability space, $(\mathcal{X},\mathcal{F})$ be a measurable space and $P$ and $Q$ be two transition probability kernels from $(\Omega,\mathcal{G})$ to $(\mathcal{X},\mathcal{F})$.\footnote{A transition probability kernel between two measurable spaces $(\Omega,\mathcal{G})$ and $(\mathcal{X},\mathcal{F})$ is a mapping $\kappa:\Omega\times\mathcal{F}\to[0,1]$ such that: (i) $\omega\mapsto\kappa(\omega,\mathcal{A})$ is a $\mathcal{G}$-measurable function for every $\mathcal{A}\in\mathcal{F}$; (ii) $\mathcal{A}\mapsto\kappa(\omega,\mathcal{A})$ is a probability measure on $(\mathcal{X},\mathcal{F})$ for every $\omega\in\Omega$.} We have
\begin{equation}
\mathbb{E}_{\mu} \big|\big|P-Q\big|\big|_{\mathsf{TV}} \leq \sqrt{1-\Big(\mathbb{E}_\mu \mathsf{F}\big(P,Q\big)\Big)^2}.\label{EQ:fidelity_vs_TV_expected}
\end{equation}
\end{lemma}

By virtue of Lemma \ref{LEMMA:fidelity_vs_TV_expected}, if $\big\{P_n\big\}_{n\in\mathbb{N}}$ and $\big\{Q_n\big\}_{n\in\mathbb{N}}$ are two sequences of Markov kernels\footnote{The formal definition is in accordance with Lemma \ref{LEMMA:fidelity_vs_TV_expected} where we replace $(\Omega,\mathcal{G},\mu)$, $(\mathcal{X},\mathcal{F})$, $P$ and $Q$ with the sequences $\big\{(\Omega_n,\mathcal{G}_n,\mu_n)\big\}_n$, $\big\{(\mathcal{X}_n,\mathcal{F}_n)\big\}_n$, $\{P_n\}_n$ and $\{Q_n\}_n$, respectively.} then
\begin{subequations}
\begin{equation}
\mathbb{E}_{\mu_n}\mathsf{F}\big(P_n,Q_n\big)\xrightarrow[n\to\infty]{}1
\end{equation}
implies
\begin{equation}
\mathbb{E}_{\mu_n}\big|\big|P_n-Q_n\big|\big|_{\mathsf{TV}}\xrightarrow[n\to\infty]{}0.
\end{equation}\label{EQ:fidelity_vs_TV_expected_sequences}
\end{subequations}


\section{Uniform Distribution Approximation Lemma}\label{SEC:uniformity_lemma}

\subsection{Marton Coding}


A Marton code involves two independent codebooks from which a pair of codewords is usually selected by means of a JTE \cite{ElGamal_Martonbound_1981}. A standard tool for the encoding error probability analysis is the MCL \cite[Lemma 8.1]{ElGamal2011}. While the JTE and the MCL are convenient for analysing reliability, security (equivocation or leakage) analysis seems cumbersome.


Several past works employ Marton coding without conditioning the security analysis on the random codebook. Attempting to repeat the steps from these derivations while conditioning the equivocation on the codebook turns out to be problematic. The principal difficulty is showing that the marginal distribution of an index chosen by the JTE is approximately uniform.\footnote{Without the conditioning, uniformity follows by symmetry.} More precisely, let the output index pair of the encoder be $(I,J)$; the corresponding alphabets are $\mathcal{I}_n$ and $\mathcal{J}_n$. Several existing proofs rely on the following relations holding true:
\begin{equation}
    H(I|\mathsf{C}_n)\geq\log|\mathcal{I}_n|-n\delta_n\quad; \quad H(J|\mathsf{C}_n)\geq\log|\mathcal{J}_n|-n\delta_n,\label{EQ:wrong_uniform_claim}
\end{equation}
where $\mathsf{C}_n$ is the random codebook\footnote{The conditioning on $\mathsf{C}_n$ is not present in many existing works. Instead, the relations \eqref{EQ:wrong_uniform_claim} were replaced with their unconditioned versions $H(I)=\log|\mathcal{I}|$ and $H(J)=\log|\mathcal{J}|$. Although these relations are true, an unconditioned analysis does not imply achievability when the codebooks are known to the eavesdropper.} and $\lim_{n\to\infty}\delta_n=0$. Proving these inequalities while using the JTE is cumbersome. A potential proof would rely on analysing the output distribution of the JTE. However, the structure of this distribution quickly makes the analysis intractable.

\subsection{Likelihood Encoder}


Our coding scheme also uses a Marton code. We circumvent the problems with the JTE by replacing it with a likelihood encoder for Marton codebooks \cite{Cuff_Song_Likelihood2014,Goldfeld_strong_secrecy_cooperation2016,Goldfeld_GP_WTC2016}. A similar encoding rule was used in \cite{Yassaee_oneshot_ISIT2013,Yassaee_fidelity_ISIT2015} under the name stochastic mutual information encoder. This encoder induces a probability distribution over the possible pairs of indices (or, equivalently, codewords). Given two independently generated bins, the probability of each codeword pair is proportional to the ratio of their joint probability (under the coding distribution) to the product of the marginal distributions. Namely, if $\mathbf{u}_i$ and $\mathbf{v}_j$ are the $i$-th and $j$-th codewords for each bin, respectively, and $Q_{U,V}$ is the coding distribution (the codebooks are i.i.d. samples of its marginals $Q_U$ and $Q_V$), then the encoder chooses $(i,j)$ with probability proportional to
\begin{equation}
    \frac{Q_{U,V}^n(\mathbf{u}_i,\mathbf{v}_j)}{Q_U^n(\mathbf{u}_i)Q_V^n(\mathbf{v}_j)}.
\end{equation}
Thus, the further the joint distribution is from the product of the marginals the more favorable the corresponding pair of codewords is.

Replacing the JTE with the likelihood encoder comes at no cost in reliability. This is because, like the JTE, if the sum of the bin rates is greater than $I(U;V)$, then the likelihood encoder chooses jointly typical codeword pairs with high probability \cite[Theorem 3]{Yassaee_fidelity_ISIT2015}. The leakage analysis, on the other hand, tremendously simplifies. This allows to derive the achievability result for the BC with privacy leakage constraints. Key to the leakage analysis is that the marginal distribution of the indices at the encoder's output are indeed approximately uniform. This relation is formulated in the next subsection and the proof is provided in Section \ref{SUBSEC:uniform_approx_proof}.


\subsection{Setup and Statement of the Lemma}

For notational convenience we formulate the setup and state the result in terms of random variables with finite alphabets. Nonetheless, as can be seen in the proof of Lemma \ref{LEMMA:uniform_approx} (Section \ref{SUBSEC:uniform_approx_proof}), the derivation is valid for random variables with general alphabets. 

Fix $Q_{W,U,V}\in\mathcal{P}(\mathcal{W}\times\mathcal{U}\times\mathcal{V})$ and for every $n\in\mathbb{N}$ define $\mathcal{I}_n\triangleq \big[1:2^{nS_1}\big]$, $\mathcal{J}_n\triangleq \big[1:2^{n S_2}\big]$ and $\mathcal{K}_n=\big[1:2^{nT}\big]$, where $S_1,S_2,T\in\mathbb{R}_+$. Let $\mathbf{W}\sim Q_W^n$ and fix $\mathbf{w}\in\mathcal{W}^n$ with $Q_W^n(\mathbf{w})>0$. Let $\mathsf{B}^{(n)}_U(\mathbf{w})\triangleq\big\{\mathbf{U}_i(\mathbf{w})\big\}_{i\in\mathcal{I}_n}$ be a random codebook that comprises $|\mathcal{I}_n|$ vectors of length $n$ that are i.i.d. according to $Q^n_{U|W=\mathbf{w}}$. Furthermore, for every $k\in\mathcal{K}_n$ let $\mathsf{B}^{(n)}_V(k,\mathbf{w})\triangleq\big\{\mathbf{V}_{j,k}(\mathbf{w})\big\}_{j\in\mathcal{J}_n}$ be a random codebook with i.i.d. codewords according to $Q^n_{V|W=\mathbf{w}}$. The codebooks in the set $\mathsf{B}^{(n)}_V(\mathbf{w})\triangleq\left\{\mathsf{B}^{(n)}_V(k,\mathbf{w})\right\}_{k\in\mathcal{K}_n}$ are conditionally independent of one another given $\mathbf{W}=\mathbf{w}$. For any $\mathbf{w}\in\mathcal{W}^n$ with $Q_W^n(\mathbf{w})>0$ we also define $\mathsf{B}_n(\mathbf{w})\triangleq\left\{\mathsf{B}^{(n)}_U(\mathbf{w}),\mathsf{B}^{(n)}_V(\mathbf{w})\right\}$ and finally we set $\mathsf{B}_n\triangleq\big\{\mathbf{W},\mathsf{B}_n(\mathbf{W})\big\}$.

A realization of $\mathsf{B}^{(n)}_U(\mathbf{w})$ or $\mathsf{B}^{(n)}_V(k,\mathbf{w})$, $k\in\mathcal{K}_n$, is denoted by $\mathcal{B}^{(n)}_U(\mathbf{w})\triangleq\big\{\mathbf{u}_i(\mathbf{w})\big\}_{i\in\mathcal{I}_n}$ and $\mathcal{B}^{(n)}_V(k,\mathbf{w})\triangleq\big\{\mathbf{v}_{j,k}(\mathbf{w})\big\}_{j\in\mathcal{J}_n}$, respectively. In accordance to the above, we also set  $\mathcal{B}^{(n)}_V(\mathbf{w})\triangleq\left\{\mathcal{B}^{(n)}_V(k,\mathbf{w})\right\}_{k\in\mathcal{K}_n}=\big\{\mathbf{v}_{j,k}(\mathbf{w})\big\}_{(j,k)\in\mathcal{J}_n\times\mathcal{K}_n}$, $\mathcal{B}_n(\mathbf{w})\triangleq\left\{\mathcal{B}^{(n)}_U(\mathbf{w}),\mathcal{B}^{(n)}_V(\mathbf{w})\right\}$ and $\mathcal{B}_n\triangleq\big\{\mathbf{w},\mathcal{B}_n(\mathbf{w})\big\}$. Letting $\mathfrak{B}_n$ denote the collection of all possible realization of $\mathsf{B}_n$, the above construction induces a PMF $\lambda\in\mathcal{P}(\mathfrak{B}_n)$ on $\mathfrak{B}_n$ that is given by
\begin{align*}
	&\lambda(\mathcal{B}_n)\\&=Q_W^n(\mathbf{w})\mspace{-5mu}\prod_{i\in\mathcal{I}_n}\mspace{-5mu}Q^n_{U|W}\big(\mathbf{u}_i(\mathbf{w})\big|\mathbf{w}\big)\mspace{-18mu}\prod_{(j,k)\in\mathcal{J}_n\times\mathcal{K}_n}\mspace{-18mu}Q^n_{V|W}\big(\mathbf{v}_{j,k}(\mathbf{w})\big|\mathbf{w}\big).\numberthis\label{EQ:lemma_codebook_PMF}
\end{align*}

Now, let $K$ be a random variable independent of $\mathsf{B}_n$ and uniformly distributed over $\mathcal{K}_n$. For each $\mathcal{B}_n\in\mathfrak{B}_n$ and $k\in\mathcal{K}_n$, the index pair $(i,j)\in\mathcal{I}_n\times\mathcal{J}_n$ is drawn according to
\begin{equation}
P_{I,J}^{(\mathcal{B}_n,k)}(i,j)=\frac{2^{i_{Q^n}\big(\mathbf{u}_i(\mathbf{w});\mathbf{v}_{j,k}(\mathbf{w})\big|\mathbf{w}\big)}}{\sum\limits_{(\bar{\ell},\bar{j})\in\mathcal{I}_n\times\mathcal{J}_n}2^{i_{Q^n}\big(\mathbf{u}_{\bar{\ell}}(\mathbf{w});\mathbf{v}_{\bar{j},k}(\mathbf{w})\big|\mathbf{w}\big)}},\label{EQ:lemma_likelihood_encoder}
\end{equation}
where
\begin{equation}
i_{Q^n}(\mathbf{u};\mathbf{v}|\mathbf{w})=\log\frac{Q^n_{U,V|W}(\mathbf{u},\mathbf{v}|\mathbf{w})}{Q^n_{U|W}(\mathbf{u}|\mathbf{w})Q^n_{V|W}(\mathbf{v}|\mathbf{w})}.
\end{equation}
$P_{I,J}^{(\mathcal{B}_n,k)}$ describes our likelihood encoder. Finally, on account of \eqref{EQ:lemma_codebook_PMF}-\eqref{EQ:lemma_likelihood_encoder} we set
\begin{equation}
P_{\mathsf{B}_n,K,I,J}(\mathcal{B}_n,k,i,j)\triangleq\lambda(\mathcal{B}_n)\frac{1}{|\mathcal{K}_n|}P_{I,J}^{(\mathcal{B}_n,k)}(i,j),\label{EQ:lemma_induced_joint_PMF}
\end{equation}
which induces a probability measure $\mathbb{P}_P$.

The following lemma specifies sufficient conditions on the sizes of the index sets for approximating the induced marginal distribution of $I$ with a uniform distribution over $\mathcal{I}_n$. To state the result let $p_{\mathcal{I}_n}^{(U)}$ be the uniform distribution over $\mathcal{I}_n$ and note that for every $\mathcal{B}_n\in\mathfrak{B}_n$
\begin{equation}
P_{I|\mathsf{B}_n}(i|\mathcal{B}_n)=\frac{1}{|\mathcal{K}_n|}\mspace{-4mu}\sum_{\substack{(j,k)\\\in\mathcal{J}_n\times\mathcal{K}_n}}\mspace{-4mu}\frac{2^{i_{Q^n}\big(\mathbf{u}_i(\mathbf{w});\mathbf{v}_{j,k}(\mathbf{w})\big|\mathbf{w}\big)}}{\sum\limits_{\substack{(\bar{\ell},\bar{j})\\\in\mathcal{I}_n\times\mathcal{J}_n}}2^{i_{Q^n}\big(\mathbf{u}_{\bar{\ell}}(\mathbf{w});\mathbf{v}_{\bar{j},k}(\mathbf{w})\big|\mathbf{w}\big)}}.\label{EQ:lemma_induced_marginal}
\end{equation}

\begin{lemma}[Uniform Approximation Lemma]\label{LEMMA:uniform_approx}
For any $Q_{W,U,V}\in\mathcal{P}(\mathcal{W}\times\mathcal{U}\times\mathcal{V})$ if 
\begin{subequations}
\begin{equation}
S_2+\min\big\{S_1,T\big\}>I_{Q_{W,U,V}}(U;V|W)
\end{equation}
then
\begin{equation}
\mathbb{E}_{\mathsf{B}_n}\Big|\Big|P_{I|\mathsf{B}_n}-p^{(U)}_{\mathcal{I}_n}\Big|\Big|_{\mathsf{TV}}\xrightarrow[n\to\infty]{}0.
\end{equation}
\end{subequations}
\end{lemma}
The Lemma is proven in Section \ref{SUBSEC:uniform_approx_proof} via an analysis of the expected fidelity between the induced marginal distribution of $I$ and the uniform distribution. Inspired by ideas from \cite{Yassaee_fidelity_ISIT2015}, we employ the Cauchy-Schwarz inequality and Jensen's inequality to show that the expected fidelity converges to 1 with the blocklengh. The result of the lemma then follows by \eqref{EQ:fidelity_vs_TV_expected_sequences}.



\section{Broadcast Channels with Privacy Leakage Constraints}\label{SEC:BC_leakage}


\subsection{Problem Setting}\label{SUBSEC:BC_leakage_def}

 The $\big(\mathcal{X},\mathcal{Y}_1,\mathcal{Y}_2,W_{Y_1,Y_2|X}:\mathcal{X}\to\mathcal{P}(\mathcal{Y}_1\times\mathcal{Y}_2)\big)$ BC with privacy leakage constraints is illustrated in Fig. \ref{FIG:general_BC_leakage}. The channel has one sender and two receivers. The sender randomly chooses a triple $(m_0,m_1,m_2)$ of indices uniformly and independently from the set $\big[1:2^{nR_0}\big]\times\big[1:2^{nR_1}\big]\times\big[1:2^{nR_2}\big]$ and maps them to a sequence $\mathbf{x}\in\mathcal{X}^n$, which is the channel input (the mapping may be random). The sequence $\mathbf{x}$ is transmitted over a BC with transition probability $W_{Y_1,Y_2|X}$. The output sequence $\mathbf{y}_j\in\mathcal{Y}^n_j$, where $j=1,2$, is received by decoder $j$. Decoder $j$ produces a pair of estimates $\big(\hat{m}_0^{(j)},\hat{m}_j\big)$ of $(m_0,m_j)$.

\begin{remark}[Specific Classes of BCs]\label{REM:BC_classes}
We sometimes specialize to the following classes of BCs:
\begin{itemize}
    \item \underline{Semi-Deterministic BCs:} A BC is SD if its channel transition matrix factors as $W_{Y_1,Y_2|X}=\mathds{1}_{\{Y_1=y_1(X)\}}W_{Y_2|X}$, where $y_1:\mathcal{X}\to\mathcal{Y}_1$ and $W_{Y_2|X}:\mathcal{X}\to\mathcal{P}(\mathcal{Y}_2)$.

    \item \underline{Physically-Degraded BCs:} A BC is PD if its channel transition matrix factors as $W_{Y_1,Y_2|X}=W_{Y_1|X}W_{Y_2|Y_1}$, where $W_{Y_1|X}:\mathcal{X}\to\mathcal{P}(\mathcal{Y}_1)$ and $W_{Y_2|Y_1}:\mathcal{Y}_1\to\mathcal{P}(\mathcal{Y}_2)$.

    \item \underline{Deterministic BCs:} A BC is deterministic if its channel transition matrix factors as $W_{Y_1,Y_2|X}=\mathds{1}_{\{Y_1=y_1(X)\}\cap\{Y_2=y_2(X)\}}$, where $y_j:\mathcal{X}\to\mathcal{Y}_j$, for $j=1,2$.
\end{itemize}

\end{remark}

\begin{definition}[Code]
An $(n,R_0,R_1,R_2)$ code $c_n$ for the BC with leakage constraints has:
\begin{enumerate}
\item Three message sets $\mathcal{M}_j^{(n)}\triangleq\big[1:2^{nR_j}\big]$, $j=0,1,2$.
\item A stochastic encoder $f^{(n)}:\mathcal{M}^{(n)}_0\times\mathcal{M}^{(n)}_1\times\mathcal{M}^{(n)}_2\to \mathcal{P}(\mathcal{X}^n)$.
%
\item Two decoding functions, $\phi^{(n)}_j: \mathcal{Y}_j^n\to \hat{\mathcal{M}}^{(n)}_{0j}$, where $\hat{\mathcal{M}}^{(n)}_{0j}\triangleq\mathcal{M}^{(n)}_0\times\mathcal{M}^{(n)}_j$, for $j=1,2$.
\end{enumerate}
\end{definition}


A code $c_n=\left(f^{(n)},\phi_1^{(n)},\phi_2^{(n)}\right)$ for the $W_{Y_1,Y_2|X}$ BC with privacy leakage constraints induces a PMF $P^{(c_n)}$ on $\mathcal{M}_0\times\mathcal{M}_1\times\mathcal{M}_2\times\mathcal{X}^n\times\mathcal{Y}_1^n\times\mathcal{Y}^n_2\times\hat{\mathcal{M}}_{01}\times\hat{\mathcal{M}}_{02}$, that is given by
\begin{align*}
    &P^{(c_n)}\Big(m_0,m_1,m_2,\mathbf{x},\mathbf{y}_1,\mathbf{y}_2,\big(\hat{m}_0^{(1)},\hat{m}_1\big),\big(\hat{m}_0^{(2)},\hat{m}_2\big)\Big)\\
    &=\prod_{j=0,1,2}\frac{1}{\big|\mathcal{M}^{(n)}_j\big|}f^{(n)}(\mathbf{x}|m_0,m_1,m_2)W^n_{Y_1,Y_2|X}(\mathbf{y}_1,\mathbf{y}_2|\mathbf{x})\\&\mspace{180mu}\times\mathds{1}_{\bigcap_{j=1,2}\big\{(\hat{m}_0^{(j)},m_j)=\phi^{(n)}_j(\mathbf{y}_j)\big\}}\numberthis\label{EQ:induced_PMF_def}.
\end{align*}
The induced PMF gives rise to the probability measure $\mathbb{P}_{P^{(c_n)}}$, which we abbreviate by $\mathbb{P}_{c_n}$. Similarly, we use the shorthand $I_{c_n}$ instead of $I_{P^{(c_n)}}$ to denote a mutual information expression taken with respect to $P^{(c_n)}$.

\begin{definition}[Average Error Probability] The average error probability for an $(n,R_0,R_1,R_2)$ code $c_n$ is
\begin{align*}
P_{\mathsf{e}}(c_n)\triangleq\mathbb{P}_{c_n}\left(\bigcup_{j=1,2}\Big\{\big(\hat{M}_0^{(j)},\hat{M}_j\big)\neq(M_0,M_j)\Big\}\right)
\numberthis\label{BC_Pe}
\end{align*}
where $\big(\hat{M}_0^{(j)},\hat{M}_j\big)=\phi_j^{(n)}(\mathbf{Y}_j)$, for $j=1,2$.
\end{definition}

\begin{definition}[Information Leakage Rate] The information leakage rate of $M_1$ to receiver 2 under an $(n,R_0,R_1,R_2)$ code $c_n$ is
\begin{subequations}
\begin{equation}
\ell_1(c_n)\triangleq\frac{1}{n}I_{c_n}(M_1;\mathbf{Y}_2).
\end{equation}
Similarly, the information leakage rate of $M_2$ to receiver 1 under $c_n$ is
\begin{equation}
\ell_2(c_n)\triangleq\frac{1}{n}I_{c_n}(M_2;\mathbf{Y}_1).
\end{equation}
\end{subequations}\label{EQ:infromation_leakage_def}%
\end{definition}

\begin{definition}[Achievable Rates] Let $(L_1,L_2)\in\mathbb{R}^2_+$. A rate triple $(R_0,R_1,R_2)\in\mathbb{R}_+^3$ is $(L_1,L_2)$-{\it{achievable}} if for any $\epsilon>0$ there exists a sufficiently large $n\in\mathbb{N}$ and an $(n,R_0,R_1,R_2)$ code $c_n$ such that
\begin{subequations}
\begin{align}
&P_{\mathsf{e}}(c_n)\leq\epsilon\label{EQ:error_prob}\\
&\ell_1(c_n)\leq L_1+\epsilon\label{EQ:achieve_leakage1}\\
&\ell_2(c_n)\leq L_2+\epsilon.\label{EQ:achieve_leakage2}
\end{align}\label{EQ:achiev_realibility_leakage}
\end{subequations}
\end{definition}

\begin{definition}[Leakage-Capacity Region]The $(L_1,L_2)$-{\it{leakage-capacity region}} $\mathcal{C}(L_1,L_2)$ is the closure of the set of the $(L_1,L_2)$-achievable rates.
\end{definition}

\begin{remark}[Inactive Leakage Constraints]\label{REM:inactive_leakage}
    Setting $L_j=R_j$, for $j=1,2$, makes \eqref{EQ:achieve_leakage1}-\eqref{EQ:achieve_leakage2} inactive and reduces the BC with privacy leakage constraints to the classic BC with a common message. This is a simple consequence of the non-negativity of entropy, which implies that $I_{c_n}(M_1;\mathbf{Y}_2)\leq nR_1$ and $I_{c_n}(M_2;\mathbf{Y}_1)\leq nR_2$ always hold. To simplify notation we write $L_j\to\infty$, $j=1,2$ to refer to leakage threshold values under which \eqref{EQ:achieve_leakage1}-\eqref{EQ:achieve_leakage2} are satisfied by default.
\end{remark}


\subsection{Leakage-Capacity Results}\label{SUBSEC:results}

This section states inner and outer bounds on the $(L_1,L_2)$-leakage-capacity region $\mathcal{C}(L_1,L_2)$ of a BC. The bounds match for SD-BCs, BCs with a degraded message set and PD-BCs, which characterizes the leakage-capacity regions for these three cases. We start with the inner bound.



In the following, the transition probability $W_{Y_1,Y_2|X}$ describing the BC stays fixed unless stated otherwise. When specifying to particular instances of BCs (see Remark \ref{REM:BC_classes}), we explicitly mention the corresponding structure of $W_{Y_1,Y_2|X}$.

\begin{theorem}[Inner Bound]\label{TM:inner_bound}
Let $\mathcal{R}_{\mathsf{I}}(L_1,L_2)$ be the closure of the union of rate triples $(R_0,R_1,R_2)\in\mathbb{R}^3_+$ satisfying:
\begin{subequations}
\begin{align}
R_0\mspace{-2mu} &\leq\mspace{-2mu} \min\Big\{I(U_0;Y_1),I(U_0;Y_2)\Big\}\label{EQ:region_inner0}\\
R_1\mspace{-2mu} &\leq\mspace{-2mu}I(U_1;Y_1|U_0)\mspace{-3mu}-\mspace{-3mu}I(U_1;U_2,Y_2|U_0)\mspace{-3mu}+\mspace{-3mu}L_1\label{EQ:region_inner11}\\
R_0\mspace{-2mu}+\mspace{-2mu}R_1\mspace{-2mu} &\leq\mspace{-2mu} I(U_1;Y_1|U_0)\mspace{-3mu}+\mspace{-3mu}\min\mspace{-3mu}\Big\{\mspace{-1.5mu}I(U_0;Y_1),\mspace{-1.5mu}I(U_0;Y_2)\mspace{-1.5mu}\Big\}\label{EQ:region_inner12}\\
R_2\mspace{-2mu} &\leq\mspace{-2mu}I(U_2;Y_2|U_0)\mspace{-3mu}-\mspace{-3mu}I(U_2;U_1,Y_1|U_0)\mspace{-3mu}+\mspace{-3mu}L_2\label{EQ:region_inner21}\\
R_0\mspace{-2mu}+\mspace{-2mu}R_2\mspace{-2mu} &\leq\mspace{-2mu} I(U_2;Y_2|U_0)\mspace{-3mu}+\mspace{-3mu}\min\mspace{-3mu}\Big\{\mspace{-1.5mu}I(U_0;Y_1),\mspace{-1.5mu}I(U_0;Y_2)\mspace{-1.5mu}\Big\}\label{EQ:region_inner22}\\
\sum_{j=0,1,2}\mspace{-4mu}R_j\mspace{-2mu} &\leq\mspace{-4mu}\begin{multlined}[t][.2\textwidth] I(U_1;Y_1|U_0)+I(U_2;Y_2|U_0)\\\mspace{-40mu}-\mspace{-2mu}I(U_1;U_2|U_0)\mspace{-2mu}+\mspace{-2mu}\min\mspace{-4mu}\Big\{\mspace{-1mu}I(U_0;\mspace{-1.5mu}Y_1),I(U_0;\mspace{-1.5mu}Y_2)\mspace{-2mu}\Big\}\end{multlined}\label{EQ:region_inner_sum1}
\end{align}\label{EQ:region_inner}%
\end{subequations}
where the union is over all PMFs $Q_{U_0,U_1,U_2,X}\in\mathcal{P}(\mathcal{U}_0\times\mathcal{U}_1\times\mathcal{U}_2\times\mathcal{X})$, each inducing a joint distribution $Q_{U_0,U_1,U_2,X}W_{Y_1,Y_2|X}$. The following inclusion holds:
\begin{equation}
\mathcal{R}_{\mathsf{I}}(L_1,L_2)\subseteq\mathcal{C}(L_1,L_2).\label{EQ:inclusion_inner}
\end{equation}
\end{theorem}

The proof of Theorem \ref{TM:inner_bound} is given in Section \ref{SUBSEC:inner_proof} and uses a leakage-adaptive Marton-like code construction. Rate-splitting is first used to decompose each private message $M_j$, $j=1,2$, into a public part $M_{0j}$ and a private part $M_{jj}$. A Marton codebook with an extra layer of bins is then constructed while treating $(M_0,M_{10},M_{20})$ as a public message and $M_{jj}$, for $j=1,2$, as private message $j$. The double-binning of the private messages permits joint encoding (outer layer) and controlling the total rate leakage to the other user (inner layer). In contrast to the classic Marton coding scheme \cite{ElGamal_Martonbound_1981} that employes a JTE, we execute joint encoding by means of the likelihood encoder from \eqref{EQ:lemma_likelihood_encoder}. Doing so doesn't affect the reliability analysis (as the likelihood encoder chooses jointly typical pairs of codewords with high probability), but it is of consequence for analysing the leakage rate. 

The leakage analysis takes into account the rate leaked due to the decoding of the public message by both users. Also, additional leakage occurs due to the joint encoding process, which introduces correlation between the private message codewords. We account for the latter by relating the bin sizes in the inner and outer coding layers to the rate of the public parts $M_{10}$ and $M_{20}$. The leakage analysis relies heavily on the structure of the likelihood encoder that lets us establish several crucial properties of our random coding experiment. The main challenge is showing that the induced marginal distribution describing the choice of the private message codewords is approximately uniform. This follows by virtue of the Uniform Approximation Lemma (Lemma \ref{LEMMA:uniform_approx}).


\begin{remark}[Relation to Marton's Region]
In \cite[Theorem 1]{GP_SemideterministicBC1980} Gelfand and Pinsker generalized Marton's inner bound \cite{Marton_BC1979} to include a common message. An alternative form of Gelfand and Pinsker's inner bound was given in \cite[Theorem 5]{Liang_Kramer_RBC2007} (see also \cite{Liang_Kramer_Proof_BCMArton2008}). This region is the best known inner bound on the capacity region of the BC with a common message. $\mathcal{R}_{\mathsf{I}}(\infty,\infty)$ recovers the Gelfand-Pinsker region since \eqref{EQ:region_inner11} and \eqref{EQ:region_inner21} are redundant. A full discussion of the special cases of $\mathcal{R}_{\mathsf{I}}(L_1,L_2)$ is given in Section \ref{SUBSEC:special_cases_SDBC}.
\end{remark}


The following corollary states a sufficient condition on the leakage thresholds $L_1$ and $L_2$ to become inactive in the bounds from \eqref{EQ:region_inner} when $R_0=0$ (i.e., no common message is present). To state the result, let $\tilde{\mathcal{R}}_{\mathsf{I}}(L_1,L_2,Q_{U_0,U_1,U_2,X})$ denote the set of rate pairs $(R_1,R_2)\in\mathbb{R}_{+}^2$ satisfying \eqref{EQ:region_inner} with $R_0=0$ when the mutual information terms are calculated with respect to $Q_{U_0,U_1,U_2,X}W_{Y_1,Y_2|X}$. Accordingly,
\begin{equation}
\tilde{\mathcal{R}}_{\mathsf{I}}(L_1,L_2)\triangleq\bigcup_{Q_{U_0,U_1,U_2,X}}\tilde{\mathcal{R}}_{\mathsf{I}}(L_1,L_2,Q_{U_0,U_1,U_2,X})\label{EQ:region_nocommon_afterunion}
\end{equation}
is the region obtained by setting $R_0=0$ in $\mathcal{R}_{\mathsf{I}}(L_1,L_2)$.

\begin{corollary}[Inactive Leakage Constraints]\label{COR:inactive_leakage}
Let $Q_{U_0,U_1,U_2,X}\in\mathcal{P}(\mathcal{U}_0\times\mathcal{U}_1\times\mathcal{U}_2\times\mathcal{X})$. For $j=1,2$ define
\begin{align*}
L_j&^\star(Q_{U_0,U_1,U_2,X})
\\&=\min\Big\{I(U_0;Y_1),I(U_0;Y_2)\Big\}+I(U_j;U_{\bar{j}},Y_{\bar{j}}|U_0),\numberthis\label{EQ:maximal_leakage_Lj}
\end{align*}
where $\bar{j}=j+(-1)^{j+1}$. The following implications hold:
\begin{enumerate}
\item If $L_1\geq L_1^\star(Q_{U_0,U_1,U_2,X})$ then $\tilde{\mathcal{R}}_{\mathsf{I}}(L_1,L_2,Q_{U_0,U_1,U_2,X})=\tilde{\mathcal{R}}_{\mathsf{I}}(\infty,L_2,Q_{U_0,U_1,U_2,X})$.
\item If $L_2\geq L_2^\star(Q_{U_0,U_1,U_2,X})$ then $\tilde{\mathcal{R}}_{\mathsf{I}}(L_1,L_2,Q_{U_0,U_1,U_2,X})=\tilde{\mathcal{R}}_{\mathsf{I}}(L_1,\infty,Q_{U_0,U_1,U_2,X})$.
\item If $L_j\geq L_j^\star(Q_{U_0,U_1,U_2,X})$, for $j=1,2$, then $\tilde{\mathcal{R}}_{\mathsf{I}}(L_1,L_2,Q_{U_0,U_1,U_2,X})=\tilde{\mathcal{R}}_{\mathsf{I}}(\infty,\infty,Q_{U_0,U_1,U_2,X})$.
\end{enumerate}
\end{corollary}

For the proof of Corollary \ref{COR:inactive_leakage} see Section \ref{SUBSEC:inactive_leakage_proof}. According to the above, if any of the leakage thresholds $L_j$, $j=1,2$ surpasses the critical value from \eqref{EQ:maximal_leakage_Lj}, then the corresponding inner bound remains unchanged if $L_j$ is further increased, and is therefore equivalent to the region where $L_j\to\infty$.

\begin{remark}[Application of Corollary \ref{COR:inactive_leakage}]\label{REM:inactive_leakage_explained}
Corollary \ref{COR:inactive_leakage} specifies a condition for $L_1$ and/or $L_2$ being inactive for each input probability. Getting a condition for the inactivity of the thresholds with respect to the entire region $\tilde{\mathcal{R}}_{\mathsf{I}}(L_1,L_2)$ from \eqref{EQ:region_nocommon_afterunion} is a more challenging task. Identifying such a condition involves identifying which input distributions achieve the boundary of $\tilde{\mathcal{R}}_{\mathsf{I}}(L_1,L_2)$. In some communication scenarios this is possible, e.g., for the MIMO Gaussian BC with or without secrecy requirements the boundary achieving distributions are Gaussian vectors \cite{Weingarten_MIMOBC2006,Poor_Shamai_Gaussian_MIMOBC_Secrecy2010,Ulukus_GaussianBCCommon2012,Chandra_Gauss_BC2014,Goldfeld_MIMOBC_Secrecy2016}. However, the structure of the optimizing distribution is unknown in general.

The merit of Corollary \ref{COR:inactive_leakage} becomes clear when explicitly calculating $\tilde{\mathcal{R}}_{\mathsf{I}}(L_1,L_2)$. One can then identify the optimizing distribution, e.g., by means of an analytical characterization or via an exhaustive search. In turn, one can calculate the maximum of $L_j^\star(Q_{U_0,U_1,U_2,X})$ over those distributions. Denoting by $L_j^\star$ this maximal value, if $L_j<L_j^\star$ then increasing $L_j$ will further shrink the region. If, on the other hand, $L_j\geq L_j^\star$, then the region remains unchanged even if $L_j$ grows. This idea is demonstrated in Section \ref{SEC:example} where we calculate the $(L_1,L_2)$-leakage-capacity region of the Blackwell BC.
\end{remark}

Next, we state an outer bound on $\mathcal{C}(L_1,L_2)$. A proof of Theorem \ref{TM:outer_bound} is given in Section \ref{SUBSEC:outer_proof}.

\begin{theorem}[Outer Bound]\label{TM:outer_bound}
Let $\mathcal{R}_{\mathsf{O}}(L_1,L_2)$ be the closure of the union of rate triples $(R_0,R_1,R_2)\in\mathbb{R}^3_+$ satisfying:
\begin{subequations}
\begin{align}
R_0\mspace{-2mu}&\leq \mspace{-1.5mu}\min\Big\{I(W;Y_1),I(W;Y_2)\Big\}\label{EQ:region_outer0}\\
R_1\mspace{-2mu}&\leq \mspace{-1.5mu}I(U;Y_1|W,V)-I(U;Y_2|W,V)+L_1\label{EQ:region_outer11}\\
R_1\mspace{-2mu}&\leq \mspace{-1.5mu}I(U;Y_1|W)-I(U;Y_2|W)+L_1\label{EQ:region_outer12}\\
R_0\mspace{-2mu}+\mspace{-2mu}R_1\mspace{-2mu}&\leq \mspace{-1.5mu}I(U;Y_1|W)\mspace{-2mu}+\mspace{-2mu}\min\mspace{-3mu}\Big\{\mspace{-1.5mu}I(W;Y_1),\mspace{-1.5mu}I(W;Y_2)\mspace{-1.5mu}\Big\}\label{EQ:region_outer13}\\
R_2\mspace{-2mu}&\leq \mspace{-1.5mu}I(V;Y_2|W,U)-I(V;Y_1|W,U)+L_2\label{EQ:region_outer21}\\
R_2\mspace{-2mu}&\leq \mspace{-1.5mu}I(V;Y_2|W)-I(V;Y_1|W)+L_2\label{EQ:region_outer22}\\
R_0\mspace{-2mu}+\mspace{-2mu}R_2\mspace{-2mu}&\leq \mspace{-1.5mu}I(V;Y_2|W)\mspace{-2mu}+\mspace{-2mu}\min\mspace{-3mu}\Big\{\mspace{-1.5mu}I(W;Y_1),\mspace{-1.5mu}I(W;Y_2)\mspace{-1.5mu}\Big\}\label{EQ:region_outer23}\\
\sum_{j=0,1,2}R_j\mspace{-2mu}&\leq\mspace{-1.5mu}\begin{multlined}[t][.3\textwidth] I(U;Y_1|W,V)+I(V;Y_2|W)\\\mspace{55mu}+\min\Big\{I(W;Y_1),I(W;Y_2)\Big\}\end{multlined}\label{EQ:region_outer_sum1}\\
\sum_{j=0,1,2}R_j\mspace{-2mu}&\leq\mspace{-1.5mu}\begin{multlined}[t][.3\textwidth] I(U;Y_1|W)+I(V;Y_2|W,U)\\\mspace{55mu}+\min\Big\{I(W;Y_1),I(W;Y_2)\Big\}\end{multlined}\label{EQ:region_outer_sum2}
\end{align}\label{EQ:region_outer}%
\end{subequations}
where the union is over all PMFs $Q_{W,U,V}Q_{X|U,V}\in\mathcal{P}(\mathcal{W}\times\mathcal{U}\times\mathcal{V}\times\mathcal{X})$, each inducing a joint distribution $Q_{W,U,V}Q_{X|U,V}W_{Y_1,Y_2|X}$. $\mathcal{R}_{\mathsf{O}}(L_1,L_2)$ is convex and the following inclusion holds:
\begin{equation}
\mathcal{C}(L_1,L_2)\subseteq\mathcal{R}_{\mathsf{O}}(L_1,L_2).\label{EQ:inclusion_outer}
\end{equation}
\end{theorem}


\begin{remark}[Relation to UVW-Outer Bound]
The best known outer bounds on the capacity region of a BC with a common message are the UVW-outer bound \cite[Bound 2]{UVW_Outer2010} and the New-Jersey outer bound \cite{NJ_Outer2008} which are equivalent. The region $\mathcal{R}_{\mathsf{O}}(\infty,\infty)$ recovers the UVW-outer bound since \eqref{EQ:region_outer11}-\eqref{EQ:region_outer12} and \eqref{EQ:region_outer21}-\eqref{EQ:region_outer22} are redundant.
\end{remark}


The inner and outer bounds in Theorems \ref{TM:inner_bound} and \ref{TM:outer_bound} are tight for SD-BCs and give rise to the following theorem.

\begin{theorem}[Leakage-Capacity - SD-BC]\label{TM:SDBC_leakage_capacity}
The $(L_1,L_2)$-leakage-capacity region $\mathcal{C}_{\mathsf{SD}}(L_1,L_2)$ of a SD-BC $\mathds{1}_{\{Y_1=y_1(X)\}}W_{Y_2|X}$ is the closure of the union of rate triples $(R_0,R_1,R_2)\in\mathbb{R}^3_+$ satisfying:
\begin{subequations}
\begin{align}
R_0 &\leq \min\Big\{I(W;Y_1),I(W;Y_2)\Big\}\label{EQ:region_SDBC0}\\
R_1 &\leq H(Y_1|W,V,Y_2)+L_1\label{EQ:region_SDBC11}\\
R_0+R_1 &\leq H(Y_1|W)+\min\mspace{-3mu}\Big\{\mspace{-1.5mu}I(W;Y_1),\mspace{-1.5mu}I(W;Y_2)\mspace{-1.5mu}\Big\}\label{EQ:region_SDBC12}\\
R_2 &\leq I(V;Y_2|W)-I(V;Y_1|W)+L_2\label{EQ:region_SDBC21}\\
R_0+R_2 &\leq I(V;Y_2|W)\mspace{-2mu}+\mspace{-2mu}\min\mspace{-3mu}\Big\{\mspace{-1.5mu}I(W;Y_1),\mspace{-1.5mu}I(W;Y_2)\mspace{-1.5mu}\Big\}\label{EQ:region_SDBC22}\\
\sum_{j=0,1,2}&\leq\mspace{-1.5mu}\begin{multlined}[t][.3\textwidth] H(Y_1|W,V)+I(V;Y_2|W)\\\mspace{58mu}+\min\Big\{I(W;Y_1),I(W;Y_2)\Big\}\end{multlined}\label{EQ:region_SDBC_sum12}
\end{align}\label{EQ:region_SDBC}%
\end{subequations}
where the union is over all PMFs $Q_{W,V,X}\in\mathcal{P}(\mathcal{W}\times\mathcal{V}\times\mathcal{X})$, each inducing a joint distribution $Q_{W,V,X}\mathds{1}_{\{Y_1=y_1(X)\}}W_{Y_2|X}$. Furthermore, $\mathcal{C}_{\mathsf{SD}}(L_1,L_2)$ is convex.
\end{theorem}
The direct part of Theorem \ref{TM:SDBC_leakage_capacity} follows from Theorem \ref{TM:inner_bound} by taking $U_0=W$, $U_1=Y_1$ and $U_2=V$, while Theorem \ref{TM:outer_bound} is used for the converse. See Section \ref{SUBSEC:SDBC_proof} for the details.

%

\begin{remark}[SD-BC Result - Special Cases]
All four cases of the SD-BC concerning secrecy (i.e., when neither, either or both messages are secret) are solved and their solutions are retrieved from $\mathcal{C}_{\mathsf{SD}}(L_1,L_2)$ by inserting the appropriate values of $L_j$, $j=1,2$. This property of $\mathcal{C}_{\mathsf{SD}}(L_1,L_2)$ is discussed in Section \ref{SUBSEC:special_cases_SDBC}.
\end{remark}

The inner and outer bounds in Theorems \ref{TM:inner_bound} and \ref{TM:outer_bound} also match when the message set is degraded, i.e., when $M_2=0$ and there is only one private message.

\begin{theorem}[Leakage-Capacity - Degraded Message Set]\label{TM:DMBC_leakage_capacity}
The $L_1$-leakage-capacity region $\mathcal{C}_{\mathsf{DM}}(L_1)$ of a BC with a degraded message set ($M_2=0$) and a privacy leakage constraint is the closure of the union of rate pairs $(R_0,R_1)\in\mathbb{R}^2_+$ satisfying:
\begin{subequations}
\begin{align}
R_0 &\leq \min\Big\{I(W;Y_1),I(W;Y_2)\Big\}\label{EQ:region_DMBC0}\\
R_1 &\leq I(U;Y_1|W)-I(U;Y_2|W)+L_1\label{EQ:region_DMBC11}\\
R_0+R_1 &\leq I(U;Y_1|W)+\min\mspace{-3mu}\big\{I(W;Y_1)\mspace{-1.5mu},\mspace{-1.5mu}I(W;Y_2)\big\}\label{EQ:region_DMBC13}
\end{align}\label{EQ:region_DMBC}%
\end{subequations}
where the union is over all PMFs $Q_{W,U}Q_{X|U}\in\mathcal{P}(\mathcal{W}\times\mathcal{U}\times\mathcal{X})$, each inducing a joint distribution $Q_{W,U}Q_{X|U}W_{Y_1,Y_2|X}$. Furthermore, $\mathcal{C}_{\mathsf{DM}}(L_1)$ is convex.
\end{theorem}

\begin{IEEEproof} The direct part follows by setting $R_2=0$, $U_0=W$, $U_1=U$ and $U_2=0$ in Theorem \ref{TM:inner_bound}. For the converse we show that $\mathcal{R}_{\mathsf{O}}(L_1,L_2)\subseteq\mathcal{C}_{\mathsf{DM}}(L_1)$. Clearly, \eqref{EQ:region_DMBC0}, \eqref{EQ:region_DMBC11} and \eqref{EQ:region_DMBC13} coincide with \eqref{EQ:region_outer0}, \eqref{EQ:region_outer12} and \eqref{EQ:region_outer13}, respectively. Dropping the rest of the inequalities from \eqref{EQ:region_outer} completes the proof.
\end{IEEEproof}

\begin{remark}[Degraded Message Set Result - Special Cases]
The BC with a degraded message set and a privacy leakage constraint captures the BC with confidential messages \cite{Csiszar_Korner_BCconfidential1978} and the BC with a degraded message set \cite{Korner_BC_DegradedMessageSet1977}. The former is obtained by taking $L_1=0$, while $L_1\to\infty$ recovers the latter. Setting $L_1=0$ or $L_1\to\infty$ into $\mathcal{C}_{\mathsf{DM}}(L_1)$ recovers the capacity regions of these special cases (see Section \ref{SUBSEC:special_cases_BC_Conf} for more details).
\end{remark}

We next characterize the leakage-capacity region of a PD-BC $W_{Y_1|X}W_{Y_2|Y_1}$ with privacy leakage constraints and without a common message ($M_0=0$). Since $X-Y_1-Y_2$ forms a Markov chain, it is impossible to achieve non-trivial leakage constraints on the message $M_2$. Accordingly, the leakage-capacity region of the PD-BC (where $X-Y_1-Y_2$) is defined only through $L_1$.

\begin{corollary}[Leakage-Capacity - PD-BC]\label{COR:PDBC_leakage_capacity}
The $L_1$-leakage-capacity region $\mathcal{C}_{\mathsf{PD}}(L_1)$ of a PD-BC $W_{Y_1|X}W_{Y_2|Y_1}$ without a common message is the closure of the union over the same domain as $\mathcal{C}_{\mathsf{DM}}(L_1)$ of rate pairs $(R_1,R_2)\in\mathbb{R}^2_+$ satisfying \eqref{EQ:region_DMBC}, while recasting $R_0$ as $R_2$ and noting that $\min\big\{I(W;Y_1),I(W;Y_2)\big\}=I(W;Y_2)$.
\end{corollary}

The proof of Corollary \ref{COR:PDBC_leakage_capacity} is similar to that of Theorem \ref{TM:DMBC_leakage_capacity} and is omitted.

\begin{remark}[Cardinality Bounds]
Cardinality bounds for the auxiliary random variables in Theorems \ref{TM:inner_bound}, \ref{TM:outer_bound}, \ref{TM:SDBC_leakage_capacity} and \ref{TM:DMBC_leakage_capacity} can be derived using the perturbation method \cite[Appendix C]{ElGamal2011} or techniques such as in \cite{UVW_Outer2010} and \cite{Nair_Matron_Bounds2013}. The computability of the derived regions is not in the scope of this work.
\end{remark}

\section{Special Cases}\label{SEC:special_cases}


\subsection{The Gelfand-Pinsker Inner Bound}\label{SUBSEC:special_cases_Marton}

Theorem \ref{TM:inner_bound} generalizes the Gelfand-Pinsker region for the BC with a common message \cite[Theorem 1]{GP_SemideterministicBC1980} to the case with privacy leakage constraints. In other words, $\mathcal{R}_{\mathsf{I}}(\infty,\infty)$ recovers the result from \cite{GP_SemideterministicBC1980}, which is tight for every BC (without secrecy) whose capacity region is known.


\subsection{UVW-Outer Bound}\label{SUBSEC:special_cases_UVW}

The New-Jersey outer bound was derived in \cite{NJ_Outer2008} and shown to be at least as good as the previously known bounds. A simpler version of this outer bound was established in \cite{UVW_Outer2010} and was named the UVW-outer bound. The UVW-outer bound is given by $\mathcal{R}_{\mathsf{O}}(\infty,\infty)$.

%

\subsection{Liu-Mari{\'c}-Spasojevi{\'c}-Yates Inner Bound}\label{SUBSEC:special_cases_Liu}

In \cite{BC_Confidential_Yates2008} an inner bound on the secrecy-capacity region of a BC $W_{Y_1,Y_2|X}$ with two confidential messages (each destined for one of the receivers and kept secret from the other) was characterized as the set of rate pairs $(R_1,R_2)\in\mathbb{R}^2_+$ satisfying:
\begin{subequations}
\begin{align}
R_1 &\leq I(U_1;Y_1|U_0)-I(U_1;U_2,Y_2|U_0)\label{EQ:region_liu_secrecy1}\\
R_2 &\leq I(U_2;Y_2|U_0)-I(U_2;U_1,Y_1|U_0)\label{EQ:region_liu_secrecy2}
\end{align}\label{EQ:region_liu_secrecy}%
\end{subequations}
where the union is over all PMFs $Q_{U_0,U_1,U_2}Q_{X|U_1,U_2}\in\mathcal{P}(\mathcal{U}_0\times\mathcal{U}_1\times\mathcal{U}_2\times\mathcal{X})$, each inducing a joint distribution $Q_{U_0,U_1,U_2}Q_{X|U_1,U_2}W_{Y_1,Y_2|X}$. This inner bound is tight for SD-BCs \cite{Semi-det_BC_secrect_two2009} and MIMO Gaussian BCs \cite{Poor_Shamai_Gaussian_MIMO_BC_Secrecy2010}. Setting $R_0=0$ in $\mathcal{R}_{\mathsf{I}}(0,0)$ recovers \eqref{EQ:region_liu_secrecy}.


\subsection{SD-BCs with and without Secrecy}\label{SUBSEC:special_cases_SDBC}

The SD-BC $\mathds{1}_{\{Y_1=y_1(X)\}}W_{Y_2|X}$ without a common message, i.e., when $R_0=0$, is solved when both, either or neither private messages are secret (see \cite{Semi-det_BC_secrect_two2009,Goldfeld_Weak_Secrecy_ISIT2015,Semi-det_BC_secrect_one2009} and \cite{GP_SemideterministicBC1980}, respectively). Setting $L_j=0$, for $j=1,2$, reduces the SD-BC with privacy leakage constraints to the problem where $M_j$ is secret. Taking $L_j\to\infty$ results in a SD-BC without a leakage constraint on $M_j$. We use Theorem \ref{TM:SDBC_leakage_capacity} to obtain the leakage-capacity region of the SD-BC without a common message.

\begin{corollary}[Leakage-Capacity - SD-BC without $\bm{M_0}$]\label{COR:SDBC_nocommon_leakage_capacity}
The $(L_1,L_2)$-leakage-capacity region $\mathcal{C}_{\mathsf{SD}}^0(L_1,L_2)$ of a SD-BC $\mathds{1}_{\{Y_1=y_1(X)\}}W_{Y_2|X}$ without a common message is the closure of the union over the domain stated in Theorem \ref{TM:SDBC_leakage_capacity} of rate pairs $(R_1,R_2)\in\mathbb{R}^2_+$ satisfying:
\begin{subequations}
\begin{align}
R_1 &\leq H(Y_1|W,V,Y_2)+L_1\label{EQ:region_nocommon_SDBC11}\\
R_1 &\leq H(Y_1|W)+\min\Big\{I(W;Y_1),I(W;Y_2)\Big\}\label{EQ:region_nocommon_SDBC12}\\
R_2 &\leq I(V;Y_2|W)-I(V;Y_1|W)+L_2\label{EQ:region_nocommon_SDBC21}\\
R_2 &\leq I(V;Y_2|W)\mspace{-2mu}+\mspace{-2mu}\min\mspace{-3mu}\Big\{\mspace{-1.5mu}I(W;Y_1),\mspace{-1.5mu}I(W;Y_2)\mspace{-1.5mu}\Big\}\label{EQ:region_nocommon_SDBC22}\\
R_1+R_2&\leq\begin{multlined}[t][.3\textwidth] H(Y_1|W,V)+I(V;Y_2|W)\\\mspace{55mu}+\min\Big\{I(W;Y_1),I(W;Y_2)\Big\}.\end{multlined}\label{EQ:region_nocommon_SDBC_sum12}
\end{align}\label{EQ:region_nocommon_SDBC}%
\end{subequations}
\end{corollary}
%


\subsubsection{Neither Message is Secret}\label{SUBSEC:special_cases_neither}

If $L_1,L_2\to\infty$, the SD-BC with privacy leakage constraints reduces to the classic case without secrecy \cite{GP_SemideterministicBC1980}. We recover $\mathcal{C}_{\mathsf{SD}}^0(\infty,\infty)$ by choosing $W=0$ so that \eqref{EQ:region_nocommon_SDBC} becomes 
\begin{subequations}
\begin{align}
R_1 &\leq H(Y_1)\label{EQ:region_neither_SDBC1}\\
R_2 &\leq I(V;Y_2)\label{EQ:region_neither_SDBC2}\\
R_1+R_2 &\leq H(Y_1|V)+I(V;Y_2)\label{EQ:region_neither_SDBC_sum}
\end{align}\label{EQ:region_neither_SDBC}%
\end{subequations}
This agrees with the discussion in Section \ref{SUBSEC:special_cases_Marton} since Marton's inner bound is tight for SD-BCs.



\subsubsection{Only $M_1$ is Secret}\label{SUBSEC:special_cases_m1only}

The SD-BC where $M_1$ is a secret is obtained by taking $L_1=0$ and $L_2\to\infty$. The secrecy-capacity region was derived in \cite[Corollary 4]{Goldfeld_Weak_Secrecy_ISIT2015} and is the closure of the union over the same domain as \eqref{EQ:region_neither_SDBC} of rate pairs $(R_1,R_2)\in\mathbb{R}^2_+$ satisfying:
\begin{subequations}
\begin{align}
R_1 &\leq H(Y_1|V,Y_2)\label{EQ:region_m1only_SDBC1}\\
R_2 &\leq I(V;Y_2).\label{EQ:region_m1only_SDBC2}
\end{align}\label{EQ:region_m1only_SDBC}
\end{subequations}

\vspace{-4mm}
\noindent To see that $\mathcal{C}^0_{\mathsf{SD}}(0,\infty)$ and \eqref{EQ:region_m1only_SDBC} match, first note that when $L_1=0$, \eqref{EQ:region_nocommon_SDBC12} is redundant due to \eqref{EQ:region_nocommon_SDBC11}. The sum rate bound \eqref{EQ:region_nocommon_SDBC_sum12} also becomes inactive as it is implied by adding \eqref{EQ:region_nocommon_SDBC11} and \eqref{EQ:region_nocommon_SDBC22}. Setting $W=0$ in $\mathcal{C}^0_{\mathsf{SD}}(0,\infty)$ now recovers \eqref{EQ:region_m1only_SDBC}.

\begin{remark}[Relation to Optimal Coding Scheme]
The optimal code for the SD-BC with a secret message $M_1$ employs no public message and relies on double-binning the codebook of $M_1$, while $M_2$ is transmitted at maximal rate and no binning of its codebook is performed. The optimality of $W=0$ in $\mathcal{C}^0_{\mathsf{SD}}(0,\infty)$ corresponds to the absence of the public messages. Furthermore, referring to the bounds in Section \ref{SUBSEC:inner_proof}, inserting $L_1=0$ and $L_2\to\infty$ into our code construction results in \eqref{EQ:achiev_rb1} and \eqref{EQ:achiev_extra_rb1_R2} becoming inactive since \eqref{EQ:achiev_extra_rb1} is the dominant constraint. Consequently, the redundancy used for correlating the transmission and ensuring security (i.e., the double-binning) is present only in the $M_1$ codebook.
\end{remark}%
%
%


\subsubsection{Only $M_2$ is Secret}\label{SUBSEC:special_cases_m2only}

The SD-BC where $M_2$ is secret is obtained by taking $L_1\to\infty$ and $L_2=0$. The secrecy-capacity region is the closure of the union of rate pairs $(R_1,R_2)\in\mathbb{R}^2_+$ satisfying:
\begin{subequations}
\begin{align}
R_1 &\leq H(Y_1)\label{EQ:region_m2only_SDBC11}\\
R_1 &\leq H(Y_1|W)+I(W;Y_2)\label{EQ:region_m2only_SDBC12}\\
R_2 &\leq I(V;Y_2|W)-I(V;Y_1|W)\label{EQ:region_m2only_SDBC2}
\end{align}\label{EQ:region_m2only_SDBC}%
\end{subequations}
where the union is over all PMFs $Q_{W,V,X}\in\mathcal{P}(\mathcal{W}\times\mathcal{V}\times\mathcal{X})$, each inducing a joint distribution $Q_{W,V,X}\mathds{1}_{\{Y_1=y_1(X)\}}W_{Y_2|X}$ \cite[Theorem 1]{Semi-det_BC_secrect_one2009}. Using Corollary \ref{COR:SDBC_nocommon_leakage_capacity}, the bounds \eqref{EQ:region_nocommon_SDBC} become
\begin{subequations}
\begin{align}
R_1 &\leq H(Y_1|W)+\min\Big\{I(W;Y_1),I(W;Y_2)\Big\}\label{EQ:region_m2only_alt_SDBC1}\\
R_2 &\leq I(V;Y_2|W)-I(V;Y_1|W)\label{EQ:region_m2only_alt_SDBC2}\\
R_1+R_2 &\leq\begin{multlined}[t][.3\textwidth] H(Y_1|W,V)+I(V;Y_2|W)\\\mspace{55mu}+\min\Big\{I(W;Y_1),I(W;Y_2)\Big\}.\end{multlined}\label{EQ:region_m2only_alt_SDBC_sum}
\end{align}\label{EQ:region_m2only_alt_SDBC}%
\end{subequations}
and \eqref{EQ:region_m2only_alt_SDBC_sum} is redundant by adding \eqref{EQ:region_m2only_alt_SDBC1} and \eqref{EQ:region_m2only_alt_SDBC2}. The regions from \eqref{EQ:region_m2only_SDBC} and \eqref{EQ:region_m2only_alt_SDBC} thus coincide.

The effect of $L_1\to\infty$ and $L_2=0$ on the bins in our coding scheme (Section \ref{SUBSEC:inner_proof}) is analogous to the one described in Section \ref{SUBSEC:special_cases_m1only}. In contrast to Section \ref{SUBSEC:special_cases_m1only}, however, here the achievability of  \eqref{EQ:region_m2only_alt_SDBC} requires a common message. Since $L_2=0$, \eqref{EQ:achiev_partial_rates_bound2} implies that the public message is a portion of $M_1$ \emph{only}. Keeping in mind that the public message is decoded by both receivers, unless $R_{20}=0$ (i.e., unless the public message contains no information about $M_2$) the secrecy constraint will be violated. 



\subsubsection{Both Messages are Secret}\label{SUBSEC:special_cases_both}

Taking $L_1=L_2=0$ recovers the SD-BC where both messages are secret. The secrecy-capacity region for this case was found in \cite[Theorem 1]{Semi-det_BC_secrect_two2009} and is the closure of the union of rate pairs $(R_1,R_2)\in\mathbb{R}^2_+$ satisfying:
\begin{subequations}
\begin{align}
R_1 &\leq H(Y_1|W,V,Y_2)\label{EQ:region_both_SDBC1}\\
R_2 &\leq I(V;Y_2|W)-I(V;Y_1|W)\label{EQ:region_both_SDBC2}
\end{align}\label{EQ:region_both_SDBC}%
\end{subequations}
where the union is over all PMFs $Q_{W,V}Q_{X|V}\in\mathcal{P}(\mathcal{W}\times\mathcal{V}\times\mathcal{X})$, each inducing a joint distribution $Q_{W,V}Q_{X|V}\mathds{1}_{\{Y_1=y_1(X)\}}W_{Y_2|X}$. The region \eqref{EQ:region_both_SDBC} coincides with $\mathcal{C}^0_{\mathsf{SD}}(0,0)$. Restricting the union in $\mathcal{C}^0_{\mathsf{SD}}(0,0)$ to encompass only PMFs that satisfy the Markov relation $W-V-X$ does not shrink the region. This is since in the proof of Theorem \ref{TM:outer_bound} we define $V_q\triangleq (M_2,W_q)$, and therefore, $X_q-V_q-W_q$ forms a Markov chain for every $q\in[1:n]$.

\begin{remark}[Relation to Optimal Coding Scheme]
The coding scheme that achieves \eqref{EQ:region_both_SDBC} uses double-binning for the codebooks of both private messages. To ensure confidentiality, the rate bounds of each message includes the penalty term $I(U_1;U_2|U_0)$. Note that without the confidentiality constraints, Marton's coding scheme \cite{Marton_BC1979} requires only that the sum-rate has that penalty term. This is evident from our scheme by setting $L_1=L_2=0$ in \eqref{EQ:achiev_partial_rates_bound2}, \eqref{EQ:achiev_extra_rb1} and \eqref{EQ:achiev_extra_rb1_R2}, which makes \eqref{EQ:achiev_rb1} redundant.
\end{remark}

\subsection{BCs with One Private Message}\label{SUBSEC:special_cases_BC_Conf}

Consider the BC with leakage constraints in which $M_2=0$; its leakage-capacity region $\mathcal{C}_{\mathsf{DM}}(L_1)$ is stated in Theorem \ref{TM:DMBC_leakage_capacity}. We show that $\mathcal{C}_{\mathsf{DM}}(L_1)$ recovers the secrecy-capacity region of the BC with confidential messages \cite{Csiszar_Korner_BCconfidential1978} and the capacity region of the BC with a degraded message set (without secrecy) \cite{Korner_BC_DegradedMessageSet1977}.\\


\subsubsection{BCs with Confidential Messages}

The secrecy-capacity region of the BC with confidential messages was derived in \cite{Csiszar_Korner_BCconfidential1978} and is the union over the same domain as in Theorem \ref{TM:DMBC_leakage_capacity} of rate pairs $(R_0,R_1)\in\mathbb{R}^2_+$ satisfying:
\begin{subequations}
\begin{align}
R_0 &\leq \min\Big\{I(W;Y_1),I(W;Y_2)\Big\}\label{EQ:region_BCConf0}\\
R_1 &\leq I(U;Y_1|W)-I(U;Y_2|W).\label{EQ:region_BCConf1}
\end{align}\label{EQ:region_BCConf}%
\end{subequations}
Inserting $L_1=0$ into the result of Theorem \ref{TM:DMBC_leakage_capacity} produces \eqref{EQ:region_BCConf}.

Our code construction (Section \ref{SUBSEC:inner_proof}) with $L_1=0$ and $U_2=0$ reduces to a superposition code for which the outer codebook (that is associated with the confidential message) is binned. This is a secrecy-capacity achieving coding scheme for the BC with confidential messages.

\begin{remark}[Wiretap Channel]
The BC with confidential messages captures the WTC by setting $M_0=0$. Thus, the WTC is also a special case of the BC with privacy leakage constraints.
\end{remark}


\subsubsection{BCs with a Degraded Message Set}

If $L_1\to\infty$, we get the BC with a degraded message set \cite{Korner_BC_DegradedMessageSet1977}. Inserting $L_1\to\infty$ into $\mathcal{C}_{\mathsf{DM}}(L_1)$ and setting $U=X$ we recover the union of rate pairs $(R_0,R_1)\in\mathbb{R}_+$ satisfying:
\begin{subequations}
\begin{align}
R_0 &\leq \min\Big\{I(W;Y_1),I(W;Y_2)\Big\}\label{EQ:region_DMBC_alt0}\\
R_0+R_1 &\leq I(X;Y_1|W)+I(W;Y_2)\label{EQ:region_DMBC_alt_sum1}\\
R_0+R_1 &\leq I(X;Y_1)\label{EQ:region_DMBC_alt_sum2}
\end{align}\label{EQ:region_DMBC_alt}%
\end{subequations}
where the union is over all PMFs $Q_{W,X}\in\mathcal{P}(\mathcal{V}\times\mathcal{X})$, each induces a joint distribution $Q_{W,X}W_{Y_1,Y_2|X}$. 

$\mathcal{C}_{\mathsf{DM}}(L_1)$ in \eqref{EQ:region_DMBC_alt} matches \cite[Theorem 7]{Liang_Kramer_RBC2007} which establishes the union over all 
PMFs
$Q_{T,U,X}\in\mathcal{P}(\mathcal{T}\times\mathcal{U}\times\mathcal{X})$ of rate pairs $(R_0,R_1)\in\mathbb{R}_+$ with
\begin{subequations}
\begin{align}
R_0 &\leq \min\Big\{I(T;Y_1),I(T;Y_2)\Big\}\label{EQ:region_DMBC_UB0}\\
R_0+R_1 &\leq I(X;Y_1|T,U)+I(T,U;Y_2)\label{EQ:region_DMBC_UB_sum1}\\
R_0+R_1 &\leq I(X;Y_1)\label{EQ:region_DMBC_UB_sum2}
\end{align}\label{EQ:region_DMBC_UB}%
\end{subequations}
as an outer bound on the capacity region of interest. The RHS of \eqref{EQ:region_DMBC_UB0} can be bounded as
\begin{equation}
\min\Big\{I(T;Y_1),I(T;Y_2)\Big\}\leq \min\Big\{I(T,U;Y_1),I(T,U;Y_2)\Big\}
\end{equation}
and relabeling $W=(T,U)$ matches \eqref{EQ:region_DMBC_alt}.


\section{Example}\label{SEC:example}

\begin{figure}[t!]
    \begin{center}
        \begin{psfrags}
            \psfragscanon
            \psfrag{A}[][][1]{$0$}
            \psfrag{B}[][][1]{$1$}
            \psfrag{C}[][][1]{$2$}
            \psfrag{D}[][][1]{\ $X$}
            \psfrag{E}[][][1]{\ \ Encoder}
            \psfrag{F}[][][1]{\ \ \ $Y_1$}
            \psfrag{G}[][][1]{$0$}
            \psfrag{H}[][][1]{$1$}
            \psfrag{I}[][][1]{$0$}
            \psfrag{J}[][][1]{$1$}
            \psfrag{K}[][][1]{\ \ \ Decoder 1}
            \psfrag{L}[][][1]{\ \ \ $Y_2$}
            \psfrag{M}[][][1]{\ \ \ Decoder 2}
            \psfrag{N}[][][1]{\ \ $R_{12}$}
            \psfrag{X}[][][1]{\ \ \ \ \ \ \ \ \ \ \ \ \ \ \ \ \ \ $I(M_2;\mathbf{Y}_1)\leq nL_2$}
            \psfrag{Y}[][][1]{\ \ \ \ \ \ \ \ \ \ \ \ \ \ \ \ \ \ $I(M_1;\mathbf{Y}_2)\leq nL_1$}
            \includegraphics[scale = .55]{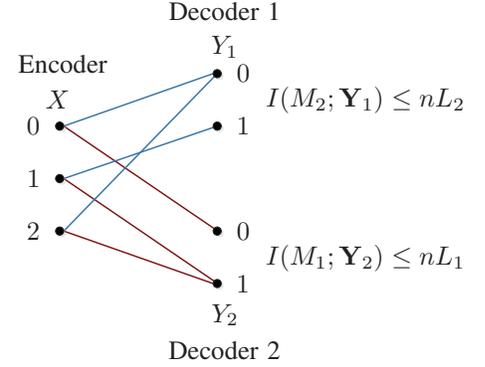}
            \caption{Blackwell BC with privacy leakage constraints.} \label{FIG:blackwell}
            \psfragscanoff
        \end{psfrags}
     \end{center}
 \end{figure}

Suppose the channel from the transmitter to receivers 1 and 2 is the BW-BC without a common message as illustrated in Fig. \ref{FIG:blackwell} \cite{vanderMeulen_Blackwell1975,Gelfand_Blackwell1977}. Using Corollary \ref{COR:SDBC_nocommon_leakage_capacity}, the $(L_1,L_2)$-leakage-capacity region of a deterministic BC (DBC) is the following.


\begin{corollary}[Leakage-Capacity - Deterministic BC]\label{COR:DBC_leakage_capacity}
The $(L_1,L_2)$-leakage-capacity region $\mathcal{C}_{\mathsf{D}}(L_1,L_2)$ of the DBC $\mathds{1}_{\{Y_1=y_1(X)\}\cap\{Y_2=y_2(X)\}}$ without a common message is the union of rate pairs $(R_1,R_2)\in\mathbb{R}^2_+$ satisfying:
\begin{subequations}
\begin{align}
R_1 &\leq \min\big\{H(Y_1)\mspace{3mu},\mspace{3mu}H(Y_1|Y_2)+L_1\big\}\label{EQ:region_DBC1}\\
R_2 &\leq \min\big\{H(Y_2)\mspace{3mu},\mspace{3mu}H(Y_2|Y_1)+L_2\big\}\label{EQ:region_DBC2}\\
R_1+R_2 &\leq H(Y_1,Y_2)\label{EQ:region_DBC_sum}
\end{align}\label{EQ:region_DBC}
\end{subequations}

\vspace{-4mm}
\noindent where the union is over all input PMFs $Q_X\in\mathcal{P}(\mathcal{X})$.
\end{corollary}

The proof of Corollary \ref{COR:DBC_leakage_capacity} is relegated to Appendix \ref{APPEN:DBC_leakage_capacity_proof}. For the BW-BC, we parametrize the input PMF $Q_X\in\mathcal{P}\big(\{0,1,2\}\big)$ in Corollary \ref{COR:DBC_leakage_capacity} as
\begin{equation}
Q_X(0)=\alpha\ ,\ Q_X(1)=\beta\ ,\ Q_X(2)=1-\alpha-\beta,\label{EQ:input_dist_param}
\end{equation}
where $\alpha,\beta\in\mathbb{R}_+$ and $\alpha+\beta\leq1$. Using \eqref{EQ:input_dist_param}, the $(L_1,L_2)$-leakage-capacity region $\mathcal{C}_{\mathsf{BW}}(L_1,L_2)$ of the BW-BC is descried as the union of rate pairs $(R_1,R_2)\in\mathbb{R}^2_+$ satisfying:
\begin{subequations}
\begin{align}
\mspace{-10mu}R_1&\leq \min\left\{H_b(\beta)\mspace{3mu},\mspace{3mu}(1\mspace{-1.5mu}-\mspace{-1.5mu}\alpha)H_b\left(\frac{\beta}{1-\alpha}\right)\mspace{-1.5mu}+\mspace{-1.5mu}L_1\right\}\label{EQ:region_blackwell_r1}\\
\mspace{-10mu}R_2&\leq \min\left\{H_b(\alpha)\mspace{3mu},\mspace{3mu}(1\mspace{-1.5mu}-\mspace{-1.5mu}\beta)H_b\left(\frac{\alpha}{1-\beta}\right)\mspace{-1.5mu}+\mspace{-1.5mu}L_2\right\}\label{EQ:region_blackwell_r2}\\
\mspace{-10mu}R_1+R_2&\leq H_b(\alpha)+(1-\alpha)H_b\left(\frac{\beta}{1-\alpha}\right)\label{EQ:region_blackwell_sum}
\end{align}\label{EQ:region_blackwell}%
\end{subequations}
where the union is over all $\alpha,\beta\in\mathbb{R}_+$ with $\alpha+\beta\leq 1$.


\begin{figure}[t!]
    \begin{center}
        \begin{psfrags}
            \psfragscanon
            \psfrag{A}[][][0.85]{$R_1$ [bits/use]}
            \psfrag{B}[][][0.85]{$R_2$ [bits/use]}
            \psfrag{C}[][][0.9]{ }
            \psfrag{D}[][][0.9]{ }
            \psfrag{E}[][][0.9]{ }
            \psfrag{W}[][][0.73]{\ \ $\mspace{-2mu}L=0$}
            \psfrag{X}[][][0.73]{\ \ \ \ \ \ \ $\mspace{-6mu}L=0.05$}
            \psfrag{Y}[][][0.73]{\ \ \ \ \ $\mspace{-2mu}L=0.1$}
            \psfrag{Z}[][][0.73]{\ \ \ \ \ \ $\mspace{-4mu}L=0.4$}
            \subfloat[]{\includegraphics[scale = 0.5]{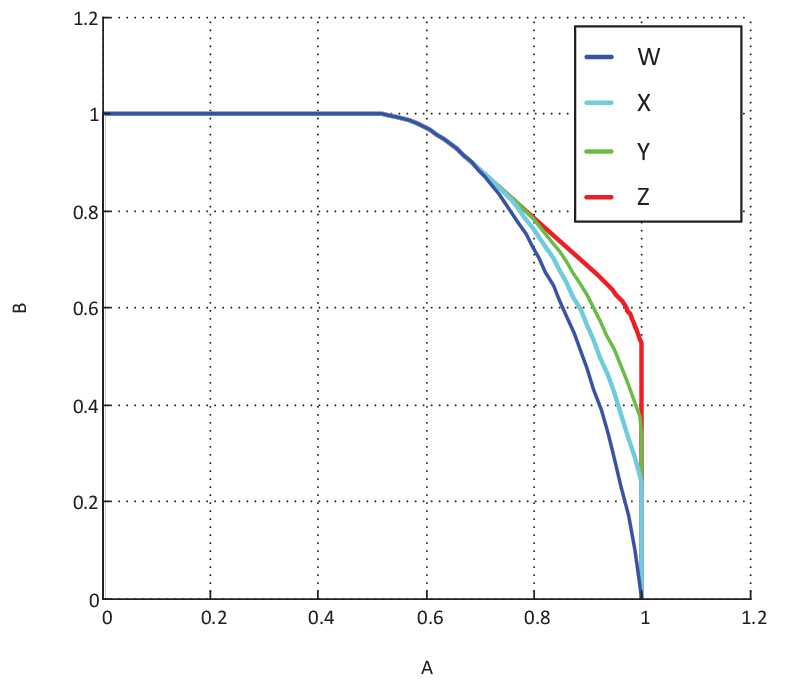}}\\
            \subfloat[]{\includegraphics[scale = 0.5]{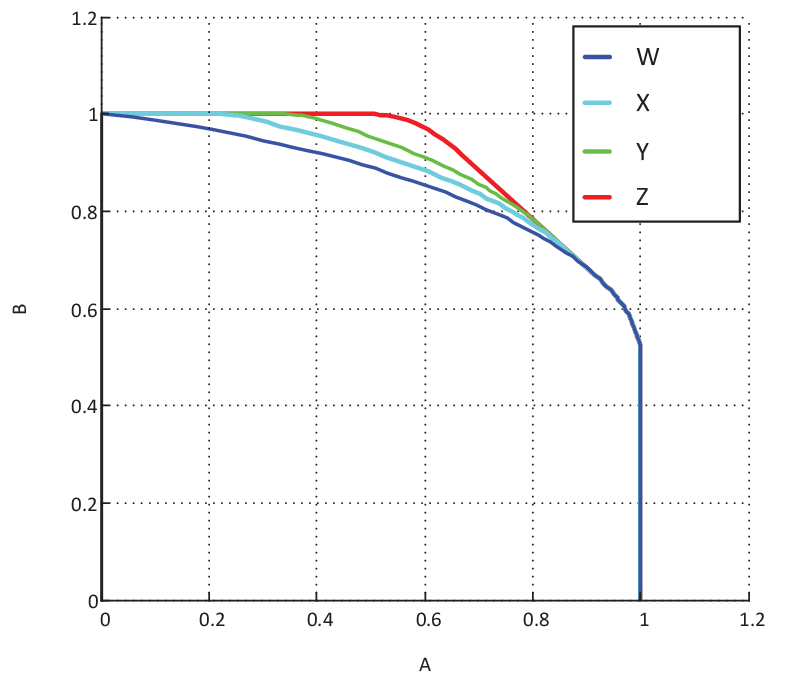}}\\
            \subfloat[]{\includegraphics[scale = 0.5]{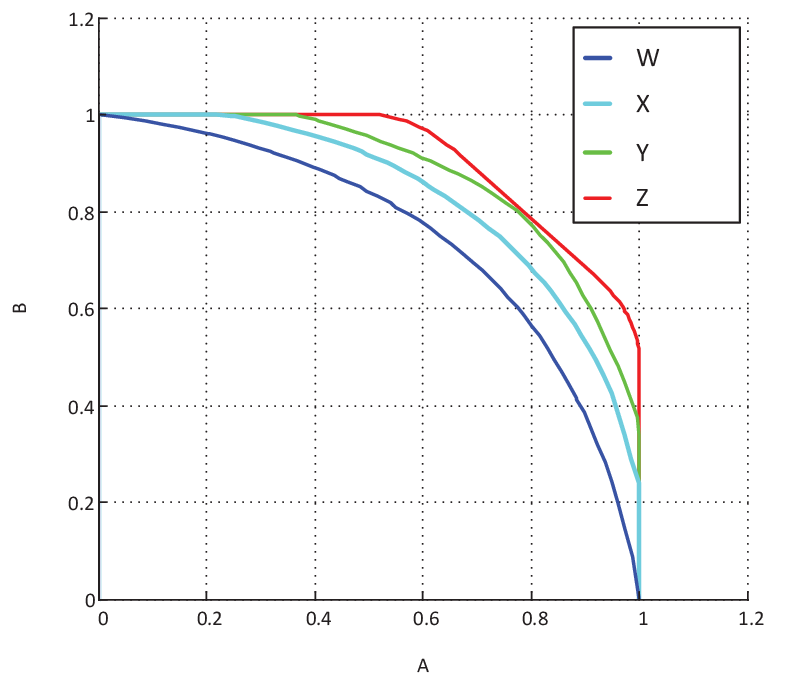}}
            \caption{$(L_1,L_2)$-leakage-capacity region of the BW-BC for three cases: (a) $L_1=L$ and $L_2\to\infty$; (b) $L_1\to\infty$ and $L_2=L$; (c) $L_1=L_2=L$.}\label{FIG:blackwell_region}
            \psfragscanoff
        \end{psfrags}
     \end{center}
 \end{figure}



\begin{figure}[t!]
    \begin{center}
        \begin{psfrags}
            \psfragscanon
            \psfrag{A}[][][0.9]{$L$ [bits/use]}
            \psfrag{B}[][][0.9]{$R_1+R_2$ [bits/use]}
            \includegraphics[scale = 0.5]{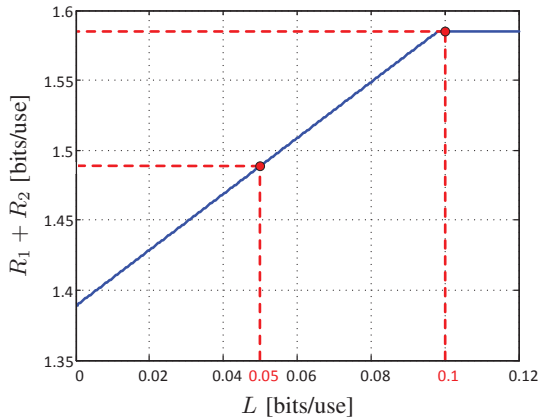}
            \caption{The sum-rate capacity versus the allowed leakage for $L_1=L_2=L$.}\label{FIG:Sum_vs_Leakage}
            \psfragscanoff
        \end{psfrags}
     \end{center}
 \end{figure}



 Fig. \ref{FIG:blackwell_region} illustrates $\mathcal{C}_{\mathsf{BW}}(L_1,L_2)$ for three cases. In Fig. \ref{FIG:blackwell_region}(a) $L_2\to\infty$ while $L_1\in\{0,0.05,0.1,0.4\}$. The blue (inner) line corresponds to $L_1=0$ and is the secrecy-capacity region of a BW-BC where $M_1$ is secret \cite[Fig. 5]{Goldfeld_Weak_Secrecy_ISIT2015}. The red (outer) line corresponds to $L_1=0.4$ (which is large enough to be thought of as $L_1\to\infty$) and depicts the capacity region of the classic BW-BC. As $L_1$ grows, the inner (blue) region converges to coincide with the outer (red) region. Fig. \ref{FIG:blackwell_region}(b) considers the opposite case, i.e., where $L_1\to\infty$ and $L_2\in\{0,0.05,0.1,0.4\}$, and is analogous to Fig. \ref{FIG:blackwell_region}(a). In Fig. \ref{FIG:blackwell_region}(c) we choose $L_1=L_2=L$, where $L\in\{0,0.05,0.1,0.4\}$, and we demonstrate the impact of two leakage constraints on the region. When $L=0$, one obtains the secrecy-capacity region of the BW-BC when both messages are confidential \cite{Semi-det_BC_secrect_two2009}. In each case, the capacity region grows with $L$ and saturates at the red (outer) region, for which neither message is secret.
Focusing on the symmetric case in Fig. \ref{FIG:blackwell_region}(c), we note that the saturation of the region at $L=0.4$ is implied by Corollary \ref{COR:inactive_leakage}. For the Blackwell BC with $L_1=L_2=L$, and some $\alpha,\beta\in\mathbb{R}_+$ with $\alpha+\beta\leq 1$, we denote by $L^\star(\alpha,\beta)$ the threshold from \eqref{EQ:maximal_leakage_Lj}, which reduces to 
\begin{equation}
L^\star(\alpha,\beta)=I(Y_1;Y_2)=H_b(\beta)-(1-\alpha)H_b\left(\frac{\beta}{1-\alpha}\right).
\end{equation}
As explained in Remark \ref{REM:inactive_leakage_explained}, for each leakage value $L$, Corollary \ref{COR:inactive_leakage} (along with some numerical calculations) can be used to tell whether a further increase of $L$ will induce a larger region or not. Accordingly, for each $L\in\{0,0.05,0.1,0.4\}$, we have calculated the maximum of $L^\star(\alpha,\beta)$ over the distributions that achieve the boundary points of the capacity region $\mathcal{C}_{\mathsf{BW}}(L,L)$. Denoting the value of the maximal $L^\star$ that corresponds to the allowed leakage $L\in\{0,0.05,0.1,0.4\}$ by $L^\star(L)$, we have
\begin{align*}
L^\star(0)&=L^\star(0.05)=0.15897\\    
L^\star(0.1)&=0.20101\\
L^\star(0.4)&=0.38317.\numberthis
\end{align*}
Observing that $L^\star(0.4)\leq L$, Corollary \ref{COR:inactive_leakage} and Remark \ref{REM:inactive_leakage_explained} imply that increasing $L$ beyond $0.4$ will not change the leakage-capacity region. Evidently, $\mathcal{C}_{\mathsf{BW}}(L,L)$ saturates at $L=0.4$. For $L\in\{0,0.05,0.1\}$, however, $L^\star(L)> L$ and consequently $\mathcal{C}_{\mathsf{BW}}(L',L')\subsetneq\mathcal{C}_{\mathsf{BW}}(L,L)$, for $L,L'\in\{0,0.05,0.1\}$ with $L'<L$.

The variation of the sum of rates $R_1+R_2$ as a function of $L$ is shown by the blue curve in Fig. \ref{FIG:Sum_vs_Leakage}; the red dashed vertical lines correspond to the values of $L$ considered in Fig. \ref{FIG:blackwell_region}. Note that for $0\leq L\leq 0.09818$, (\ref{EQ:region_blackwell_sum}) is inactive, and therefore, $R_1+R_2$ is bounded by the summation of (\ref{EQ:region_blackwell_r1}) and (\ref{EQ:region_blackwell_r2}). Thus, for $0\leq L\leq 0.09818$, the sum of rates $R_1+R_2$ increases linearly with $L$. For $L>0.09818$, the bound in (\ref{EQ:region_blackwell_sum}) is no longer redundant, and because it is independent of $L$, the sum rate saturates.


\begin{figure}[t!]
    \begin{center}
        \begin{psfrags}
            \psfragscanon
            \psfrag{A}[][][1]{$R_1$}
            \psfrag{G}[][][1]{\ \ $R_2$}
            \psfrag{D}[][][1]{$0$}
            \psfrag{E}[][][1]{$\ \ \ \ \ \ \ \ \ \ \ \ \ \ \ \ \ H(Y_2|Y_1)$}
            \psfrag{F}[][][1]{$\ \ \ \ \ H(Y_2)$}
            \psfrag{B}[][][1]{$\ \ \ \ H(Y_1)$}
            \psfrag{I}[][][1]{$\ \ \ \ \ \ \ \ \ \ \ \ \ H(Y_1|Y_2)$}
            \hspace{-6mm}\includegraphics[scale = 0.6]{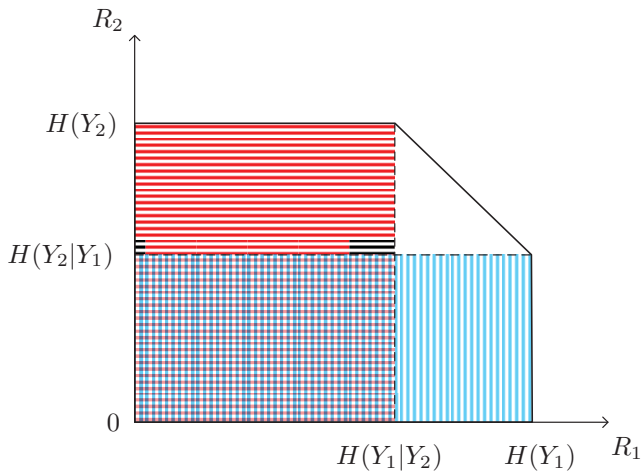}
            \caption{The pentagons/rectangles whose union produces the capacity region of a BW-BC for different secrecy cases: The outer pentagon corresponds to the case without secrecy; the red and blue rectangles correspond to $L_1=0$ and $L_2=0$, respectively; the inner rectangle corresponds to $L_1=L_2=0$.}\label{FIG:secrecy_regions}
            \psfragscanoff
        \end{psfrags}
     \end{center}
 \end{figure}


\begin{figure*}[!t]%
\setcounter{equation}{51}
\begin{align*}
    &\mathbb{E}_{\mathsf{B}_n(\mathbf{w})}\mathsf{F}\left(P_{I|\mathsf{B}_n},p^{(U)}_{\mathcal{I}_n}\right)\\
    &\stackrel{(a)}\geq|\mathcal{I}_n|^{\frac{1}{2}}|\mathcal{J}_n||\mathcal{K}_n|^{\frac{1}{2}}\mathbb{E}_{\mathsf{B}_n(\mathbf{w})}\left[2^{i\big(\mathbf{U}_1(\mathbf{w});\mathbf{V}_{1,1}(\mathbf{w})\big|\mathbf{w}\big)}\left(\sum\limits_{(\bar{\ell},\bar{j})}2^{i\big(\mathbf{U}_{\bar{\ell}}(\mathbf{w});\mathbf{V}_{\bar{j},1}\big|\mathbf{w}\big)}\right)^{\mspace{-6mu}-\frac{1}{2}}\left(\sum\limits_{(\tilde{j},\tilde{k})}\mspace{-3.5mu}2^{i\big(\mathbf{U}_1(\mathbf{w});\mathbf{V}_{\tilde{j},\tilde{k}}(\mathbf{w})\big)}\right)^{\mspace{-6mu}-\frac{1}{2}}\right]\\
    &\begin{multlined}[b][.99\textwidth]\stackrel{(b)}=|\mathcal{I}_n|^{\frac{1}{2}}|\mathcal{J}_n||\mathcal{K}_n|^{\frac{1}{2}}\mathbb{E}_{\mathbf{U}_1(\mathbf{w}),\mathbf{V}_{1,1}(\mathbf{w})}\vast[2^{i\big(\mathbf{U}_1(\mathbf{w}),\mathbf{V}_{1,1}(\mathbf{w})\big|\mathbf{w}\big)}\\\times\mathbb{E}_{\mathsf{B}_n(\mathbf{w})|\mathbf{U}_1(\mathbf{w}),\mathbf{V}_{1,1}(\mathbf{w})}\vast\{\left(\sum\limits_{(\bar{\ell},\bar{j})}2^{i\big(\mathbf{U}_{\bar{\ell}}(\mathbf{w});\mathbf{V}_{\bar{j},1}(\mathbf{w})\big|\mathbf{w}\big)}\right)^{\mspace{-6mu}-\frac{1}{2}}\left(\sum\limits_{(\tilde{j},\tilde{k})}2^{i\big(\mathbf{U}_1(\mathbf{w});\mathbf{V}_{\tilde{j},\tilde{k}}(\mathbf{w})\big|\mathbf{w}\big)}\right)^{\mspace{-6mu}-\frac{1}{2}}\vast\}\vast]\end{multlined}\\
    &\begin{multlined}[b][.99\textwidth]\stackrel{(c)}\geq|\mathcal{I}_n|^{\frac{1}{2}}|\mathcal{J}_n||\mathcal{K}_n|^{\frac{1}{2}}\mathbb{E}_{\mathbf{U}_1(\mathbf{w}),\mathbf{V}_{1,1}(\mathbf{w})}\vasti[2^{i\big(\mathbf{U}_1(\mathbf{w});\mathbf{V}_{1,1}(\mathbf{w})\big|\mathbf{w}\big)}\left(2^{i\big(\mathbf{U}_1(\mathbf{w});\mathbf{V}_{1,1}(\mathbf{w})\big|\mathbf{w}\big)}+|\mathcal{I}_n||\mathcal{J}_n|-1\right)^{\mspace{-6mu}-\frac{1}{2}}\\\times\left(2^{i\big(\mathbf{U}_1(\mathbf{w});\mathbf{V}_{1,1}(\mathbf{w})\big|\mathbf{w}\big)}+|\mathcal{J}_n||\mathcal{K}_n|-1\right)^{\mspace{-6mu}-\frac{1}{2}}\vasti]\end{multlined}\\
    &\stackrel{(d)}>\mathbb{E}_{Q_{U|W=\mathbf{w}}^nQ_{V|W=\mathbf{w}}^n}\left[2^{i(\mathbf{U};\mathbf{V}|\mathbf{w})}\left(1+\big(|\mathcal{I}_n||\mathcal{J}_n|\big)^{-1}2^{i(\mathbf{U};\mathbf{V}|\mathbf{w})}\right)^{\mspace{-6mu}-\frac{1}{2}}\left(1+\big(|\mathcal{J}_n||\mathcal{K}_n|\big)^{-1}2^{i(\mathbf{U};\mathbf{V}|\mathbf{w})}\right)^{\mspace{-6mu}-\frac{1}{2}}\right]\\
    &\stackrel{(e)}=\mathbb{E}_{Q_{U,V|W=\mathbf{w}}^n}\left[\left(1+\big(|\mathcal{I}_n||\mathcal{J}_n|\big)^{-1}2^{i(\mathbf{U};\mathbf{V}|\mathbf{w})}\right)^{\mspace{-6mu}-\frac{1}{2}}\left(1+\big(|\mathcal{J}_n||\mathcal{K}_n|\big)^{-1}2^{i(\mathbf{U};\mathbf{V}|\mathbf{w})}\right)^{\mspace{-6mu}-\frac{1}{2}}\right]\numberthis\label{EQ:lemma_proof_fidelity_expected}
\end{align*}
\hrulefill
\end{figure*}
\setcounter{equation}{46}

The regions in Fig. \ref{FIG:blackwell_region} are a union of rectangles or pentagons, each corresponds to a different input PMF $Q_X\in\mathcal{P}\big(\{0,1,2\}\big)$. In Fig. \ref{FIG:secrecy_regions} we illustrate a typical structure of these rectangles and pentagons for a fixed $Q_X$ at the extreme values of $L_1$ and $L_2$. When both $L_1$ and $L_2$ are sufficiently large, the leakage constraints degenerate and the classic BW-BC is obtained. Its capacity region (the red (outer) line in, e.g., Fig. \ref{FIG:blackwell_region}(c)) is a union of the pentagons depicted in Fig. \ref{FIG:secrecy_regions}. The secrecy-capacity region for $L_1=0$ and $L_2\to\infty$ (depicted by the blue line in Fig. \ref{FIG:blackwell_region}(a)) is a union of the red rectangles in Fig. \ref{FIG:secrecy_regions}. Similarly, when $L_2=0$ and $L_1\to\infty$ the secrecy-capacity region is a union of the blue rectangles in Fig. \ref{FIG:secrecy_regions}. Finally, if $L_1=L_2=0$ and both messages are secret, the secrecy-capacity region of the BW-BC is the union of the dark rectangles in Fig. \ref{FIG:secrecy_regions}, i.e., the intersection of the blue and the red regions. Fig. \ref{FIG:secrecy_regions} highlights that as $L_1$ and/or $L_2$ decrease, the underlying pentagons/rectangles (the union of which produces the admissible rate region) shrink, which results in a smaller region.


\section{Proofs}\label{SEC:proofs}
 

\subsection{Proof of Lemma \ref{LEMMA:uniform_approx}}\label{SUBSEC:uniform_approx_proof}

We derives sufficient conditions for
\begin{equation}
\mathbb{E}_{\mathsf{B}_n}\mathsf{F}\left(P_{I|\mathsf{B}_n},p^{(U)}_{\mathcal{I}_n}\right)\xrightarrow[n\to\infty]{}1
\end{equation}
which implies Lemma \ref{LEMMA:uniform_approx} using \eqref{EQ:fidelity_vs_TV_expected_sequences}.

First, for each $\mathcal{B}_n\in\mathfrak{B}_n$ the fidelity between the induced and the desired (uniform) distribution is
\begin{align*}
&\mathsf{F}\left(P_{I|\mathsf{B}_n=\mathcal{B}_n},p^{(U)}_{\mathcal{I}_n}\right)\\
       &\mspace{30mu}=\sum_\ell\left(\frac{1}{|\mathcal{I}_n||\mathcal{K}_n|}\sum_{(j,k)}\frac{2^{i\big(\mathbf{u}_\ell(\mathbf{w});\mathbf{v}_{j,k}(\mathbf{w})\big|\mathbf{w}\big)}}{\sum\limits_{(\bar{\ell},\bar{j})}2^{i\big(\mathbf{u}_{\bar{\ell}}(\mathbf{w});\mathbf{v}_{\bar{j},k}(\mathbf{w})\big|\mathbf{w}\big)}}\right)^{\mspace{-5mu}\frac{1}{2}}\mspace{-6mu}\numberthis\label{EQ:lemma_proof_fidelity}
\end{align*}
where as in \eqref{EQ:lemma_induced_marginal}, the information density is taken with respect to $Q_{U,V}^n$. Now, by the Cauchy-Schwarz inequality we have the following bound for $\big\{a_{j,k}\big\},\big\{b_{j,k}\big\}\subset \mathbb{R}_+$:
\begin{equation}
\sum_{j,k}\sqrt{a_{j,k} b_{j,k}}\leq\left(\sum_{j,k}a_{j,k}\right)^{\frac{1}{2}}\left(\sum_{j,k} b_{j,k}\right)^{\frac{1}{2}}.\label{EQ:lemma_proof_CS_Ineq}
\end{equation}
Using this on each of the summands from the right-hand side (RHS) of \eqref{EQ:lemma_proof_fidelity} with 
\begin{subequations}
\begin{equation}
a_{j,k}=\frac{1}{|\mathcal{I}_n||\mathcal{K}_n|}\frac{2^{i\big(\mathbf{u}_\ell(\mathbf{w});\mathbf{v}_{j,k}(\mathbf{w})\big|\mathbf{w}\big)}}{\sum\limits_{(\bar{\ell},\bar{j})}2^{i\big(\mathbf{u}_{\bar{\ell}}(\mathbf{w});\mathbf{v}_{\bar{j},k}(\mathbf{w})\big|\mathbf{w}\big)}}
\end{equation}
and
\begin{equation}
b_{j,k}=\frac{2^{i\big(\mathbf{u}_\ell(\mathbf{w});\mathbf{v}_{j,k}(\mathbf{w})\big|\mathbf{w}\big)}}{\sum\limits_{(\bar{\ell},\bar{k})}2^{i\big(\mathbf{u}_{\bar{\ell}}(\mathbf{w});\mathbf{v}_{\bar{j},\bar{k}}(\mathbf{w})\big|\mathbf{w}\big)}}
\end{equation}
\end{subequations}
we obtain
\begin{align*}
    F&\left(P_{I|\mathsf{B}_n=\mathcal{B}_n},p^{(U)}_{\mathcal{I}_n}\right)\\
    &\geq\sum_{(\ell,j,k)}\frac{2^{i\big(\mathbf{u}_\ell(\mathbf{w});\mathbf{v}_{j,k}(\mathbf{w})\big|\mathbf{w}\big)}}{\left(\mspace{-3.5mu}|\mathcal{I}_n||\mathcal{K}_n|\sum\limits_{(\bar{\ell},\bar{j})}2^{i\big(\mathbf{u}_{\bar{\ell}}(\mathbf{w});\mathbf{v}_{\bar{j},k}(\mathbf{w})\big|\mathbf{w}\big)}\right)^{\mspace{-6mu}\frac{1}{2}}}\\
    &\mspace{140mu}\times\frac{1}{\left(\sum\limits_{(\bar{j},\bar{k})}2^{i\big(\mathbf{u}_{\bar{\ell}}(\mathbf{w});\mathbf{v}_{\bar{j},\bar{k}}(\mathbf{w})\big|\mathbf{w}\big)}\right)^{\mspace{-6mu}\frac{1}{2}}}.\numberthis\label{EQ:lemma_proof_fidelity_CS}
\end{align*}

\begin{figure*}[!t]
\setcounter{equation}{60}
\begin{equation}
\mu(\mathcal{C}_n)=\prod_{m_p\in\mathcal{M}_p}Q^n_{U_0}\big(\mathbf{u}_0(m_p)\big)\prod_{j=1,2}\prod_{\substack{\big(m^{(j)}_p,m_{jj},w_j,i_j\big)\\\in\mathcal{M}_p\times\mathcal{M}_{jj}\times\mathcal{W}_j\times\mathcal{I}_j}}Q^n_{U_j|U_0}\Big(\mathbf{u}_j\big(m_p^{(j)},m_{jj},w_j,i_j\big)\Big|\mathbf{u}_0(m_p^{(j)})\Big)\label{EQ:codebook_probability}
\end{equation}
\hrulefill
\end{figure*}

\setcounter{equation}{50}
For any $\mathbf{w}\in\mathcal{W}^n$ with $Q_W^n(\mathbf{w})>0$, we evaluate the conditional expectation of the fidelity given $\mathbf{W}=\mathbf{w}$ as given in \eqref{EQ:lemma_proof_fidelity_expected} at the top of this page. First note that with respect to the notation from Section \ref{SEC:uniformity_lemma}, we have
\setcounter{equation}{52}
\begin{equation}
\mathbb{E}_{\mathsf{B}_n|\mathbf{W}=\mathbf{w}}\mathsf{F}\left(P_{I|\mathsf{B}_n},p^{(U)}_{\mathcal{I}_n}\right)=\mathbb{E}_{\mathsf{B}_n(\mathbf{w})}\mathsf{F}\left(P_{I|\mathsf{B}_n},p^{(U)}_{\mathcal{I}_n}\right).
\end{equation}
Now consider the following justifications for the steps of \eqref{EQ:lemma_proof_fidelity_expected}:
(a) uses \eqref{EQ:lemma_proof_fidelity_CS} and the symmetry of the random codebook;\\
(b) is the law of total expectation;\\
(c) uses Jensen's inequality for the two-valued convex function $f:(x,y)\mapsto (xy)^{-\frac{1}{2}}$ and the relation
\begin{equation}
\mathbb{E}_{\mathsf{B}_n(\mathbf{w})|\mathbf{U}_1(\mathbf{w}),\mathbf{V}_{1,1}(\mathbf{w})}2^{i\big(\mathbf{U}_{\bar{\ell}}(\mathbf{w});\mathbf{V}_{\bar{j},\bar{k}}(\mathbf{w})\big|\mathbf{w}\big)}=1
\end{equation}
which holds for any $(\bar{\ell},\bar{j},\bar{k})\neq(1,1,1)$ (see \cite{Kramer_EffectiveSecrecy2014,Yassaee_fidelity_ISIT2015} for a similar derivation);\\
(d) is by increasing each term in the parenthesis by 1;\\
(e) is because $\big(\mathbf{U}_1(\mathbf{w}),\mathbf{V}_{1,1}(\mathbf{w}))\sim Q_{U|W=\mathbf{w}}^nQ_{V|W=\mathbf{w}}^n$.

Taking an expectation over $\mathbf{W}$ of both sides of \eqref{EQ:lemma_proof_fidelity_expected}, while making use of the law of total expectation and of the monotonicity of expectation, gives
\begin{align*}
&\mathbb{E}_{\mathsf{B}_n}\mathsf{F}\left(P_{I|\mathsf{B}_n},p^{(U)}_{\mathcal{I}_n}\right)\\
&=\mathbb{E}_\mathbf{W}\mathbb{E}_{\mathsf{B}_n|\mathbf{W}}\mathsf{F}\left(P_{I|\mathsf{B}_n},p^{(U)}_{\mathcal{I}_n}\right)\\
&\begin{multlined}[b][.41\textwidth]\geq\mathbb{E}_{Q_{W,U,V}^n}\bigg[\left(1+\big(|\mathcal{I}_n||\mathcal{J}_n|\big)^{-1}2^{i(\mathbf{U};\mathbf{V}|\mathbf{W})}\right)^{\mspace{-6mu}-\frac{1}{2}}\\\times\left(1+\big(|\mathcal{J}_n||\mathcal{K}_n|\big)^{-1}2^{i(\mathbf{U};\mathbf{V}|\mathbf{W})}\right)^{\mspace{-6mu}-\frac{1}{2}}\bigg]\end{multlined}.\numberthis\label{EQ:lemma_proof_fidelity_expected_final}
\end{align*}
Finally, note that 
\begin{equation}
\frac{1}{n}\mathbb{E}_{Q_{W,U,V}^n}2^{i_{Q_{W,U,V}}(\mathbf{U};\mathbf{V}|\mathbf{W})}=I_{Q_{W,U,V}}(U;V|W).
\end{equation}
Therefore, by the weak law of large number for any $\zeta>0$ there exists a sequence $\{\delta_n\}_{n\in\mathbb{N}}$ with $\lim_{n\to\infty}\delta_n=0$, such that
\begin{equation}
\mathbb{P}\mspace{-2mu}_{Q_{W,U,V}^n}\mspace{-4mu}\Bigg(\mspace{-3mu}\left|\frac{1}{n}i_{Q^n_{W,U,V}}\mspace{-1.5mu}(\mspace{-1mu}\mathbf{U}\mspace{-1mu};\mspace{-2mu}\mathbf{V}|\mathbf{W}\mspace{-1mu})\mspace{-2.5mu}-\mspace{-2.5mu}I_{Q_{W,U,V}}\mspace{-2mu}(\mspace{-1mu}U\mspace{-1mu};\mspace{-2mu}V|W\mspace{-1mu})\right|\mspace{-3mu}>\mspace{-3mu}\zeta\mspace{-3mu}\Bigg)\mspace{-4mu}\leq\mspace{-2mu}\delta_n,\label{EQ:lemma_proof_LLN}
\end{equation}
for all $n\in\mathbb{N}$. Combining \eqref{EQ:lemma_proof_fidelity_expected_final} and \eqref{EQ:lemma_proof_LLN} we see that as long as
\begin{equation}
S_2+\min\big\{S_1,T\big\}>I_{Q_{W,U,V}}(U;V|W)+\zeta
\end{equation}
then
\begin{equation}
    \mathbb{E}_{\mathsf{B}_n}\mathsf{F}\left(P_{I|\mathsf{B}_n},p^{(U)}_{\mathcal{I}_n}\right)\nearrow 1
\end{equation}
as $n\to\infty$. The relation \eqref{EQ:fidelity_vs_TV_expected_sequences}  now establishes the result of Lemma \ref{LEMMA:uniform_approx}.



\subsection{Proof of Theorem \ref{TM:inner_bound}}\label{SUBSEC:inner_proof}


%

Fix $n\in\mathbb{N}$, $(L_1,L_2)\in\mathbb{R}^2_+$, $\epsilon,\delta>0$, a PMF $Q_{U_0,U_1,U_2,X}\in\mathcal{P}(\mathcal{U}_0\times\mathcal{U}_1\times\mathcal{U}_2\times\mathcal{X})$ and denote $Q_{U_0,U_1,U_2,X,Y_1,Y_2}\triangleq Q_{U_0,U_1,U_2,X}W_{Y_1,Y_2|X}$. In the following we omit the blocklength $n$ from our notations of the ets of indices, e.g., we write $\mathcal{M}_0$ instead of $\mathcal{M}_0^{(n)}$. Furthermore, we assume that quantities of the form $2^{nR}$, where $n\in\mathbb{N}$ and $R\in\mathbb{R}_+$, are integers.\footnote{Otherwise simple modifications of some of the subsequent expressions using floor and ceiling operations are required.}



\textbf{Message Splitting:} Split each message $m_j\in\mathcal{M}_j$, $j=1,2$, into a pair of messages denoted by $(m_{j0},m_{jj})$. The triple $m_p\triangleq(m_0,m_{10},m_{20})$ is referred to as a \emph{public message} while $m_{jj}$, $j=1,2$, serves as \emph{private message} $j$. The rates associated with $m_{j0}$ and $m_{jj}$, $j=1,2$, are denoted by $R_{j0}$ and $R_{jj}$, while the corresponding alphabets are $\mathcal{M}_{j0}$ and $\mathcal{M}_{jj}$, respectively. The partial rates $R_{j0}$ and $R_{jj}$, $j=1,2$, satisfy
\begin{subequations}
\begin{align}
&R_j=R_{j0}+R_{jj}\label{EQ:achiev_partial_rates_sum}\\
&0\leq R_{j0}\leq R_j\label{EQ:achiev_partial_rates_bound1}\\
&R_{j0}\leq L_j.\label{EQ:achiev_partial_rates_bound2}
\end{align}\label{EQ:achiev_partial_rates}%
\end{subequations}
Let $M_{j0}$ and $M_{jj}$ be independent random variables uniformly distributed over $\mathcal{M}_{j0}$ and $\mathcal{M}_{jj}$, respectively. We use the notations $M_p\triangleq(M_0,M_{10},M_{20})$, $\mathcal{M}_p\triangleq\mathcal{M}_0\times\mathcal{M}_{10}\times\mathcal{M}_{20}$ and $R_p\triangleq R_0+R_{10}+R_{20}$. Note that $M_p$ is uniformly distributed over $\mathcal{M}_p$ and that $|\mathcal{M}_p|=2^{nR_p}$. Moreover, let $(W_1,W_2)$ be a pair of independent random variables, where $W_j$, $j=1,2$, is uniformly distributed over $\mathcal{W}_j=\big[1:2^{n\tilde{R}_j}\big]$ and independent of $(M_0,M_1,M_2)$ (which implies their independence of $(M_p,M_{11},M_{22})$ as well).

\begin{figure*}[!t]
\setcounter{equation}{64}
\begin{align*}
P^{(\mathcal{C}_n)}\Big(m_p,&m_{11},m_{22},w_1,w_2,i_1,i_2,\mathbf{u}_0,\mathbf{u}_1,\mathbf{u}_2,\mathbf{x},\mathbf{y}_1,\mathbf{y}_2,\big(\hat{m}_0,\hat{m}_1\big),\big(\hat{m}_0,\hat{m}_2\big)\Big)=\\& 2^{-n(R_p+R_{11}+R_{11}+\tilde{R}_1+\tilde{R}_2)} P_\mathsf{LE}^{(\mathcal{C}_n)}(i_1,i_2|m_p,m_{11},m_{22},w_1,w_2)\mathds{1}_{\big\{\mathbf{u}_0=\mathbf{u}_0(m_p)\big\}\cap\bigcap\limits_{j=1,2}\big\{\mathbf{u}_j=\mathbf{u}_j(m_p,m_{jj},w_j,i_j)\big\}}\\&\mspace{100mu}\times Q^n_{X|U_0,U_1,U_2}(\mathbf{x}|\mathbf{u}_0,\mathbf{u}_1,\mathbf{u}_2) W^n_{Y_1,Y_2|X}(\mathbf{y}_1,\mathbf{y}_2|\mathbf{x})\mathds{1}_{\bigcap\limits_{j=1,2}\big\{\big(\hat{m}_0,\hat{m}_j\big)=\phi_j^{(\mathcal{C}_n)}(\mathbf{y}_j)\big\}}\numberthis\label{EQ:achiev_induced_PMF_code}
\end{align*}
\hrulefill
\begin{align*}
P\Big(\mathcal{C}_n,m_p,m_{11},&m_{22},w_1,w_2,i_1,i_2,\mathbf{u}_0,\mathbf{u}_1,\mathbf{u}_2,\mathbf{x},\mathbf{y}_1,\mathbf{y}_2,\big(\hat{m}_0,\hat{m}_1\big),\big(\hat{m}_0,\hat{m}_2\big)\Big)\\&=\mu(\mathcal{C}_n)P^{(\mathcal{C}_n)}\Big(m_p,m_{11},m_{22},w_1,w_2,s_1,s_2,\mathbf{u}_0,\mathbf{u}_1,\mathbf{u}_2,\mathbf{x},\mathbf{y}_1,\mathbf{y}_2,\big(\hat{m}_0,\hat{m}_1\big),\big(\hat{m}_0,\hat{m}_2\big)\Big)\numberthis\label{EQ:achiev_induced_PMF}
\end{align*}
\hrulefill
\end{figure*}
\setcounter{equation}{61}

\textbf{Codebook $\bm{\mathcal{C}_n}$:} Let $\mathsf{C}^{(n)}_0\triangleq\big\{\mathbf{U}_0(m_p)\big\}_{m_p\in\mathcal{M}_p}$ be a random public message codebook that comprises $2^{nR_p}$ i.i.d. random vectors $\mathbf{U}_0(m_p)$, each distributed according to $Q_{U_0}^n$. A realization of $\mathsf{C}^{(n)}_0$ is denoted by $\mathcal{C}^{(n)}_0\triangleq\big\{\mathbf{u}_0(m_p)\big\}_{m_p\in\mathcal{M}_p}$.


Fix a public message codebook $\mathcal{C}^{(n)}_0$. For every $m_p\in\mathcal{M}_p$ and $j=1,2$, let $\mathsf{C}^{(n)}_j(m_p)\triangleq\big\{\mathbf{U}_j(m_p,m_{jj},w_j,i_j)\big\}_{(m_{jj},w_j,i_j)\in\mathcal{M}_{jj}\times\mathcal{W}_j\times\mathcal{I}_j}$, where $(m_{jj},w_j,i_j)\in\mathcal{M}_{jj}\times\mathcal{W}_j\times\mathcal{I}_j$ and $\mathcal{I}_j\triangleq\big[1:2^{nR'_j}\big]$, be a random codebook of private messages $j$, consisting of conditionally independent random vectors each distributed according to
$Q^n_{U_j|U_0=\mathbf{u}_0(m_p)}$. A realization of $\mathsf{C}^{(n)}_j(m_p)$ is denoted by $\mathcal{C}^{(n)}_j(m_p)\triangleq\big\{\mathbf{u}_j(m_p,m_{jj},w_j,i_j)\big\}_{(m_{jj},w_j,i_j)\in\mathcal{M}_{jj}\times\mathcal{W}_j\times\mathcal{I}_j}$.

We denote $\mathsf{C}^{(n)}_j\triangleq\Big\{\mathsf{C}^{(n)}_j(m_p)\Big\}_{m_p\in\mathcal{M}_p}$, and its realization by $\mathcal{C}^{(n)}_j$. A random codebook is denoted by $\mathsf{C}_n=\Big\{\mathsf{C}^{(n)}_0,\mathsf{C}^{(n)}_1,\mathsf{C}^{(n)}_2\Big\}$, while
$\mathcal{C}_n=\Big\{\mathcal{C}^{(n)}_0,\mathcal{C}^{(n)}_1,\mathcal{C}^{(n)}_2\Big\}$ denotes a fixed codebook (a possible outcome of $\mathsf{C}_n$). Denoting the set of all possible realizations of $\mathsf{C}_n$ by $\mathfrak{C}_n$, the above codebook construction induces a PMF $\mu\in\mathcal{P}(\mathfrak{C}_n)$ over the codebook ensemble. For every $\mathcal{C}_n\in\mathfrak{C}_n$, we have \eqref{EQ:codebook_probability} at the top of this page.

For a fixed codebook $\mathcal{C}_n\in\mathfrak{C}_n$ we now describe its associated encoding function $f^{(\mathcal{C}_n)}$ and decoding functions $\phi^{(\mathcal{C}_n)}_j$, for $j=1,2$.

\textbf{Encoder $\bm{f^{(\mathcal{C}_n)}}$:} Fix a codebook $\mathcal{C}_n\in\mathfrak{C}_n$. To transmit the message pair $(m_0,m_1,m_2)$ the encoder transforms it into the triple $\big(m_p,m_{11},m_{22})$, and draws $W_j$ uniformly from $\mathcal{W}_j$, $j=1,2$; denote the realization of $W_j$ by $w_j\in\mathcal{W}_j$. Given $(m_p,m_{11},m_{22},w_1,w_2)$, a pair of indices $(i_1,i_2)\in\mathcal{I}_1\times\mathcal{I}_2$ is randomly selected by the likelihood encoder according to
\begin{align*}
&P^{(\mathcal{C}_n)}_\mathsf{LE}(i_1,i_2|m_p,m_{11},m_{22},w_1,w_2)&\\
&\triangleq\frac{2^{i_{Q^n}\big(\mathbf{u}_1(m_p,m_{11},w_1,i_1);\mathbf{u}_2(m_p,m_{22},w_2,i_2)\big|\mathbf{u}_0(m_p)\big)}}{\sum\limits_{\substack{(i'_1,i'_2)\\\in\mathcal{I}_1\times\mathcal{I}_2}}2^{i_{Q^n}\big(\mathbf{u}_1(m_p,m_{11},w_1,i'_1);\mathbf{u}_2(m_p,m_{22},w_2,i'_2)\big|\mathbf{u}_0(m_p)\big)}}\numberthis\label{EQ:SMC_encoder}
\end{align*}
where $i_{Q^n}$ stands for the information density with respect to the conditional product distribution $Q^n_{U_1,U_2|U_0}$ (and its marginals). The structure of $P^{(\mathcal{C}_n)}_\mathsf{LE}$ adheres to the setup of Lemma \ref{LEMMA:uniform_approx} from Section \ref{SEC:uniformity_lemma} and, in particular, to the stochastic choice of indices therein as described in \eqref{EQ:lemma_likelihood_encoder}. Replacing the commonly used joint typicality encoder with $P^{(\mathcal{C}_n)}_\mathsf{LE}$, we are able to establish several important properties of the chosen codewords and their induced distribution.

Let $(i_1,i_2)$ be the selected pair of indices. The channel input sequence is randomly generated according to the conditional product distribution
\begin{equation*}
Q^n_{X|U_0=\mathbf{u}_0(m_p),U_1=\mathbf{u}_1(m_p,m_{11},w_1,i_1),U_2=\mathbf{u}_2(m_p,m_{22},w_2,i_2)}.
\end{equation*}



\textbf{Decoder $\bm{\phi_j^{(\mathcal{C}_n)}}$:} Decoder $j=1,2$ operates in two stages. First, it searches for a unique $\hat{m}_p\in\mathcal{M}_p$ such that
\begin{equation}
\Big(\mathbf{u}_0(\hat{m}_p),\mathbf{y}_j\Big)\in\mathcal{T}_\delta^{n}(Q_{U_0,Y_j}).\label{EQ:achiev_decoder0j}
\end{equation}
If no such unique index is found, set $\phi^{(\mathcal{C}_n)}_j=(1,1)$. Otherwise, having $\hat{m}_p\in\mathcal{M}_p$, Decoder $j=1,2$ proceeds by looking for a unique pair $(\hat{m}_{jj},\hat{w}_j)\in\mathcal{M}_{jj}\times\mathcal{W}_j$ for which there exists an index $\hat{i}_j\in\mathcal{I}_j$ such that
\begin{equation}
\Big(\mathbf{u}_0(\hat{m}_p),\mathbf{u}_j(\hat{m}_p,\hat{m}_{jj},\hat{w}_j,\hat{i}_j),\mathbf{y}_j\Big)\in\mathcal{T}_\delta^{n}(Q_{U_0,U_j,Y_j}).\label{EQ:achiev_decoderj}
\end{equation}
Recall that each $m_p\in\mathcal{M}_p$ specifies a triple $(m_0,m_{10},m_{20})\in\mathcal{M}_0\times\mathcal{M}_{10}\times\mathcal{M}_{20}$. If the second stage is also executed successfully the decoder has a triple $(\hat{m}_p,\hat{m}_{jj},\hat{w}_j)\in\mathcal{M}_p\times\mathcal{M}_{jj}\times\mathcal{W}_j$ with $\hat{m}_p$ and $(\hat{m}_{jj},\hat{w}_j)$ being the unique indices satisfying \eqref{EQ:achiev_decoder0j} and \eqref{EQ:achiev_decoderj}, respectively. In this case we set $\phi^{(\mathcal{C}_n)}_j(\mathbf{y}_j)=\big(\hat{m}_0,\hat{m}_j\big)$, where $\hat{m}_j$ is assembled from $(\hat{m}_{j0},\hat{m}_{jj})$; otherwise, set $\phi^{(\mathcal{C}_n)}_j=(1,1)$.

\textbf{Induced Code and Joint Distribution:} The triple $\left(f^{(\mathcal{C}_n)},\phi^{(\mathcal{C}_n)}_1,\phi^{(\mathcal{C}_n)}_2\right)$ defined with respect to the codebook $\mathcal{C}_n\in\mathfrak{C}_n$ constitutes an $(n,R_0,R_1,R_2)$ code $c_n$ for the BC with privacy leakage constraints. Thus, for every codebook $\mathcal{C}_n\in\mathfrak{C}_n$, the induced joint distribution is given in \eqref{EQ:achiev_induced_PMF_code} at the top of this page, where the random variables $\mathbf{U}_0$, $\mathbf{U}_1$ and $\mathbf{U}_2$ are the chosen codewords at the conclusion of the encoding process (from which the input $\mathbf{X}$ to the BC is generated).

Taking the random codebook generation into account, we also have \eqref{EQ:achiev_induced_PMF} at the top of this page, where $\mu\in\mathcal{P}(\mathfrak{C}_n)$ is described in \eqref{EQ:codebook_probability}. The PMF $P$ induces a probability measure $\mathbb{P}\triangleq\mathbb{P}_P$, with respect to which the subsequent analysis is performed. Specifically, all the multi-letter information measures in the sequel are taken with respect to $P$ from \eqref{EQ:achiev_induced_PMF}, while single-letter information terms are calculated with respect to $Q_{U_0,U_1,U_2,X,Y_1,Y_2}$.

\setcounter{equation}{66}

\textbf{Average Error Probability Analysis:} The output sequences of $P^{(\mathcal{C}_n)}_\mathsf{LE}$ from \eqref{EQ:SMC_encoder} are jointly typical with high probability as long as the sum of the rates of the product bin is greater than the mutual information between the coding random variables \cite[Theorem 3]{Yassaee_fidelity_ISIT2015}. The rest of the error probability analysis goes through via classic joint typicality arguments. The details of the analysis are relegated to Appendix \ref{APPEN:error_analysis}, where it is shown that
\begin{equation}
\mathbb{E}P_{\mathsf{e}}(\mathsf{C}_n)\leq \eta(n,\delta,\delta')\label{EQ:error_prob_decays}
\end{equation}
where $\delta'\in(0,\delta)$ and $\lim_{n\to\infty}\eta(n,\delta,\delta')=0$ for all $0<\delta'<\delta$, if
\begin{subequations}
\begin{align}
R_1'+R_2'&>I(U_1;U_2|U_0)\label{EQ:achiev_rb1}\\
R_0+R_{10}+R_{20}&<I(U_0;Y_1)-\tau_\delta\label{EQ:achiev_rb3}\\
R_0+R_{10}+R_{20}&<I(U_0;Y_2)-\tau_\delta\label{EQ:achiev_rb5}\\
R_{11}+\tilde{R}_1+R'_1&< I(U_1;Y_1|U_0)-\tau_\delta\label{EQ:achiev_rb2}\\
R_{22}+\tilde{R}_2+R'_2 &< I(U_2;Y_2|U_0)-\tau_\delta\label{EQ:achiev_rb4}
\end{align}\label{EQ:achiev_rb}%
\end{subequations}
with $\tau_\delta\to 0$ as $\delta\to 0$. Furthermore, setting $\eta_n\triangleq\eta(n,\delta_n,\delta'_n)$ where $\{\delta_n\}_{n\in\mathbb{N}}$ and $\{\delta'_n\}_{n\in\mathbb{N}}$ are sequences that converge sufficiently slowly to zero as $n$ grows, we have $\lim_{n\to\infty}\eta_n=0$. To clarify, the $\delta'$ that appears in \eqref{EQ:error_prob_decays} and in upper bounds below is a consequence of the Conditional Typicality Lemma \cite[Section 2.5]{ElGamal2011}. This lemma considers conditioning on sequences that are jointly letter-typical with respect to a slightly smaller gap than the original $\delta$. This smaller gap is $\delta'$.

\textbf{Properties for Leakage Analysis:} In contrast to previous works, we do not analyse the expected leakages of the random code. Instead, we establish certain properties that the random code possesses and then extract a specific sequence of codes that satisfies these properties as well as reliability. It is then shown that the extracted sequence of codes admits the leakage constraints. 

By symmetry, we consider only the properties required for the analysis of the rate-leakage from $M_1$ to the 2nd receiver. The corresponding derivations for $M_2$ follows similar lines and the resulting rate constraints match up to changing some indices.


We first need a decodability property. Specifically, Decoder 2 should be able to decode $(W_1,I_1)$ with a low error probability based on $(M_p,M_{11},M_{22},W_2,I_2,\mathbf{Y}_2)$. We consider a decoding rule based on a joint typicality test: Decoder 2 searches for a unique pair $(\check{w}_1,\check{i}_1)\in\mathcal{W}_1\times\mathcal{I}_1$ such that
\begin{align*}
\Big(\mathbf{u}_0(m_p),\mathbf{u}_1(m_p,m_{11},\check{w}_1,\check{i}_1),&\mathbf{u}_2(m_p,m_{22},w_2,i_2),\mathbf{y}_2\Big)\\&\in\mathcal{T}_\delta^{n}(Q_{U_0,U_1,U_2,Y_2}).\numberthis\label{EQ:leakage_testj}
\end{align*} 
For a fixed codebook $\mathcal{C}_n\in\mathfrak{C}_n$ (which specifies a code $c_n$), let $P_1^{(\mathsf{Leak)}}(\mathcal{C}_n)$ denote the probability that Decoder 2 fails in this decoding process. As explained in Appendix \ref{APPEN:error_analysis}, we have
\begin{equation}
\mathbb{E}P_1^{(\mathsf{Leak)}}(\mathsf{C}_n)\leq \kappa(n,\delta,\delta')\label{EQ:leakage_error_prob_decays}
\end{equation}
where $\delta'\in(0,\delta)$ and $\lim_{n\to\infty}\kappa(n,\delta,\delta')=0$ for all $0<\delta'<\delta$, if
\begin{subequations}
    \begin{align}
        \tilde{R}_1&<I(U_1;Y_2|U_0,U_2)-\xi_\delta\label{EQ:achiev_rb6a}\\
        \tilde{R}_1+R'_1&<I(U_1;U_2,Y_2|U_0)-\xi_\delta\label{EQ:achiev_rb6b}
    \end{align}\label{EQ:achiev_rb6}%
\end{subequations}
with $\xi_\delta\to 0$ as $\delta\to 0$. Again, by allowing $\delta$ and $\delta'$ from \eqref{EQ:leakage_error_prob_decays} to converge to zero sufficiently slow with $n$, $\kappa(n,\delta,\delta')$ may be replaced by a $\kappa_n$ with $\lim_{n\to\infty}\kappa_n=0$.


We are now ready to state Lemmas \ref{LEMMA:1}-\ref{LEMMA:0}. Proofs are given in Appendices \ref{APPEN:lemma1_proof}-\ref{APPEN:lemma0_proof}.

\begin{lemma}\label{LEMMA:1}
If \eqref{EQ:achiev_rb6} is valid with $\xi_\delta\to 0$ as $\delta\to 0$, then there exists $\zeta_1(n,\delta,\delta')$, where $\delta'\in(0,\delta)$, such that
\begin{equation}
H(W_1,I_1|M_p,M_{11},M_{22},W_2,I_2,\mathbf{Y}_2,\mathsf{C}_n)\leq n\zeta_1(n,\delta,\delta')\label{EQ:lemma2_ineq}
\end{equation}
and $\lim_{n\to\infty}\zeta_1(n,\delta,\delta')=0$ for all $0<\delta'<\delta$. Furthermore, setting $\zeta_{1,n}\triangleq\zeta_1(n,\delta_n,\delta'_n)$ where $\{\delta_n\}_{n\in\mathbb{N}}$ and $\{\delta'_n\}_{n\in\mathbb{N}}$ are sequences that decay sufficiently slow to zero as $n$ grows, we have $\lim_{n\to\infty}\zeta_{1,n}=0$.
\end{lemma}

\begin{lemma}\label{LEMMA:2}
There exist $\zeta_2(n,\delta,\delta')$ that satisfies the same properties as $\zeta_1(n,\delta,\delta')$ from Lemma
\ref{LEMMA:1}, such that
\begin{equation}
I(\mathbf{U}_1;\mathbf{Y}_2|\mathbf{U}_0,\mathbf{U}_2,\mathsf{C}_n)\leq nI(U_1;Y_2|U_0,U_2)+n\zeta_2(n,\delta,\delta').\label{EQ:lemma1_ineq2}
\end{equation}
\end{lemma}

\begin{lemma}\label{LEMMA:0}
There exists $\zeta_3(n,\delta,\delta')$ that satisfies the same properties as $\zeta_1(n,\delta,\delta')$ from Lemma
\ref{LEMMA:1}, such that
\begin{equation}
I(\mathbf{U}_1;M_{22},W_2,I_2|M_p,\mathsf{C}_n)\leq nI(U_1;U_2|U_0)+n\zeta_3(n,\delta,\delta').\label{EQ:lemma1_ineq1}
\end{equation}
\end{lemma}

The Uniform Approximation Lemma from Section \ref{SEC:uniformity_lemma} further implies that if 
\begin{equation}
R'_2+\min\big\{R'_1,R_{22}+\tilde{R}_2\big\}>I(U_1;U_2|U_0)+\delta\label{EQ:achiev_rb_uniformity}
\end{equation}
then there exist $\zeta_{4,n}$ with $\lim_{n\to\infty}\zeta_{4,n}$, such that
\begin{equation}
\mathbb{E}_{\mathsf{C}_n}\Big|\Big|P_{M_p,M_{11},W_1,I_1|\mathsf{C}_n}-p^{(U)}_{\mathcal{M}_p\times\mathcal{M}_{11}\times\mathcal{W}_1\times\mathcal{I}_1}\Big|\Big|_{\mathsf{TV}}\leq\zeta_{4,n}\label{EQ:leakage_analysis_uniform_TV}
\end{equation}
where $p^{(U)}_{\mathcal{M}_p\times\mathcal{M}_{11}\times\mathcal{W}_1\times\mathcal{I}_1}$ is the uniform distribution on $\mathcal{M}_p\times\mathcal{M}_{11}\times\mathcal{W}_1\times\mathcal{I}_1$. To see this, observe that by symmetry we have
\begin{align*}
    &\mathbb{E}_{\mathsf{C}_n}\Big|\Big|P_{M_p,M_{11},W_1,I_1|\mathsf{C}_n}-p^{(U)}_{\mathcal{M}_p\times\mathcal{M}_{11}\times\mathcal{W}_1\times\mathcal{I}_1}\Big|\Big|_{\mathsf{TV}}\\
    &\begin{multlined}[b][.48\textwidth]=\sum_{\substack{(m_p,m_{11},w_1)\\\in\mathcal{M}_p\times\mathcal{M}_{11}\times\mathcal{W}_1}}\frac{1}{|\mathcal{M}_p||\mathcal{M}_{11}||\mathcal{W}_1|}\\\times\mathbb{E}_{\mathsf{C}_n}\Big|\Big|P_{I_1|M_p=m_p,M_{11}=m_{11},W_1=w_1,\mathsf{C}_n}-p^{(U)}_{\mathcal{I}_1}\Big|\Big|_{\mathsf{TV}}\end{multlined}\\
    &=\mathbb{E}_{\mathsf{C}_n}\Big|\Big|P_{I_1|M_p=1,M_{11}=1,W_1=1,\mathsf{C}_n}-p^{(U)}_{\mathcal{I}_1}\Big|\Big|_{\mathsf{TV}}\numberthis.
\end{align*}
Note that $(M_p,M_{11},W_1)=(1,1,1)$ fixes a single $u_1$-bin (comprising $2^{nR_1'}$ codewords), while the pair $(M_{22},W_2)$ (of total rate $R_{22}+\tilde{R}_2$) uniformly chooses a $u_2$-bin (comprising $2^{nR_2'}$ codewords). Lemma \ref{LEMMA:uniform_approx} now gives the desired relation because bins are generated conditionally independent given $\mathbf{U}_0$ and the chosen codeword pair is drawn according to $P^{(\mathcal{C}_n)}_\mathsf{LE}$ from \eqref{EQ:SMC_encoder} which adheres to the structure of~\eqref{EQ:lemma_likelihood_encoder}.

We now invoke the Selection Lemma \cite[Lemma 5]{Goldfeld_WTCII_semantic2016} to extract a specific sequence of codes that satisfies several desired properties. We restate this lemma next.

\begin{lemma}[Selection Lemma]\label{LEMMA:selection_lemma}
Let $\big\{A_n\big\}_{n\in\mathbb{N}}$ be a sequence of random variables, where $A_n$ takes values in $\mathcal{A}_n$. Let $\left\{f_n^{(1)},f_n^{(2)},\ldots,f_n^{(\mathrm{J})}\right\}_{n\in\mathbb{N}}$ be a collection of $\mathrm{J}<\infty$ sequences of bounded functions $f_n^{(i)}:\mathcal{A}_n\to\mathbb{R}_+$, $j\in[1:\mathrm{J}\mspace{2mu}]$. If
\begin{subequations}
\begin{equation}
\mathbb{E}f_n^{(j)}(A_n)\xrightarrow[n\to\infty]{}0,\quad\forall j\in[1:\mathrm{J}\mspace{2mu}],
\end{equation}
then there exists a sequence $\{a_n\}_{n\in\mathbb{N}}$, where $a_n\in\mathcal{A}_n$ for every $n\in\mathbb{N}$, such that
\begin{equation}
f_n^{(j)}(a_n)\xrightarrow[n\to\infty]{}0,\quad\forall j\in[1:\mathrm{J}\mspace{2mu}].
\end{equation}
\end{subequations}
\end{lemma}

Consider the sequence of random codes $\big\{\mathsf{C}_n\big\}_{n\in\mathbb{N}}$, the functions\footnote{We slightly abuse notation in the definition of $f_n^{(1)}$ because $P_\mathsf{e}$ is actually a function of the code $c_n$ rather than the codebook $\mathcal{C}_n$. However, since $\mathcal{C}_n$ uniquely defines $c_n$ we prefer this presentation for the sake of simplicity.}
\begin{subequations}
\begin{align}
&f_n^{(1)}\mspace{-1mu}(\mathcal{C}_n\mspace{-1mu})\mspace{-2mu}\triangleq\mspace{-2mu} P_{\mathsf{e}}(\mathcal{C}_n)\\
&f_n^{(2)}\mspace{-1mu}(\mathcal{C}_n\mspace{-1mu})\mspace{-2mu}\triangleq\mspace{-2mu} H(W_1,I_1|M_p,M_{11},M_{22},W_2,I_2,\mathbf{Y}_2,\mathsf{C}_n=\mathcal{C}_n)\\
&\begin{multlined}[b][.415\textwidth]f_n^{(3)}\mspace{-1mu}(\mathcal{C}_n\mspace{-1mu})\mspace{-2mu}\triangleq\mspace{-2mu}
\frac{1}{n}I(\mathbf{U}_1;\mathbf{Y}_2|\mathbf{U}_0,\mathbf{U}_2,\mathsf{C}_n=\mathcal{C}_n)\\-I(U_1;Y_2|U_0,U_2)\end{multlined}\\
&\begin{multlined}[b][.415\textwidth]f_n^{(4)}\mspace{-1mu}(\mathcal{C}_n\mspace{-1mu})\mspace{-2mu}\triangleq\mspace{-2mu}
\frac{1}{n}I(\mathbf{U}_1;M_{22},W_2,I_2|M_p,\mathsf{C}_n=\mathcal{C}_n)\\-I(U_1;U_2|U_0)\end{multlined}\\
&\begin{multlined}[b][.415\textwidth]f_n^{(5)}\mspace{-1mu}(\mathcal{C}_n\mspace{-1mu})\mspace{-2mu}\triangleq\mspace{-2mu}\Big|\Big|P_{M_p,M_{11},W_1,I_1|\mathsf{C}_n=\mathcal{C}_n}\mspace{-3mu}-p^{(U)}_{\mathcal{M}_p\times\mathcal{M}_{11}\times\mathcal{W}_1\times\mathcal{I}_1}\Big|\Big|_{\mathsf{TV}}\end{multlined}
\end{align}
\end{subequations}
as well as the functions $f_n^{(6)}$, $f_n^{(7)}$ and $f_n^{(8)}$ that correspond to $f_n^{(2)}$, $f_n^{(3)}$ and $f_n^{(4)}$, respectively, with respect to the analysis for $M_2$. We also impose constraints on the rates that arise from repeating the above steps for $M_2$. Namely, we set
\begin{subequations}
\begin{align}
    \tilde{R}_2&<I(U_2;Y_1|U_0,U_1)-\xi(\delta)\label{EQ:achiev_rb6_R2a}\\
    \tilde{R}_2+R'_2&<I(U_2;U_1,Y_1|U_0)-\xi(\delta)\label{EQ:achiev_rb6_R2b}
\end{align}
and
\begin{equation}
R'_1+\min\big\{R'_2,R_{11}+\tilde{R}_1\big\}>I(U_1;U_2|U_0)+\delta\label{EQ:achiev_rb_uniformity_R2}
\end{equation}\label{EQ:achiev_rb_R2}%
\end{subequations}
in accordance with \eqref{EQ:achiev_rb6} and \eqref{EQ:achiev_rb_uniformity}, respectively. This implies that results analog to those of Lemmas \ref{LEMMA:1}-\ref{LEMMA:0} and \eqref{EQ:leakage_analysis_uniform_TV} hold for $M_2$.

Replacing $\delta$ and $\delta'$ in the definitions of $\eta_j(n,\delta,\delta')$, for $j=1,2,3$, with $\{\delta_n\}_{n\in\mathbb{N}}$ and $\{\delta_n'\}_{n\in\mathbb{N}}$ that decay to zero sufficiently slow, we have
\begin{equation}
\mathbb{E}_{\mathsf{C}_n}f_n^{(j)}(\mathsf{C}_n)\xrightarrow[n\to\infty]{}0,\quad j\in[1:7].
\end{equation}
Lemma \ref{LEMMA:selection_lemma} now implies the existence of a sequence of codebooks $\big\{\mathcal{C}_n\big\}_{n\in\mathbb{N}}$ (each inducing an $(n,R_1,R_1,R_2)$ code $c_n$) and another sequence of numbers $\{\eta_n\}_{n\in\mathbb{N}}$ with $\lim_{n\to\infty}\eta_n=0$, such that for $j\in[1:7]$ we have
\begin{equation}
f_n^{(j)}(\mathcal{C}_n)\leq\eta_n, \quad\forall n\in\mathbb{N}.\label{EQ:existence_sequence}
\end{equation}

\textbf{Leakage Analysis of $\bm{M_1}$ Under $\bm{\mathcal{C}_n}$:} All subsequent information measures are calculated with respect to $P^{(\mathcal{C}_n)}$ from \eqref{EQ:achiev_induced_PMF_code}. We emphasize this by using $H_{\mathcal{C}_n}$ and $I_{\mathcal{C}_n}$ as the notation of such entropy or mutual information terms, respectively. 

First, because $f^{(5)}_n(\mathcal{C}_n)\xrightarrow[n\to\infty]{}0$ and by the continuity of entropy, there exists a sequence $\{\theta_n\}_{n\in\mathbb{N}}$ with $\lim_{n\to\infty}\theta_n =0$, such that
\begin{equation}
    \Big|H_{\mathcal{C}_n}(M_{11},W_1,I_1|M_p) - \log\big(|\mathcal{M}_{11}||\mathcal{W}_1||\mathcal{I}_1|\big)\Big|\leq \theta_n\label{EQ:uniform_entropy}
\end{equation}
for every $n\in\mathbb{N}$. Next, since
\begin{equation}
\ell_1(c_n)=\frac{1}{n}I_{\mathcal{C}_n}(M_1;\mathbf{Y}_2)=R_1-\frac{1}{n}H_{\mathcal{C}_n}(M_1|\mathbf{Y}_2)\label{EQ:leakage_UB1}
\end{equation}
we can upper bound the leakage of $M_1$ to the second receiver by lower bounding the conditional entropy term from the RHS of \eqref{EQ:leakage_UB1}. We have
\begin{align*}
&H_{\mathcal{C}_n}(M_1|\mathbf{Y}_2)\\
    &\stackrel{(a)}\geq H_{\mathcal{C}_n}(M_{11}|M_p,M_{22},W_2,I_2,\mathbf{Y}_2)\\                   &\begin{multlined}[b][.48\textwidth]= H_{\mathcal{C}_n}(M_{11},\mathbf{Y}_2|M_p,M_{22},W_2,I_2)\\-H_{\mathcal{C}_n}(\mathbf{Y}_2|M_p,M_{22},W_2,I_2)\end{multlined}\\
    &\stackrel{(b)}{\geq} H_{\mathcal{C}_n}(M_{11},\mathbf{Y}_2|M_p,M_{22},W_2,I_2)-H_{\mathcal{C}_n}(\mathbf{Y}_2|\mathbf{U}_0,\mathbf{U}_2)\\
    &\begin{multlined}[b][.475\textwidth]=H_{\mathcal{C}_n}(M_{11},W_1,I_1,\mathbf{Y}_2|M_p,M_{22},W_2,I_2)\\-H_{\mathcal{C}_n}(W_1,I_1|M_p,M_{11},M_{22},W_2,I_2,\mathbf{Y}_2)\\-H(\mathbf{Y}_2|\mathbf{U}_0,\mathbf{U}_2)\end{multlined}\\
    &\begin{multlined}[b][.475\textwidth]=H_{\mathcal{C}_n}(M_{11},W_1,I_1|M_p,M_{22},W_2,I_2)\\+H_{\mathcal{C}_n}(\mathbf{Y}_2|M_p,M_{11},W_1,I_1,M_{22},W_2,I_2)\\\mspace{45mu}-H_{\mathcal{C}_n}(W_1,I_1|M_p,M_{11},M_{22},W_2,I_2,\mathbf{Y}_2)\\-H_{\mathcal{C}_n}(\mathbf{Y}_2|\mathbf{U}_0,\mathbf{U}_2)\end{multlined}\\
    &\begin{multlined}[b][.475\textwidth]\stackrel{(c)}=H_{\mathcal{C}_n}(M_{11},W_1,I_1|M_p,M_{22},W_2,I_2)\\-H_{\mathcal{C}_n}(W_1,I_1|M_p,M_{11},M_{22},W_2,I_2,\mathbf{Y}_2)\\-I_{\mathcal{C}_n}(\mathbf{U}_1;\mathbf{Y}_2|\mathbf{U}_0,\mathbf{U}_2)\end{multlined}\\
    &\begin{multlined}[b][.43\textwidth]= H_{\mathcal{C}_n}(M_{11},W_1,I_1|M_p)-I_{\mathcal{C}_n}(\mathbf{U}_1;M_{22},W_2,I_2|M_p)\\-H_{\mathcal{C}_n}(W_1,I_1|M_p,M_{11},M_{22},W_2,I_2,\mathbf{Y}_2)\\-I_{\mathcal{C}_n}(\mathbf{U}_1;\mathbf{Y}_2|\mathbf{U}_0,\mathbf{U}_2)\end{multlined}\numberthis\label{EQ:secrecy_analysis}
\end{align*}
where:\\
(a) is because conditioning cannot increase entropy and since $M_1$ corresponds to the pair $(M_{10},M_{11})$ while $M_p=(M_0,M_{10},M_{20})$;\\
(b) follows because $\mathbf{U}_0$ and $\mathbf{U}_2$ are specified by $(M_p,M_{22},W_2,I_2)$ and since conditioning cannot increase entropy;\\
(c) uses the deterministic relations stated in (b) along with $\mathbf{U}_1$ being determined by $(M_p,M_{11},W_1,I_1)$ and the Markov relation $\mathbf{Y}_2-(\mathbf{U}_0,\mathbf{U}_1,\mathbf{U}_2)-(M_p,M_{11},W_1,I_1,M_{22},W_2,I_2)$.

Inserting \eqref{EQ:existence_sequence} (for $j\in[2:4]$), \eqref{EQ:uniform_entropy} and $R_{11}=R_1-R_{10}$ into \eqref{EQ:secrecy_analysis} further gives
\begin{align*}
&H_{\mathcal{C}_n}(M_1|\mathbf{Y}_2)\\&\geq n\big(R_1-R_{10}+\tilde{R}_1+R'_1-I(U_1;U_2,Y_2|U_0)-3\eta_n-\theta_n\big)\\
&\stackrel{(a)}\geq nR_1 -n\big(L_1+3\eta_n+\theta_n\big)
\end{align*}
where (a) follows by taking
\begin{subequations}
\begin{align}
\tilde{R}_1+R'_1-R_{10}&> I(U_1;U_2,Y_2|U_0)-L_1\label{EQ:achiev_extra_rb3}\\
R_1'+L_1-R_{10}&>I(U_1;U_2|U_0)\label{EQ:achiev_extra_rb1}.
\end{align}\label{EQ:achiev_extra_rb_m1}%
\end{subequations}
The bound in \eqref{EQ:achiev_extra_rb1} ensures the feasibility of an $\tilde{R}_1>0$ that satisfies \eqref{EQ:achiev_rb6a} and \eqref{EQ:achiev_extra_rb3} simultaneously. The corresponding rate bounds for the analysis of $\ell_2(c_n)$ are
\begin{subequations}
\begin{align}
\tilde{R}_2+R'_2-R_{20}&> I(U_2;U_1,Y_1|U_0)-L_2\label{EQ:achiev_extra_rb3_R2}\\
R_2'+L_2-R_{20}&>I(U_1;U_2|U_0)\label{EQ:achiev_extra_rb1_R2}.
\end{align}\label{EQ:achiev_extra_rb_m1_R2}%
\end{subequations}
Recalling that $\eta_n$ and $\theta_n$ can be made arbitrarily small with $n$, there exists $n_0(\epsilon)\in\mathbb{N}$, such that for all $n>n_0(\epsilon)$ 
\begin{subequations}
\begin{align}
&P_{\mathsf{e}}(c_n)\leq\epsilon\label{EQ:error_prob_proof}\\
&\ell_1(c_n)\leq L_1+\epsilon\label{EQ:achieve_leakage1_proof}\\
&\ell_2(c_n)\leq L_2+\epsilon.\label{EQ:achieve_leakage2_proof}
\end{align}\label{EQ:achiev_realibility_leakage_proof}
\end{subequations}
as required. 

Our last step is to apply FME on \eqref{EQ:achiev_rb}, \eqref{EQ:achiev_rb_uniformity}, \eqref{EQ:achiev_rb_R2} and  \eqref{EQ:achiev_extra_rb_m1}-\eqref{EQ:achiev_extra_rb_m1_R2}, while using \eqref{EQ:achiev_partial_rates} and the non-negativity of the involved terms, to eliminate $R_{j0}$, $R_j'$ and $\tilde{R}_j$, for $j=1,2$. Since all the above linear inequalities have constant coefficients, the FME can be performed by a computer program, e.g., by the FME-IT software \cite{FME&ITIP}. This shows the sufficiency of \eqref{EQ:region_inner}.

\subsection{Proof of Corollary \ref{COR:inactive_leakage}}\label{SUBSEC:inactive_leakage_proof}

Fix $(L_1,L_2)\in\mathbb{R}^2_+$ and $Q_{U_0,U_1,U_2,X}\in\mathcal{P}(\mathcal{U}_0\times\mathcal{U}_1\times\mathcal{U}_2\times\mathcal{X})$. The rate bounds describing $\tilde{\mathcal{R}}_{\mathsf{I}}(L_1,L_2,Q_{U_0,U_1,U_2,X})$ are:
\begin{subequations}
\begin{align}
R_1\mspace{-2mu} &\leq\mspace{-2mu} I(U_1;Y_1|U_0)\mspace{-3mu}-\mspace{-3mu}I(U_1;U_2,Y_2|U_0)\mspace{-3mu}+\mspace{-3mu}L_1\label{EQ:region_nocommon_inner11}\\
R_1\mspace{-2mu} &\leq\mspace{-2mu} I(U_1;Y_1|U_0)\mspace{-3mu}+\mspace{-3mu}\min\mspace{-3mu}\Big\{\mspace{-1.5mu}I(U_0;Y_1),\mspace{-1.5mu}I(U_0;Y_2)\mspace{-1.5mu}\Big\}\label{EQ:region_nocommon_inner12}\\
R_2\mspace{-2mu} &\leq\mspace{-2mu} I(U_2;Y_2|U_0)\mspace{-3mu}-\mspace{-3mu}I(U_2;U_1,Y_1|U_0)\mspace{-3mu}+\mspace{-3mu}L_2\label{EQ:region_nocommon_inner21}\\
R_2\mspace{-2mu} &\leq\mspace{-2mu} I(U_2;Y_2|U_0)\mspace{-3mu}+\mspace{-3mu}\min\mspace{-3mu}\Big\{\mspace{-1.5mu}I(U_0;Y_1),\mspace{-1.5mu}I(U_0;Y_2)\mspace{-1.5mu}\Big\}\label{EQ:region_nocommon_inner22}\\
R_1\mspace{-2mu}+\mspace{-2mu}R_2\mspace{-2mu} &\leq\mspace{-2mu} I(U_1;Y_1|U_0)+I(U_2;Y_2|U_0)\nonumber\\&-\mspace{-2mu}I(U_1;U_2|U_0)\mspace{-2.5mu}+\mspace{-2mu}\min\mspace{-4mu}\Big\{\mspace{-1mu}I(U_0;\mspace{-1.5mu}Y_1),I(U_0;\mspace{-1.5mu}Y_2)\mspace{-2mu}\Big\}\label{EQ:region_nocommon_inner_sum1}.
\end{align}\label{EQ:region_nocommon_inner}
\end{subequations}
To prove the first claim, assume that $L_1\geq L_1^\star(Q_{U_0,U_1,U_2,X})$. Consequently, the term inside the positive part function from the RHS of \eqref{EQ:region_nocommon_inner11} is non-negative as it satisfies
\begin{align*}
I(U_1;&Y_1|U_0)-I(U_1;U_2,Y_2|U_0)+L_1\\
&\geq I(U_1;Y_1|U_0)+\min\Big\{I(U_0;Y_1),I(U_0;Y_2)\Big\},\numberthis
\end{align*}
which makes \eqref{EQ:region_nocommon_inner11} inactive due to \eqref{EQ:region_nocommon_inner12}, and therefore, $\tilde{\mathcal{R}}_{\mathsf{O}}(L_1,L_2,Q_{U_0,U_1,U_2,X})=\tilde{\mathcal{R}}_{\mathsf{O}}(\infty,L_2,Q_{U_0,U_1,U_2,X})$.

An analogous argument with respect to $L_2$ proves the second claim (essentially by showing that if $L_2\geq L_2^\star$ then \eqref{EQ:region_nocommon_inner21} is inactive due \eqref{EQ:region_nocommon_inner22}). The third claim follows by combining both preceding arguments.

\subsection{Proof of Theorem \ref{TM:outer_bound}}\label{SUBSEC:outer_proof}

We show that given an $(L_1,L_2)$-achievable rate triple $(R_0,R_1,R_2)$, there is a PMF $Q_{W,U,V,X}\in\mathcal{P}(\mathcal{W}\times\mathcal{U}\times\mathcal{V}\times\mathcal{X})$, such that \eqref{EQ:region_outer} holds when the information measures are calculated with respect to $Q_{W,U,V,X}W_{Y_1,Y_2|X}$. Due to the symmetric structure of the rate bounds defining  $\mathcal{R}_{\mathsf{O}}(L_1,L_2)$, we present only the derivation of \eqref{EQ:region_outer0}-\eqref{EQ:region_outer13} and \eqref{EQ:region_outer_sum1}. The other inequalities from \eqref{EQ:region_outer} are established by similar arguments.

Since $(R_0,R_1,R_2)$ is $(L_1,L_2)$-achievable, for every $\epsilon>0$ there is a sufficiently large $n\in\mathbb{N}$ and an $(n,R_0,R_1,R_2)$ code $c_n$ for which \eqref{EQ:achiev_realibility_leakage} holds. We note that all subsequent entropy and mutual information terms are calculated with respect to the PMF from \eqref{EQ:induced_PMF_def} that is specified by $c_n$.

Fix $\epsilon>0$ and find the corresponding blocklength $n\in\mathbb{N}$. By Fano's inequality we have
\begin{equation}
H(M_0,M_j|Y_j^n)\leq 1+n\epsilon R_j\triangleq n\delta_{n,\epsilon}^{(j)},\quad j=1,2.\label{EQ:outer_Fano}
\end{equation}
Define $\delta_{n,\epsilon}=\max\big\{\delta_{n,\epsilon}^{(1)},\delta_{n,\epsilon}^{(2)}\big\}$. Next, by (\ref{EQ:achieve_leakage1}), we write
\begin{align*}
n(L_1+\epsilon)&\geq I(M_1;Y_2^n)\\
   &=I(M_1;M_0,M_2,Y_2^n)-I(M_1;M_0,M_2|Y_2^n)\\
   &\stackrel{(a)}\geq I(M_1;Y_2^n|M_0,M_2)-H(M_0,M_2|Y_2^n)\\
   &\stackrel{(b)}\geq I(M_1;Y_2^n|M_0,M_2)-n\delta_{n,\epsilon}\numberthis\label{EQ:outer_secrecy_cond_info}
\end{align*}
where (a) uses the independence of $M_1$ and $(M_0,M_2)$ and the non-negativity of entropy, while (b) is by \eqref{EQ:outer_Fano}. \eqref{EQ:outer_secrecy_cond_info} implies
\begin{equation}
I(M_1;Y_2^n|M_0,M_2)\leq nL_1+n(\epsilon+\delta_{n,\epsilon}).\label{EQ:outer_secrecy_cond_info2}
\end{equation}
Similarly, we have
\begin{equation}
I(M_1;Y_2^n|M_0)\leq nL_1+n(\epsilon+\delta_{n,\epsilon}).\label{EQ:outer_secrecy_cond_info3}
\end{equation}

 The common message rate $R_0$ satisfies
\begin{subequations}
\begin{align*}
nR_0&=H(M_0)\\
    &\stackrel{(a)}\leq I(M_0;Y_1^n)+n\delta_{n,\epsilon}\\
    &=\sum_{i=1}^nI(M_0;Y_{1,i}|Y_1^{i-1})+n\delta_{n,\epsilon}\\
    &\leq\sum_{i=1}^nI(M_0,Y_1^{i-1};Y_{1,i})+n\delta_{n,\epsilon}\numberthis\label{EQ:outer_0UB_final1a}\\
    &\stackrel{(b)}\leq\sum_{i=1}^nI(W_i;Y_{1,i})+n\delta_{n,\epsilon}\numberthis\label{EQ:outer_0UB_final1b}
\end{align*}
\end{subequations}
where (a) uses \eqref{EQ:outer_Fano} and (b) defines $W_i\triangleq(M_0,Y_1^{i-1},Y_{2,i+1}^n)$. By reversing the roles of $Y_1^n$ and $Y_2^n$ and repeating similar steps, we also have
\begin{subequations}
\begin{align}
nR_0&\leq\sum_{i=1}^nI(M_0,Y_{2,i+1}^n;Y_{2,i})+n\delta_{n,\epsilon}\label{EQ:outer_0UB_final2a}\\
    &\leq \sum_{i=1}^nI(W_i;Y_{2,i})+n\delta_{n,\epsilon}.\label{EQ:outer_0UB_final2b}
\end{align}
\end{subequations}
For $R_1$, it follows that
\begin{align*}
&nR_1\\
    &=H(M_1|M_0,M_2)\\
    &\stackrel{(a)}\leq I(M_1;Y_1^n|M_0,M_2)-I(M_1;Y_2^n|M_0,M_2)+nL_1+n\xi_{n,\epsilon}\\
    &\begin{multlined}[b][.48\textwidth]\stackrel{(b)}=\sum_{i=1}^n\Big[I(M_1;Y_1^i,Y_{2,i+1}^n|M_0,M_2)\\-I(M_1;Y_1^{i-1},Y_{2,i}^n|M_0,M_2)\Big]+nL_1+n\xi_{n,\epsilon}\end{multlined}\\
    &\begin{multlined}[b][.48\textwidth]=\sum_{i=1}^n\Big[I(M_1;Y_{1,i}|M_2,W_i)-I(M_1;Y_{2,i}|M_2,W_i)\Big]\\+nL_1+n\xi_{n,\epsilon}\end{multlined}\\
    &\begin{multlined}[b][.48\textwidth]\stackrel{(c)}=\sum_{i=1}^n\Big[I(U_i;Y_{1,i}|W_i,V_i)-I(U_i;Y_{2,i}|W_i,V_i)\Big]\\+nL_1+n\xi_{n,\epsilon}\end{multlined}\numberthis\label{EQ:outer_1UB_final1}
\end{align*}
where (a) uses \eqref{EQ:outer_Fano} and \eqref{EQ:outer_secrecy_cond_info} and $\xi_{n,\epsilon}=2\delta_{n,\epsilon}+\epsilon$, (b) follows from a telescoping identity \cite[Eqs. (9) and (11)]{Kramer_telescopic2011}, and (c) uses $U_i\triangleq (M_1,W_i)$ and $V_i\triangleq (M_2,W_i)$.

$R_1$ is also upper bounded as
\begin{align*}
  nR_1&=H(M_1|M_0)\\
      &\stackrel{(a)}\leq I(M_1;Y_1^n|M_0)-I(M_1;Y_2^n|M_0)+nL_1+n\xi_{n,\epsilon}\\
      &\stackrel{(b)}=\sum_{i=1}^n\Big[I(M_1;Y_1^i,Y_{2,i+1}^n|M_0)-I(M_1;Y_1^{i-1},Y_{2,i}^n|M_0)\Big]\\&\mspace{305mu}+nL_1+n\xi_{n,\epsilon}\\      &\stackrel{(c)}=\sum_{i=1}^n\Big[I(U_i;Y_{1,i}|W_i)-I(U_i;Y_{2,i}|W_i)\Big]+nL_1+n\xi_{n,\epsilon}\numberthis\label{EQ:outer_1UB_final2}
\end{align*}
where (a) is by \eqref{EQ:outer_Fano} and \eqref{EQ:outer_secrecy_cond_info3}, (b) uses a telescoping identity, while (c) follows by the definition of $(W_i,U_i)$.

For the sum $R_0+R_1$, we have
\begin{align*}
n(R_0+R_1)&= H(M_0,M_1)\\
          &\stackrel{(a)}\leq I(M_0,M_1;Y_1^n)+n\delta_{n,\epsilon}\\
          &\stackrel{(b)}\leq \sum_{i=1}^n I(W_i,U_i;Y_{1,i})+n\delta_{n,\epsilon}\numberthis\label{EQ:outer_1UB_final3}
\end{align*}
where (a) follows from (\ref{EQ:outer_Fano}) and (b) follows by the definition of $(W_i,U_i)$. Moreover, consider
\begin{align*}
&n(R_0+R_1)\\
          &= H(M_1|M_0)+H(M_0)\\
          &\stackrel{(a)}\leq I(M_1;Y_1^n|M_0)+I(M_0;Y_2^n)+n\delta_{n,\epsilon}\\
          &\leq \sum_{i=1}^n \Big[I(M_1,Y_{2,i+1}^n;Y_{1,i}|M_0,Y_1^{i-1})+I(M_0;Y_{2,i}|Y_{2,i+1}^n)\Big]\\
          &\mspace{385mu}+n\delta_{n,\epsilon}\\
          &\begin{multlined}[b][.48\textwidth]=\sum_{i=1}^n \Big[I(U_i;Y_{1,i}|W_i)+I(Y_{2,i+1}^n;Y_{1,i}|M_0,Y_1^{i-1})\\+I(M_0;Y_{2,i}|Y_{2,i+1}^n)\Big]+n\delta_{n,\epsilon}\end{multlined}\\
          &\begin{multlined}[b][.48\textwidth]\stackrel{(b)}= \sum_{i=1}^n \Big[I(U_i;Y_{1,i}|W_i)+I(Y_1^{i-1};Y_{2,i}|M_0,Y_{2,i+1}^n)\\+I(M_0;Y_{2,i}|Y_{2,i+1}^n)\Big]+n\delta_{n,\epsilon}\end{multlined}\\
          &\stackrel{(c)}\leq \sum_{i=1}^n \Big[I(U_i;Y_{1,i}|W_i)+I(W_i;Y_{2,i})\Big]+n\delta_{n,\epsilon}\numberthis\label{EQ:outer_1UB_final4}
\end{align*}
where (a) is by \eqref{EQ:outer_Fano}, (b) is Csisz{\'a}r's sum identity, while (c) uses the definition of $(W_i,U_i)$.

\begin{figure*}[!t]
\setcounter{equation}{107}
\begin{subequations}
\begin{align}
nR_2&\leq\sum_{i=1}^n\Big[I(V_i;Y_{2,i}|W_i,U_i)-I(V_i;Y_{1,i}|W_i,U_i)\Big]+nL_2+n\xi_{n,\epsilon}\label{EQ:outer_2UB_final1}\\
nR_2&\leq\sum_{i=1}^n\Big[I(V_i;Y_{2,i}|W_i)-I(V_i;Y_{1,i}|W_i)\Big]+nL_2+n\xi_{n,\epsilon}\label{EQ:outer_2UB_final2}\\
n(R_0+R_2)&\leq\sum_{i=1}^n I(W_i,V_i;Y_{2,i})+n\delta_{n,\epsilon}\label{EQ:outer_02UB_final1}\\
n(R_0+R_2)&\leq\sum_{i=1}^n \Big[I(V_i;Y_{2,i}|W_i)+I(W_i;Y_{1,i})\Big]+n\delta_{n,\epsilon}\label{EQ:outer_02UB_final2}\\
n(R_0+R_1+R_2)&\leq\sum_{i=1}^n \Big[I(U_i;Y_{1,i}|W_i)+I(V_i;Y_{2,i}|W_i,U_i)+I(W_i;Y_{1,i})\Big]+3n\delta_{n,\epsilon}\label{EQ:outer_sumUB_final3}\\
n(R_0+R_1+R_2)&\leq\sum_{i=1}^n \Big[I(U_i;Y_{1,i}|W_i)+I(V_i;Y_{2,i}|W_i,U_i)+I(W_i;Y_{2,i})\Big]+3n\delta_{n,\epsilon}\label{EQ:outer_sumUB_final4}
\end{align}
\end{subequations}
\hrulefill
\end{figure*}

\setcounter{equation}{102}

 To bound the sum $R_0+R_1+R_2$, we start by writing
\begin{align*}
&H(M_1|M_0,M_2)\\
&\stackrel{(a)}\leq I(M_1;Y_1^n|M_0,M_2)+n\delta_{n,\epsilon}\\
              &=\sum_{i=1}^nI(M_1;Y_{1,i}|M_0,M_2,Y_1^{i-1})+n\delta_{n,\epsilon}\\
              &\leq\sum_{i=1}^nI(M_1,Y_{2,i+1}^n;Y_{1,i}|M_0,M_2,Y_1^{i-1})+n\delta_{n,\epsilon}\\
              &\stackrel{(b)}= \sum_{i=1}^n \Big[I(U_i;Y_{1,i}|W_i,V_i)+I(Y_{2,i+1}^n;Y_{1,i}|M_0,M_2,Y_1^{i-1})\Big]\\&\mspace{375mu}+n\delta_{n,\epsilon}\numberthis\label{EQ:outer_sumUB_temp1|02}
\end{align*}
where (a) uses \eqref{EQ:outer_Fano} and (b) follows by the definition of $(W_i,U_i,V_i)$. Moreover, we have
\begin{align*}
&H(M_2|M_0)\\
          &\stackrel{(a)}\leq I(M_2;Y_2^n|M_0)+n\delta_{n,\epsilon}\\
          &\begin{multlined}[b][.48\textwidth]\stackrel{(b)}=\sum_{i=1}^n\Big[I(M_2;Y_{2,i}^n|M_0,Y_1^{i-1})-I(M_2;Y_{2,i+1}^n|M_0,Y_1^i)\Big]\\+n\delta_{n,\epsilon}\end{multlined}\\
          &\begin{multlined}[b][.48\textwidth]\stackrel{(c)}=\sum_{i=1}^n\Big[I(M_2;Y_{2,i+1}^n|M_0,Y_1^{i-1})+I(V_i;Y_{2,i}|W_i)\\-I(M_2;Y_{1,i},Y_{2,i+1}^n|M_0,Y_1^{i-1})\\+I(M_2;Y_{1,i}|M_0,Y_1^{i-1})\Big]+n\delta_{n,\epsilon}\end{multlined}\\
          &\begin{multlined}[b][.48\textwidth]\stackrel{(d)}=\sum_{i=1}^n\Big[I(V_i;Y_{2,i}|W_i)-I(V_i;Y_{1,i}|W_i)\\+I(M_2;Y_{1,i}|M_0,Y_1^{i-1})\Big]+n\delta_{n,\epsilon}\end{multlined}\numberthis\label{EQ:outer_sumUB_temp2|0}
\end{align*}
where:\\
  (a) follows from (\ref{EQ:outer_Fano});\\
  (b) is a telescoping identity;\\
  (c) is by the mutual information chain rule and the definition of $(V_i,U_i)$;\\
  (d) uses the mutual information chain rule again.
 Combining \eqref{EQ:outer_sumUB_temp1|02} and \eqref{EQ:outer_sumUB_temp2|0} yields
\begin{subequations}
\begin{align*}
&n(R_1+R_2)\\
          &\leq \sum_{i=1}^n \Big[I(U_i;Y_{1,i}|W_i,V_i)+I(V_i;Y_{2,i}|W_i)-I(V_i;Y_{1,i}|W_i)\\
          &\mspace{130mu}+I(M_2,Y_{2,i+1}^n;Y_{1,i}|M_0,Y_1^{i-1})\Big]+2n\delta_{n,\epsilon}\\
          &\begin{multlined}[b][.48\textwidth]=\sum_{i=1}^n \Big[I(U_i;Y_{1,i}|W_i,V_i)+I(V_i;Y_{2,i}|W_i)\\+\mspace{-1.5mu}I(Y_{2,i+1}^n;Y_{1,i}|M_0,Y_1^{i-1})\Big]\mspace{-1.5mu}+\mspace{-1.5mu}2n\delta_{n,\epsilon}\end{multlined}.\numberthis\label{EQ:outer_sumUB_temp12|0a}
\end{align*}
Applying Csisz{\'a}r's sum identity on the last term in \eqref{EQ:outer_sumUB_temp12|0a} gives
\begin{align*}
n(R_1+R_2)&=\sum_{i=1}^n\Big[I(U_i;Y_{1,i}|W_i,V_i)+I(V_i;Y_{2,i}|W_i)\\
&\mspace{15mu}+I(Y_1^{i-1};Y_{2,i}|M_0,Y_{2,i+1}^n)\Big]+2n\delta_{n,\epsilon}.\numberthis\label{EQ:outer_sumUB_temp12|0b}
\end{align*}
\end{subequations}
Combining \eqref{EQ:outer_0UB_final1a} with \eqref{EQ:outer_sumUB_temp12|0a} and \eqref{EQ:outer_0UB_final2a} with \eqref{EQ:outer_sumUB_temp12|0b} yields
\begin{align*}
n(R_0+R_1+R_2)&\leq\sum_{i=1}^n \Big[I(U_i;Y_{1,i}|W_i,V_i)\\
&\mspace{-45mu}+I(V_i;Y_{2,i}|W_i)+I(W_i;Y_{1,i})\Big]+3n\delta_{n,\epsilon}\numberthis\label{EQ:outer_sumUB_final1}
\end{align*}
and
\begin{align*}
n(R_0+R_1+R_2)\leq\sum_{i=1}^n \Big[I(&U_i;Y_{1,i}|W_i,V_i)+I(V_i;Y_{2,i}|W_i)\\
&\mspace{5mu}+I(W_i;Y_{2,i})\Big]+3n\delta_{n,\epsilon},\numberthis\label{EQ:outer_sumUB_final2}
\end{align*}
respectively.

By repeating similar steps, we obtain bounds related to the remaining rate bounds in \eqref{EQ:region_outer} as given in \eqref{EQ:outer_2UB_final1}-\eqref{EQ:outer_sumUB_final4} at the top of this page.

The bounds are rewritten by introducing a time-sharing random variable $Q$ that is uniformly distributed over the set $[1:n]$ and is independent of all the other random variables whose distribution is described in \eqref{EQ:induced_PMF_def}. For instance, the bound \eqref{EQ:outer_1UB_final1} is rewritten as
\setcounter{equation}{108}
\begin{align*}
  R_1&\leq\frac{1}{n}\sum_{q=1}^n\Big[I(U_q;Y_{1,q}|W_q,V_q)-I(U_q;Y_{2,q}|W_q,V_q)\Big]\\
  &\mspace{325mu}+L_1+\xi_{n,\epsilon}\\
     &\begin{multlined}[b][.445\textwidth]=\sum_{i=q}^n\mathbb{P}\big(Q=q\big)\Big[I(U_Q;Y_{1,Q}|W_Q,V_Q,Q=q)\\-I(U_Q;Y_{2,Q}|W_Q,V_Q,Q=q)\Big]+L_1+\xi_{n,\epsilon}\end{multlined}\\
     &\begin{multlined}[b][.45\textwidth]\leq I(U_Q;Y_{1,Q}|W_Q,V_Q,Q)-I(U_Q;Y_{2,Q}|W_Q,V_Q,Q)\\+L_1+n\xi_{n,\epsilon}\end{multlined}\numberthis\label{EQ:outer_1UBQ_final1}
\end{align*}
Denote $Y_1\triangleq Y_{1,Q},\ Y_2\triangleq Y_{2,Q},\ W\triangleq (W_Q,Q)$, $U\triangleq (U_Q,Q)$ and $V\triangleq (V_Q,Q)$. We thus have the bounds from \eqref{EQ:region_outer} with the added terms $\delta_{n,\epsilon}$ and $\xi_{n,\epsilon}$, which can be made arbitrarily small by increasing the blocklength $n$ while decreasing $\epsilon$.

To complete the converse proof note that since the channel is memoryless and without feedback, and because $U_q=(M_1,W_q)$ and $V_q=(M_2,W_q)$, the chain
\begin{equation}
(Y_{1,q},Y_{2,q})-X_q-(U_q,V_q)-W_q\label{EQ:outer_Markov}
\end{equation}
is Markov for every $q\in[1:n]$. This implies that $(Y_1,Y_2)-X-(U,V)-W$ forms a Markov chain, which establishes Theorem \ref{TM:outer_bound}.

\subsection{Proof of Theorem \ref{TM:SDBC_leakage_capacity}}\label{SUBSEC:SDBC_proof}

The direct part of Theorem \ref{TM:SDBC_leakage_capacity} follows by setting $U_0=W$, $U_1=Y_1$ and $U_2=V$ into $\mathcal{C}_{\mathsf{SD}}(L_1,L_2)$, which establishes its inclusion in $\mathcal{R}_{\mathsf{I}}(L_1,L_2)$.

For the converse we prove the reverse inclusion, i.e.,  $\mathcal{R}_{\mathsf{O}}(L_1,L_2)\subseteq\mathcal{C}_{\mathsf{SD}}(L_1,L_2)$. First we remove the restriction from $\mathcal{R}_\mathsf{O}(L_1,L_2)$ that $X-(U,V)-W$ forms a Markov chain; this can only increase the region. Fix a PMF $Q_{W,U,V,X}\in\mathcal{P}(\mathcal{W}\times\mathcal{U}\times\mathcal{V}\times\mathcal{X})$, which induces a joint distribution $Q_{W,U,V,X}\mathds{1}_{\{Y_1=y_1(X)\}}W_{Y_2|X}$, and let $Q_{W,V,Y_1,X}W_{Y_2|X}$ be its marginal PMF of $(W,V,Y_1,X,Y_2)$. Each of the bounds defining $\mathcal{R}_\mathsf{O}(L_1,L_2)$ are evaluated with respect to $Q_{W,U,V,X}\mathds{1}_{\{Y_1=y_1(X)\}}W_{Y_2|X}$, while the information terms from $\mathcal{C}_\mathsf{SD}(L_1,L_2)$ are taken with respect to $Q_{W,V,Y_1,X}W_{Y_2|X}$.

We start by noting that \eqref{EQ:region_SDBC0} and \eqref{EQ:region_outer0} are the same. Next, the RHS of \eqref{EQ:region_outer11} is upper bounded by the RHS of \eqref{EQ:region_SDBC11} since
\begin{align*}
R_1&\leq I(U;Y_1|W,V)-I(U;Y_2|W,V)+L_1\\
   &=H(Y_1|W,V)-H(Y_1|W,V,U)-I(U;Y_2|W,V)+L_1\\
   &\stackrel{(a)}\leq\mspace{-4.5mu} H(Y_1|W,V)\mspace{-2mu}-\mspace{-2mu}I(Y_1;Y_2|W,V,U)\mspace{-2mu}-\mspace{-2mu}I(U;Y_2|W,V)\mspace{-2mu}+\mspace{-2mu}L_1\\
   &=H(Y_1|W,V)-I(U,Y_1;Y_2|W,V)+L_1\\
   &\stackrel{(b)}\leq H(Y_1|W,V,Y_2)+L_1\numberthis\label{EQ:SDBC_proof_bound1}
\end{align*}
where (a) is by the non-negativity of entropy and (b) is because conditioning cannot increase entropy.

For \eqref{EQ:region_outer13} we clearly have
\begin{align*}
R_0+R_1&\leq I(U;Y_1|W)+\min\Big\{I(W;Y_1),I(W;Y_2)\Big\}\\
&\leq H(Y_1|W)+\min\Big\{I(W;Y_1),I(W;Y_2)\Big\},
\numberthis\label{EQ:SDBC_proof_bound01a}
\end{align*}
which coincides with \eqref{EQ:region_SDBC12}. Furthermore, inequalities \eqref{EQ:region_SDBC21} and \eqref{EQ:region_SDBC22} are the same as \eqref{EQ:region_outer22}
and \eqref{EQ:region_outer23}, respectively. For the sum of rates, the RHS of
\eqref{EQ:region_SDBC_sum12} upper bounds that of \eqref{EQ:region_outer_sum1} because
\begin{equation}
I(U;Y_1|W,V)\leq H(Y_1|W,V).
\end{equation}

Removing the other bounds from \eqref{EQ:region_outer} can only increase $\mathcal{R}_{\mathsf{O}}(L_1,L_2)$, which shows its inclusion in $\mathcal{C}_{\mathsf{SD}}(L_1,L_2)$. This characterizes $\mathcal{C}_{\mathsf{SD}}(L_1,L_2)$ as the $(L_1,L_2)$-leakage-capacity region of the SD-BC.

\begin{figure*}[!t]
\setcounter{equation}{117}
\begin{equation}
\mathcal{E}=\Big\{\big(\mathbf{U}_0(1),\mathbf{U}_1(1,1,1,I_1),\mathbf{U}_2(1,1,1,I_2)\big)\notin\mathcal{T}_{\delta'}^{n}(Q_{U_0,U_1,U_2})\Big\}\label{EQ:analysis_event_encoding}
\end{equation}
\hrulefill
\begin{subequations}
\begin{equation}
\mathcal{D}_0=\Big\{\big(\mathbf{U}_0(1),\mathbf{U}_1(1,1,1,I_1),\mathbf{U}_2(1,1,1,I_2),\mathbf{Y}_1,\mathbf{Y}_2\big)\in\mathcal{T}_\delta^{n}(Q_{U_0,U_1,U_2,Y_1,Y_2})\Big\}\label{EQ:analysis_event_LLN}
\end{equation}
\hrulefill
\begin{equation}
\mathcal{D}^{(j)}_0(m_p)=\Big\{\big(\mathbf{U}_0(m_p),\mathbf{Y}_j\big)\in\mathcal{T}_\delta^{n}(Q_{U_0,Y_j})\Big\}\label{EQ:analysis_event0}
\end{equation}
\hrulefill
\begin{equation}
\mathcal{D}^{(j)}_1(m_{jj},w_j,i_j)=\Big\{\big(\mathbf{U}_0(1),\mathbf{U}_j(1,m_{jj},w_j,i_j),\mathbf{Y}_j\big)\in\mathcal{T}_\delta^{n}(Q_{U_0,U_j,Y_j})\Big\}\label{EQ:analysis_event1}
\end{equation}\label{EQ:decoding_errors}
\end{subequations}
\hrulefill
\end{figure*}

\setcounter{equation}{113}

\section{Summary and Concluding Remarks}\label{SEC:summary}

We considered the BC with privacy leakage constraints. Under this model, all four scenarios concerning secrecy (i.e., when both, either or neither of the private messages are secret) are special cases by appropriate choices for the leakage thresholds. Inner and outer bounds on the leakage-capacity region were derived and shown to be tight for SD and PD BCs, as well as for BCs with a degraded message set. The coding strategy that achieved the inner bound is based on a Marton-like codebook construction with a common message supplemented by an extra layer of binning. Splitting each private message into a public and a private part, a public message that comprises the public parts and the common message was constructed. To correlate the codewords for the private parts, we used the likelihood encoder. Its simple structure enabled a rigorous analysis of performance for the proposed scheme. Theorem \ref{TM:inner_bound} fixes a weakness of previous work by letting the eavesdropper know the codebook. The main tool needed was the likelihood encoder (Lemma \ref{LEMMA:uniform_approx}). 


Our results include various past works as special cases. Large leakage thresholds reduce our inner and outer bounds to Marton's inner bound with a common message \cite{GP_SemideterministicBC1980} and the UVW-outer bound \cite{UVW_Outer2010}, respectively. The leakage-capacity region of the SD-BC without a common message recovers the capacity regions where both \cite{Semi-det_BC_secrect_two2009}, either \cite{Goldfeld_Weak_Secrecy_ISIT2015,Semi-det_BC_secrect_one2009}, or neither \cite{GP_SemideterministicBC1980} private message is secret. The result for the BC with a degraded message set and a privacy leakage constraint captures the capacity regions for the BC with confidential messages \cite{Csiszar_Korner_BCconfidential1978} and the BC with a degraded message set (without secrecy) \cite{Korner_BC_DegradedMessageSet1977}. Furthermore, we derived conditions on the allowed leakage values that differentiates whether a further increase of each leakage threshold induces a larger inner bound or not. The conditions effectively let one (numerically) calculate privacy leakage threshold values above which the inner bound saturates. This idea was visualized by means of a BW-BC example that showed the transition of the leakage-capacity region from secrecy-capacity regions for different scenarios to the capacity region without secrecy.


\section*{Acknowledgements}

The authors would like to thank the Associate Editor and the anonymous reviewers for helping to improve the presentation of this paper. We also kindly thank Ido B. Gattegno for his work on the FME-IT software \cite{FME&ITIP} that assisted us with technical details of proofs.

\appendices

\section{Proof of Corollary \ref{COR:DBC_leakage_capacity}}\label{APPEN:DBC_leakage_capacity_proof}

The region $\mathcal{C}_{\mathsf{D}}(L_1,L_2)$ is obtained from $\mathcal{C}_{\mathsf{SD}}^0(L_1,L_2)$ by setting $W=0$ and $V=Y_2$, which implies that $\mathcal{C}_{\mathsf{D}}(L_1,L_2)\subseteq\mathcal{C}_{\mathsf{SD}}^0(L_1,L_2)$. For the converse, the RHS of \eqref{EQ:region_nocommon_SDBC11} is upper bounded by
\begin{equation}
R_1\leq H(Y_1|W,V,Y_2)+L_1\leq H(Y_1|Y_2)+L_1.
\end{equation}
For \eqref{EQ:region_nocommon_SDBC21}, we have
\begin{align*}
I(V;Y_2|W)-I(V;&Y_1|W)+L_2\\
&\leq I(V;Y_1,Y_2|W)-I(V;Y_1|W)+L_2\\
                         &=I(V;Y_2|W,Y_1)+L_2\\
                         &\leq H(Y_2|Y_1)+L_2.\numberthis
\end{align*}
The RHSs of \eqref{EQ:region_nocommon_SDBC12} and \eqref{EQ:region_nocommon_SDBC22} are clearly upper bounded as
\begin{equation}
H(Y_j|W)\mspace{-1.5mu}+\mspace{-1.5mu}\min\mspace{-3mu}\Big\{I(W;Y_1),I(W;Y_2)\Big\}\mspace{-1.5mu}\leq\mspace{-1.5mu} H(Y_j),\mspace{-4mu}\quad j=1,2.
 \end{equation}
Finally, \eqref{EQ:region_DBC_sum} is implied by \eqref{EQ:region_nocommon_SDBC_sum12} since
\begin{align*}
&R_1+R_2\\
&\leq H(Y_1|W,V)+I(V;Y_2|W)+\min\big\{I(W;Y_1),I(W;Y_2)\big\}\\
       &\leq H(Y_1|W,V)+I(W,V;Y_2)\\
       &\leq H(Y_1,Y_2|W,V)+I(W,V;Y_1,Y_2)\\
       &= H(Y_1,Y_2).\numberthis
\end{align*}


\section{Error Probability Analysis for the Proof of Theorem \ref{TM:inner_bound}}\label{APPEN:error_analysis}



By the symmetry of the codebook construction with respect to $(M_p,M_{11},W_1,M_{22},W_2)$ and due to their uniformity, we may assume that $(M_p,M_{11},W_1,M_{22},W_2)=(1,1,1,1,1)$.

\textbf{Encoding errors:} Fix any $\delta'\in(0,\delta)$. An encoding error event is described as given in \eqref{EQ:analysis_event_encoding} at the top of this page. 

\textbf{Decoding errors:} To account for decoding errors, define \eqref{EQ:decoding_errors} from the top of this page, where $j=1,2$.
\setcounter{equation}{119}


For any event $\mathcal{A}$ from the $\sigma$-algebra over which $\mathbb{P}$ is defined, denote $\mathbb{P}_1=\mathbb{P}\big(\mathcal{A}\big|M_p=1,M_{11}=1,W_1=1,M_{22}=1,W_2=1\big)$. By the union bound, the expected error probability is bounded as in \eqref{EQ:analysis_error_prob_UB}, given at the top of the next page.\footnote{As in Section \ref{SUBSEC:inner_proof}, we slightly abuse notation in writing $\mathbb{E}P_\mathsf{e}(\mathsf{C}_n)$ because $P_e$ is actually a function of the code $c_n$ rather than the codebook $\mathcal{C}_n$. We favor this notation for its simplicity and remind the reader that $\mathcal{C}_n$ uniquely defines $c_n$.} Note that with respect to the notation in \eqref{EQ:analysis_error_prob_UB}, $P_0^{[1]}$ is the probability of an encoding error, while $P_j^{[k]}$, for $k\in[0:3]$, are the decoding errors of Decoder $j$. We proceed with the following steps:
\begin{figure*}[!t]

\begin{align*}
&\mathbb{E}P_{\mathsf{e}}(\mathsf{C}_n)\\
&\leq\mathbb{P}_1\vast(\mathcal{E}\cup\mathcal{D}_0^c\cup\bigcup_{j=1,2}\vast\{\mathcal{D}^{(j)}_0(1)^c\cup\left\{\bigcup_{\tilde{m}_p\neq 1}\mathcal{D}^{(j)}_0(\tilde{m}_p)\right\}\cup\mathcal{D}^{(j)}_1(1,1,I_j)^c\cup\left\{\bigcup_{(\tilde{m}_{jj},\tilde{w}_j)\neq(1,1)}\mspace{-15mu}\mathcal{D}^{(j)}_0(\tilde{m}_{jj},\tilde{w}_j,I_j)\right\}\vast\}\vast)\\
&\begin{multlined}[b][.9\textwidth]\leq\underbrace{\mathbb{P}_1\big(\mathcal{E}\big)}_{P_0^{[1]}}+\underbrace{\mathbb{P}_1\big(\mathcal{D}_0^c\cap\mathcal{E}^c\big)}_{P_0^{[2]}}+\sum_{j=1,2}\Vast[\underbrace{\mathbb{P}_1\Big(\mathcal{D}^{(j)}_0(1)^c\cap\mathcal{D}_0\Big)}_{P_j^{[0]}}+\underbrace{\mathbb{P}_1\Big(\mathcal{D}^{(j)}_1(1,1,I_j)^c\cap\mathcal{D}_0\Big)}_{P_j^{[1]}}+\underbrace{\mathbb{P}_1\left(\bigcup_{\tilde{m}_p\neq1}\mathcal{D}^{(j)}_0(\tilde{m}_p)\right)}_{P_J^{[2]}}\\+\underbrace{\mathbb{P}_1\left(\bigcup_{\substack{(\tilde{m}_{jj},\tilde{w}_j)\neq(1,1),\\\tilde{i}_j\in\mathcal{I}_j}}\mathcal{D}^{(j)}_1(\tilde{m}_{jj},\tilde{w}_j,\tilde{i}_j)\right)}_{P_j^{[3]}}\Vast]\end{multlined}\numberthis\label{EQ:analysis_error_prob_UB}
\end{align*}
\hrulefill
\end{figure*}



\begin{enumerate}


\item By \cite[Theorem 3]{Yassaee_fidelity_ISIT2015}, $P_0^{[1]}\to 0$ as $n\to\infty$ if
\begin{equation}
R'_1+R'_2>I(U_1;U_2|U_0).\label{EQ:analysis_covering_rb}
\end{equation}

\item The Conditional Typicality Lemma \cite[Section 2.5]{ElGamal2011} implies that $P_0^{[2]}\to 0$ as $n$ grows. More precisely, there exists a function $\beta(n,\delta,\delta')$ with $\lim_{n\to\infty}\beta(n,\delta,\delta')=0$ for any $0<\delta'<\delta$, such that $P_0^{[2]}\leq \beta(n,\delta,\delta')$. Furthermore, replacing $\delta$ and $\delta'$ with properly chosen decaying sequences $\{\delta_n\}_{n\in\mathbb{N}}$ and $\{\delta'_n\}_{n\in\mathbb{N}}$, respectively, and setting $\beta_n\triangleq\beta(n,\delta_n,\delta'_n)$, we have $\lim_{n\to\infty}\beta_n=0$.

\item The definitions in \eqref{EQ:decoding_errors} clearly give $P_j^{[0]}=P_j^{[1]}=0$, for $j=1,2$ and every $n\in\mathbb{N}$. This is since $\left\{\mathcal{D}^{(j)}_0(1)^c\cap\mathcal{D}_0\right\}=\left\{\mathcal{D}^{(j)}_1(1,1,I_j)^c\cap\mathcal{D}_0\right\}=\emptyset$, for $j=1,2$.


\item
For $P_j^{[2]}$, $j=1,2$, we have
\begin{align*}
P_j^{[2]}&\stackrel{(a)}\leq\sum_{\tilde{m}_p\neq1}2^{-n\big(I(U_0;Y_j)-\tau^{[2]}_j(\delta)\big)}\\
         &\leq2^{nR_p}2^{-n\big(I(U_0;Y_j)-\tau^{[2]}_j(\delta)\big)}\\
         &=2^{n\big(R_p-I(U_0;Y_j)+\tau^{[2]}_j(\delta)\big)}\numberthis
\end{align*}
where (a) follows since $\mathbf{U}_0(\tilde{m}_p)$ is independent of $\mathbf{Y}_j$, for any $\tilde{m}_p\neq 1$. Thus, for $P_j^{[2]}$ to vanish as $n\to\infty$, we take:
\begin{equation}
     R_p<I(U_0;Y_j)-\tau^{[2]}_j(\delta),\quad j=1,2.\label{EQ:analysis_RB1}
\end{equation}
where $\tau^{[2]}_j(\delta)\to 0$ as $\delta\to 0$.


\item
For $P_j^{[3]}$, $j=1,2$, we have
\begin{align*}
P_j^{[3]}&\stackrel{(a)}\leq\sum_{\substack{(\tilde{m}_{jj},\tilde{w}_j)\neq (1,1),\\\tilde{i}_j\in\mathcal{I}_j}}2^{-n\big(I(U_j;Y_j|U_0)-\tau^{[3]}_j(\delta)\big)}\\
         &\leq2^{n(R_{jj}+R'_j+\tilde{R}_j)}2^{-n\big(I(U_j;Y_j|U_0)-\tau^{[3]}_j(\delta)\big)}\\
         &=2^{n\big(R_{jj}+R'_j+\tilde{R}_j-I(U_j;Y_j|U_0)+\tau^{[3]}_j(\delta)\big)}\numberthis
\end{align*}
where (a) follows since $\mathbf{U}_j(1,\tilde{m}_{jj},\tilde{w}_j,\tilde{i}_j)$ is independent of $\mathbf{Y}_j$, for any $(\tilde{m}_{jj},\tilde{w}_j)\neq (1,1)$ and $\tilde{i}_j\in\mathcal{I}_j$, while both of them are drawn conditioned on $\mathbf{U}_0(1)$. We have $P_j^{[3]}\to 0$ as $n\to\infty$ if
\begin{equation}
     R_{jj}+R'_j+\tilde{R}_j<I(U_j;Y_j|U_0)-\tau^{[3]}_j(\delta),\quad j=1,2,\label{EQ:analysis_RB2}
\end{equation}
where, as before, $\tau^{[3]}_j(\delta)\to 0$ as $\delta\to 0$.

\end{enumerate}

Summarizing the above results, while substituting $R_p=R_0+R_{10}+R_{20}$ and setting
\begin{equation}
\tau_{\delta}\triangleq\max\left\{\tau^{[k]}_j(\delta)\right\}_{\substack{j=1,2,\\k=2,3}}
\end{equation}
we find that 
\begin{equation}
\mathbb{E}P_{\mathsf{e}}(\mathsf{C}_n)\leq \eta(n,\delta,\delta'),\quad\forall n\in\mathbb{N},
\end{equation}
where $\lim_{n\to\infty}\eta(n,\delta,\delta')=0$ for all $0<\delta'<\delta$, if the conditions in \eqref{EQ:achiev_rb} are met. As mentioned before, if we replace $\delta$ and $\delta'$ with properly chosen sequences $\{\delta_n\}_{n\in\mathbb{N}}$ and $\{\delta'_n\}_{n\in\mathbb{N}}$, respectively, that decay sufficiently slowly to zero and set $\eta_n\triangleq\eta(n,\delta_n,\delta'_n)$, we have $\lim_{n\to\infty}\eta_n=0$.

\subsection{Leakage Associated Errors}
This subsection shows how \eqref{EQ:achiev_rb6} ensures $\mathbb{E}\lambda^{(1)}_{m_{11}}(\mathsf{C}_n)\to 0$ as $n\to\infty$, for any $m_{11}\in\mathcal{M}_{11}$. As before, by the symmetry of the underlying random code with respect to the messages, we have
\begin{equation}
\mathbb{E}\lambda^{(1)}_{m_{11}}(\mathsf{C}_n)=\mathbb{E}\lambda^{(1)}_1(\mathsf{C}_n),\quad\forall m_{11}\in\mathcal{M}_{11},\label{EQ:leakage_error_symmetry}
\end{equation}
and we may further assume that $(M_p,W_1,M_{22},W_2)=(1,1,1,1)$. By arguments similar to those presented in the encoding and decoding error probability analysis, one can verify that \eqref{EQ:achiev_rb6} implies the existence of a function $\kappa(n,\delta)$ with $\lim_{n\to\infty}\kappa(n,\delta)=0$ for any $\delta>0$, such that $\mathbb{E}\lambda^{(1)}_1(\mathsf{C}_n)\leq \kappa(n,\delta)$. Furthermore, replacing $\delta$ with a sequence $\{\delta_n\}_{n\in\mathbb{N}}$ that decays sufficiently slow to zero as $n$ grows and setting $\kappa_n\triangleq\kappa(n,\delta_n)$, we have $\kappa_n\to 0$ as $n\to\infty$.

This essentially follows by the law of large numbers and the Conditional Typicality Lemma that ensure the joint typicality of the transmitted sequences and the outputs. 
If $\tilde{w}_1$ is incorrect but $I_1$ is the true index chosen by the likelihood encoder, $\mathbf{U}_1(1,1,\tilde{w}_1,I_1)$ is conditionally independent $\mathbf{Y}_2$ given $(\mathbf{U}_0(1),\mathbf{U}_2(1,1,1,I_2))$. The correlation between $\mathbf{U}_0(1)$, $\mathbf{U}_1(1,1,\tilde{w}_1,I_1)$ and $\mathbf{U}_2(1,1,1,I_2)$ is a consequence of the likelihood encoder's operation. Since the search space in this case is of size $2^{n\tilde{R}_1}$, taking
\begin{subequations}
\begin{equation}
\tilde{R}_1<I(U_1;Y_2|U_0,U_2)-\xi(\delta)
\end{equation}
where $\xi(\delta)\to0$ as $\delta\to 0$, results in a vanishing probability of the event that this $u_1$-sequence satisfies the typicality test from \eqref{EQ:leakage_testj}. 

Furthermore, if $\tilde{w}_1$ and $\tilde{i}_1$ are both incorrect,  $\mathbf{U}_1(1,1,\tilde{w}_1,\tilde{i}_1)$ is conditionally independent of  $\big(\mathbf{U}_2(1,1,1,I_2),\mathbf{Y}_2\big)$ given $\mathbf{U}_0(1)$. The search space is now of size $2^{n(\tilde{R}_1+R'_1)}$, and therefore, taking
\begin{equation}
\tilde{R}_1+R'_1<I(U_1;U_2,Y_2|U_0)-\xi(\delta)\label{EQ:leakage_error_dominate}
\end{equation}
\end{subequations}
implies a vanishing probability of this second error event.

Finally, note that the error event where $W_1=1$ is correct but $\tilde{i}_1$ is wrong has arbitrarily small probability if $R'_1<I(U_1;U_2,Y_2|U_0)-\xi(\delta)$ (the structure of the mutual information term is the same as in \eqref{EQ:leakage_error_dominate} because an incorrect $\tilde{i}_1$ produces the same statistical relations as an incorrect pair $(\tilde{w}_1,\tilde{i}_1)$). Evidently, the latter   constraint is redundant due to \eqref{EQ:leakage_error_dominate}.

\begin{figure*}[!t]
\setcounter{equation}{136}
\begin{align*}
H(&\mathbf{Y}_1|\mathbf{U}_0,\mathbf{U}_1,\mathbf{U}_2,E=0,\mathsf{C}_n)\\&\stackrel{(a)}=H(\mathbf{Y}_1|\mathbf{U}_0,\mathbf{U}_1,\mathbf{U}_2,E=0)\\
&=\mspace{-10mu}\sum_{\substack{(\mathbf{u}_0,\mathbf{u}_1,\mathbf{u}_2)\\\in\mathcal{T}_\delta^n(Q_{U_0,U_1,U_2})}}\mspace{-20mu}P_{\mathbf{U}_0,\mathbf{U}_1,\mathbf{U}_2|E}(\mathbf{u}_0,\mathbf{u}_1,\mathbf{u}_2|0)H(\mathbf{Y}_2|\mathbf{U}_0=\mathbf{u}_0,\mathbf{U}_1=\mathbf{u}_1,\mathbf{U}_2=\mathbf{u}_2,E=0)\\
&=\mspace{-10mu}\sum_{\substack{(\mathbf{u}_0,\mathbf{u}_1,\mathbf{u}_2)\\\in\mathcal{T}_\delta^n(Q_{U_0,U_1,U_2})}}\mspace{-20mu}P_{\mathbf{U}_0,\mathbf{U}_1,\mathbf{U}_2|E}(\mathbf{u}_0,\mathbf{u}_1,\mathbf{u}_2|0)\sum_{i=1}^nH(Y_{2,i}|\mathbf{U}_0=\mathbf{u}_0,\mathbf{U}_1=\mathbf{u}_1,\mathbf{U}_2=\mathbf{u}_2,Y_2^{i-1},E=0)\\
&\stackrel{(b)}=\mspace{-10mu}\sum_{\substack{(\mathbf{u}_0,\mathbf{u}_1,\mathbf{u}_2)\\\in\mathcal{T}_\delta^n(Q_{U_0,U_1,U_2})}}\mspace{-20mu}P_{\mathbf{U}_0,\mathbf{U}_1,\mathbf{U}_2|E}(\mathbf{u}_0,\mathbf{u}_1,\mathbf{u}_2|0)\sum_{i=1}^nH(Y_{2,i}|U_{0,i}=u_{0,i},U_{1,i}=u_{1,i},U_{2,i}=u_{2,i})\\
&=\mspace{-10mu}\sum_{\substack{(\mathbf{u}_0,\mathbf{u}_1,\mathbf{u}_2)\\\in\mathcal{T}_\delta^n(Q_{U_0,U_1,U_2})}}\mspace{-20mu}P_{\mathbf{U}_0,\mathbf{U}_1,\mathbf{U}_2|E}(\mathbf{u}_0,\mathbf{u}_1,\mathbf{u}_2|0)\mspace{-10mu}\sum_{\substack{(u_0,u_1,u_2)\\\in\mathcal{U}_0\times\mathcal{U}_1\times\mathcal{U}_2}}\mspace{-10mu}\nu_{\mathbf{u}_0,\mathbf{u}_1,\mathbf{u}_2}(u_0,u_1,u_2)H(Y_2|U_0=u_0,U_1=u_1,U_2=u_2)\\
&\stackrel{(c)}\geq n\cdot \sum_{\substack{(\mathbf{u}_0,\mathbf{u}_1,\mathbf{u}_2)\\\in\mathcal{T}_\delta^n(Q_{U_0,U_1,U_2})}}\mspace{-20mu}P_{\mathbf{U}_0,\mathbf{U}_1,\mathbf{U}_2|E}(\mathbf{u}_0,\mathbf{u}_1,\mathbf{u}_2|0)(1-\delta) H(Y_2|U_0,U_1,U_2)\\
&=n(1-\delta)H(Y_2|U_0,U_1,U_2)\numberthis\label{EQ:lemma1_proof_LB1}
\end{align*}
\hrulefill
\end{figure*}
\setcounter{equation}{129}






\section{Proof of Lemma \ref{LEMMA:1}}\label{APPEN:lemma1_proof}
Recall that $P_1^{(\mathsf{Leak)}}(\mathcal{C}_n)$ denotes the error probability in decoding $(W_1,I_1)$ from $(M_p,M_{11},M_{22},W_2,I_2,\mathbf{Y}_2)$ by means of the typicality test from \eqref{EQ:leakage_testj} with respect to the fixed code $\mathcal{C}_n\in\mathfrak{C}_n$. The analysis in Appendix \ref{APPEN:error_analysis} shows that as long as \eqref{EQ:achiev_rb6} holds, we have
\begin{equation}
\mathbb{E}P_1^{(\mathsf{Leak)}}(\mathsf{C}_n)\leq \kappa(n,\delta,\delta')\label{EQ:lemma2_proof_UB}
\end{equation}
where $\lim_{n\to\infty}\kappa(n,\delta,\delta')=0$ for all $0<\delta'<\delta$. As a consequence, we have
\begin{align*}
    H(W_1,I_1|&M_p,M_{11},M_{22},W_2,I_2,\mathbf{Y}_2,\mathsf{C}_n)\\
    &\leq H(W_1,I_1|M_p,M_{11},M_{22},W_2,I_2,\mathbf{Y}_2)\\
    & \stackrel{(a)}\leq 1+n\cdot\kappa(n,\delta,\delta') n(\tilde{R}_1+R'_1)\numberthis
    \end{align*}
where (a) is because conditioning cannot increase entropy, while (b) uses Fano's inequality and \eqref{EQ:lemma2_proof_UB}. Setting $\zeta_1(n,\delta,\delta')\triangleq  \frac{1}{n}+\kappa(n,\delta,\delta') (\tilde{R}_1+R'_1)$ completes the proof.


\section{Proof of Lemma \ref{LEMMA:2}}\label{APPEN:lemma2_proof}

Define the indicator function $E=\mathds{1}_\mathcal{A}$, where
\begin{equation}
\mathcal{A}=\Big\{(\mathbf{U}_0,\mathbf{U}_1,\mathbf{U}_2,\mathbf{Y}_2)\notin\mathcal{T}_\delta^{n}(Q_{U_0,U_1,U_2,Y_2})\Big\}
\end{equation}
and note that $\mathbb{P}\big(E=1\big)\leq \mathbb{P}_1\big(\mathcal{E}\big)+\mathbb{P}_1\big(\mathcal{D}_0^c\cap\mathcal{E}^c\big)$, where $\mathcal{E}$ and $\mathcal{D}_0$ are defined in \eqref{EQ:analysis_event_encoding} and \eqref{EQ:analysis_event_LLN}, respectively, from Appendix \ref{APPEN:error_analysis}. The analysis in Appendix \ref{APPEN:error_analysis} shows the existence of a function $\tilde{\beta}(n,\delta,\delta')$, such that 
\begin{equation}
\mathbb{P}\big(E=1\big)=\mathbb{P}_1\big(\mathcal{E}\big)+\mathbb{P}_1\big(\mathcal{D}_0^c\cap\mathcal{E}^c\big)\leq \tilde{\beta}(n,\delta,\delta')\label{EQ:lemma2_atypical_prob}
\end{equation}
where $0<\delta'<\delta$ and $\lim_{n\to\infty}\tilde{\beta}(n,\delta,\delta')=0$ for all such values of $\delta$ and $\delta'$. Furthermore, $\lim_{n\to\infty}\tilde{\beta}(n,\delta_n,\delta'_n)=0$ for sequences $\{\delta_n\}_{n\in\mathbb{N}}$ and $\{\delta'_n\}_{n\in\mathbb{N}}$ that decay sufficiently slow to zero  with $n$.

We now expand the mutual information term from the LHS of \eqref{EQ:lemma1_ineq2} as follows
\begin{align*}
&I(\mathbf{U}_1;\mathbf{Y}_2|\mathbf{U}_0,\mathbf{U}_2,\mathsf{C}_n)\\
    &\leq I(\mathbf{U}_1,E;\mathbf{Y}_2|\mathbf{U}_0,\mathbf{U}_2,\mathsf{C}_n)\\
    &=I(E;\mathbf{U}_2|\mathbf{U}_0,,\mathbf{U}_2,\mathsf{C}_n)+I(\mathbf{U}_1;\mathbf{Y}_2|\mathbf{U}_0,\mathbf{U}_2,E,\mathsf{C}_n)\\
    &\stackrel{(a)}\leq 1+\sum_{j=0}^1\mathbb{P}\big(E=j\big)I(\mathbf{U}_1;\mathbf{Y}_2|\mathbf{U}_0,\mathbf{U}_2,E=j,\mathsf{C}_n).\numberthis\label{EQ:lemma1_proof_target}
\end{align*}
where (a) is because $E$ is binary and the entropy function is non-negative. Note that
\begin{align*}
\mathbb{P}\big(E=1\big)&I(\mathbf{U}_1;\mathbf{Y}_2|\mathbf{U}_0,\mathbf{U}_2,E=1,\mathsf{C}_n)\\
    &\leq\mathbb{P}\big(E=1\big)H(\mathbf{Y}_2|E=1,\mathsf{C}_n)\\
    &\leq \tilde{\beta}(n,\delta,\delta')\cdot n\log|\mathcal{Y}_2|\numberthis\label{EQ:lemma1_proof_UB1}.
\end{align*}
where (a) uses \eqref{EQ:lemma2_atypical_prob}.

\begin{figure*}[!t]
\setcounter{equation}{142}
\begin{equation}
\tilde{\mathcal{D}}\triangleq\Big\{\exists (\tilde{m}_p,\tilde{m}_{22},\tilde{w}_2,\tilde{i}_2)\neq (M_p,M_{22},W_2,I_2),\quad \mathbf{U}_0(\tilde{m}_p)=\mathbf{U}_0\ \ \mbox{and}\ \ \mathbf{U}_2(\tilde{m}_p,\tilde{m}_{22},\tilde{w}_2,\tilde{i}_2)=\mathbf{U}_2\Big\}\label{EQ:tilde_D}
\end{equation}
\hrulefill
\end{figure*}
\setcounter{equation}{135}

For the mutual information term conditioned on $E=0$, we first have
\begin{align*}
&H(\mathbf{Y}_2|\mathbf{U}_0,\mathbf{U}_2,E=0,\mathsf{C}_n)\\
    &\leq H(\mathbf{Y}_2|\mathbf{U}_0,\mathbf{U}_2,E=0)\\
    &=\mspace{-20mu}\sum_{\substack{(\mathbf{u}_0,\mathbf{u}_2)\\\in\mathcal{T}_\delta^n(Q_{U_0,U_2})}}\mspace{-25mu}\mspace{-3mu}Q_{\mathbf{U}_0,\mathbf{U}_2|E}(\mathbf{u}_0,\mathbf{u}_2|0)H(\mathbf{Y}_2|\mathbf{U}_0\mspace{-3mu}=\mspace{-3mu}\mathbf{u}_0,\mathbf{U}_2\mspace{-3mu}=\mspace{-3mu}\mathbf{u}_2,E\mspace{-3mu}=\mspace{-3mu}0)\\
    &\leq nH(Y_2|U_0,U_2)\numberthis\label{EQ:lemma1_proof_UB2}
\end{align*}
where the last inequality is because for every $(\mathbf{u}_0,\mathbf{u}_2)\in\mathcal{T}_\delta^n(Q_{U_0,U_2})$ the support of the conditional PMF $P_{\mathbf{Y}_2|\mathbf{U}_0=\mathbf{u}_0,\mathbf{U}_2=\mathbf{u}_2,E=0}$ is upper bounded by the size of the conditional typical set $\mathcal{T}_\delta^n(Q_{U_0,U_2,Y_2}|\mathbf{u}_0,\mathbf{u}_2)$, which is upper bounded by $2^{nH(Y_2|U_0,U_2)(1+\delta)}$. This step also relies on the entropy being maximized by the uniform distribution and the logarithm being a monotonically increasing function.

For the other (subtracted) entropy term, we have \eqref{EQ:lemma1_proof_LB1} given at the top of this page, where (a) is because $\mathbf{Y}_2-(\mathbf{U}_0,\mathbf{U}_1,\mathbf{U}_2)-\mathsf{C}_n$ forms a Markov chain, (b) follows since given $(U_{0,i},U_{1,i},U_{2,i})$, $Y_{2,i}$ is independent of all other random variables, while (c) is by the definition of letter-typical sequences from \eqref{EQ:typical_set_def}.

\setcounter{equation}{137}
Inserting \eqref{EQ:lemma1_proof_UB1}, \eqref{EQ:lemma1_proof_UB2} and \eqref{EQ:lemma1_proof_LB1} into \eqref{EQ:lemma1_proof_target} gives
\begin{equation}
I(\mathbf{U}_1;\mathbf{Y}_2|\mathbf{U}_0,\mathbf{U}_2,\mathsf{C}_n)\leq nI(U_1;Y_2|U_0,U_2)+n\zeta_2(n,\delta,\delta')
\end{equation}
where
\begin{equation}
\zeta_2(n,\delta,\delta')=\frac{1}{n}+\delta H(Y_2|U_0,U_1,U_2)+\tilde{\beta}(n,\delta,\delta')\numberthis\label{EQ:lemma2_proof_done}
\end{equation}
as needed.


\section{Proof of Lemma \ref{LEMMA:0}}\label{APPEN:lemma0_proof}

Rewriting the mutual information term of interest as a difference of entropies, we have
\begin{align*}
I(&\mathbf{U}_1;M_{22},W_2,I_2|M_p,\mathsf{C}_n)\\
&=H(\mathbf{U}_1|M_p,\mathsf{C}_n)-H(\mathbf{U}_1|M_p,M_{22},W_2,I_2,\mathsf{C}_n).\numberthis\label{EQ:lemma0_proof_target}
\end{align*}
Since $\mathbf{U}_0$ is defined by $(M_p,\mathsf{C}_n)$ we clearly have,
\begin{equation}
H(\mathbf{U}_1|M_p,\mathsf{C}_n)\leq H(\mathbf{U}_1|\mathbf{U}_0,\mathsf{C}_n).\label{EQ:lemma0_proof_UB1}
\end{equation}
Next, since $(M_p,M_{22},W_2,I_2,\mathsf{C}_n)$ determines both $\mathbf{U}_0$ and $\mathbf{U}_2$, we write the subtracted entropy term as
\begin{align*}
&H(\mathbf{U}_1|M_p,M_{22},W_2,I_2,\mathsf{C}_n)\\
&=H(\mathbf{U}_1|\mathbf{U}_0,\mathbf{U}_2,M_p,M_{22},W_2,I_2,\mathsf{C}_n)\\
&=H(\mathbf{U}_1|\mathbf{U}_0,\mathbf{U}_2,\mspace{-1mu}\mathsf{C}_n)\mspace{-3mu}-\mspace{-3mu}I(\mathbf{U}_1;M_p,M_{22},W_2,I_2|\mathbf{U}_0,\mathbf{U}_2,\mspace{-1mu}\mathsf{C}_n)\\
&\geq H(\mathbf{U}_1|\mathbf{U}_0,\mathbf{U}_2,\mathsf{C}_n)-H(M_p,M_{22},W_2,I_2|\mathbf{U}_0,\mathbf{U}_2,\mathsf{C}_n).\numberthis\label{EQ:lemma0_proof_LB1}
\end{align*}

We now upper bound $H(M_p,M_{22},W_2,I_2|\mathbf{U}_0,\mathbf{U}_2,\mathsf{C}_n)$ by a vanishing term times the blocklength $n$. Let $F=\mathds{1}_{\tilde{\mathcal{D}}}$ be the indicator function to the event $\tilde{\mathcal{D}}$ defined in  \eqref{EQ:tilde_D} at the top of this page. Standard error probability analysis of random codes shows that
\setcounter{equation}{143}
\begin{equation}
\mathbb{P}\big(F=1\big)=\mathbb{P}\big(\tilde{\mathcal{D}}\big)\leq 2^{n\big(R_p+R_{22}+\tilde{R}_2+R'_2-H(U_0,U_2)+\tilde{\alpha}(\delta)\big)}
\end{equation}
where $\tilde{\alpha}(\delta)\to 0$ as $\delta\to 0$. Consequently, taking
\begin{equation}
R_p+R_{22}+\tilde{R}_2+R'_2<H(U_0,U_2)-\tilde{\alpha}(\delta)\label{EQ:lemma0_proof_UB}
\end{equation}
results in $\mathbb{P}\big(F=1\big)\leq \tilde{\kappa}(n,\delta)$ with $\lim_{n\to\infty}\tilde{\kappa}(n,\delta)=0$ for every $\delta>0$. Next, note that \eqref{EQ:lemma0_proof_UB} holds on account of \eqref{EQ:achiev_rb5} and \eqref{EQ:achiev_rb4} (adding \eqref{EQ:achiev_rb5} and \eqref{EQ:achiev_rb4} results in a tighter bound on the same rates) and consider the following:
\begin{align*}
    &H(M_p,M_{22},W_2,I_2|\mathbf{U}_0,\mathbf{U}_2,\mathsf{C}_n)\\
    &\leq H(M_p,M_{22},W_2,I_2|\mathbf{U}_0,\mathbf{U}_2)\\
    &\stackrel{(a)}\leq 1+ H(M_p,M_{22},W_2,I_2|\mathbf{U}_0,\mathbf{U}_2,F)\\
    &\begin{multlined}[b][.45\textwidth]= 1+\mathbb{P}\big(F=0\big)H(M_p,M_{22},W_2,I_2|\mathbf{U}_0,\mathbf{U}_2,F=0)\\+\mathbb{P}\big(F=1\big)H(M_p,M_{22},W_2,I_2|\mathbf{U}_0,\mathbf{U}_2,F=1)\end{multlined}\\
    &\begin{multlined}[b][.45\textwidth]\stackrel{(b)}\leq 1+H(M_p,M_{22},W_2,I_2|\mathbf{U}_0,\mathbf{U}_2,F=0)\\+\mathbb{P}\big(F=1\big)\cdot n(R_p+R_{22}+\tilde{R}_2+R'_2)\end{multlined}\\
    &\stackrel{(c)}\leq 2\Big[1+n\cdot\tilde{\kappa}(n,\delta)(R_p+R_{22}+\tilde{R}_2+R'_2)\Big]\\
    &= n\tilde{\zeta}_3^{(1)}(n,\delta)\numberthis\label{EQ:lemma0_proof_UB2}
\end{align*}
where 
\begin{equation}
\tilde{\zeta}_3^{(1)}(n,\delta)\triangleq \frac{1}{n}+\tilde{\kappa}(n,\delta)(R_p+R_{22}+\tilde{R}_2+R'_2).
\end{equation}
In the above derivation (a) follows because the uniform distribution maximizes entropy and since $F$ is binary, (b) upper bounds the first entropy term by the logarithm of the support size, while (c) uses Fano's inequality. 

Inserting \eqref{EQ:lemma0_proof_UB1}, \eqref{EQ:lemma0_proof_LB1} and \eqref{EQ:lemma0_proof_UB2} into \eqref{EQ:lemma0_proof_target} gives
\begin{equation}
I(\mathbf{U}_1;\mspace{-1.5mu}M_{22},\mspace{-1.5mu}W_2,I_2|M_p,\mspace{-1.5mu}\mathsf{C}_n\mspace{-1mu})\mspace{-2mu}\leq\mspace{-2mu} I(\mathbf{U}_1;\mspace{-1.5mu}\mathbf{U}_2|\mathbf{U}_0,\mspace{-1.5mu}\mathsf{C}_n\mspace{-1mu})+n\tilde{\zeta}_3^{(1)}(n,\mspace{-1.5mu}\delta).\label{EQ:lemma0_final1}
\end{equation}
To complete the proof, it suffices to show that there exists a function $\tilde{\zeta}_3^{(2)}(n,\delta,\delta')$ that satisfies the same properties as $\zeta_3(n,\delta,\delta')$ from the statement of Lemma \ref{LEMMA:0} for which
\begin{equation}
I(\mathbf{U}_1;\mathbf{U}_2|\mathbf{U}_0,\mathsf{C}_n)\leq n I(U_1;U_2|U_0)+n\tilde{\zeta}_3^{(2)}(n,\delta,\delta').\label{EQ:lemma0_final2}
\end{equation}
This can be established by arguments similar to those presented in the proof of Lemma \ref{LEMMA:2} and we therefore omit the details. Combining \eqref{EQ:lemma0_final1} with \eqref{EQ:lemma0_final2} and setting $\zeta_3(n,\delta,\delta')\triangleq \tilde{\zeta}_3^{(1)}(n,\delta)+\tilde{\zeta}_3^{(2)}(n,\delta,\delta')$ completes the proof.

\bibliographystyle{unsrt}
\bibliographystyle{IEEEtran}
\bibliography{ref}

\begin{IEEEbiographynophoto}{Ziv Goldfeld}
(S'13) received his B.Sc.\@ (summa cum laude) and M.Sc.\@ (summa cum laude) degrees in Electrical and Computer Engineering from the Ben-Gurion University, Israel, in 2012 and 2014, respectively. He is currently a 
student in the direct Ph.D. program for honor students in Electrical and Computer Engineering at that same institution.

Between 2003 and 2006, he served in the intelligence corps of the Israeli Defense Forces.

Ziv is a recipient of several awards, among them are the Dean's List Award, the Basor Fellowship, the Lev-Zion fellowship, IEEEI-2014 best student paper award, a Minerva Short-Term Research Grant (MRG), and a Feder Family Award in the national student contest for outstanding research work in the field of communications technology.
\end{IEEEbiographynophoto}

\begin{IEEEbiographynophoto}{Gerhard Kramer}
(S'91-M'94-SM'08-F'10) received the Dr. sc. techn. (Doktor der technischen Wissenschaften) degree from the Swiss Federal Institute of Technology (ETH), Zurich, in 1998.

From 1998 to 2000, he was with Endora Tech AG, Basel, Switzerland, as a Communications Engineering Consultant. From 2000 to 2008, he was with Bell Labs, Alcatel-Lucent, Murray Hill, NJ, as a Member of Technical Staff. He joined the University of Southern California (USC), Los Angeles, in 2009. Since 2010, he has been a Professor and Head of the Institute for Communications Engineering at the Technical University of Munich (TUM), Munich, Germany.

Dr. Kramer served as the 2013 President of the IEEE Information Theory Society. He has won several awards for his work and teaching, including an Alexander von Humboldt Professorship in 2010 and a Lecturer Award from the Student Association of the TUM Electrical and Computer Engineering Department in 2015. He has been a member of the Bavarian Academy of Sciences and Humanities since 2015.
\end{IEEEbiographynophoto}

\begin{IEEEbiographynophoto}{Haim H. Permuter}
(M'08-SM'13) received his B.Sc.\@ (summa cum laude) and M.Sc.\@ (summa cum laude) degrees in Electrical and Computer Engineering from the Ben-Gurion University, Israel, in 1997 and 2003, respectively, and the Ph.D. degree in Electrical Engineering from Stanford University, California in 2008.

Between 1997 and 2004, he was an officer at a research and development unit of the Israeli Defense Forces. Since 2009 he is with the department of Electrical and Computer Engineering at Ben-Gurion University where he is currently an associate professor.

Prof. Permuter is a recipient of several awards, among them the Fullbright Fellowship, the Stanford Graduate Fellowship (SGF), Allon Fellowship, and and the U.S.-Israel Binational Science Foundation Bergmann Memorial Award. Haim is currently serving on the editorial board of the IEEE Transactions on Information Theory.
\end{IEEEbiographynophoto}
\end{document}